\documentclass[12pt]{article}
\usepackage{amsmath,amsthm}
\usepackage{eufrak}
\usepackage[pdftex]{graphicx}

\newcommand{\ar}{\renewcommand{\arraystretch}{1}} 
\DeclareMathAlphabet{\bb}{U}{msb}{m}{n} \gdef\C{\bb C}    \gdef\dS{\bb S} \gdef\R{\bb R}
\gdef\K{\bb K} \gdef\BH{\bb H} \gdef\F{\bb F} 

 \DeclareMathOperator{\spin}{{\bf
Spin}} \DeclareMathOperator{\pin}{{\bf Pin}}

\DeclareMathOperator{\Sym}{Sym}

 \DeclareMathOperator{\SL}{SL}
\DeclareMathOperator{\SO}{SO}\DeclareMathOperator{\SU}{SU}
\DeclareMathOperator{\Sp}{Sp}

\newcommand{\scr}{\scriptstyle}

\newcommand{\cA}{\mathcal{A}}

\newcommand{\cP}{{\cal P}}

\newcommand{\sA}{{\sf A}}
\newcommand{\sB}{{\sf B}}

\newcommand{\sJ}{{\sf J}}

\newcommand{\sH}{{\sf H}}
\newcommand{\sT}{{\sf T}}

\newcommand{\sV}{{\sf V}}
\newcommand{\sX}{{\sf X}}
\newcommand{\sY}{{\sf Y}}
\newcommand{\sK}{{\sf K}}

\newcommand{\bsH}{{\boldsymbol{\sf H}}}
\newcommand{\bsJ}{{\boldsymbol{\sf J}}}
\newcommand{\bsK}{{\boldsymbol{\sf K}}}
\newcommand{\bsL}{{\boldsymbol{\sf L}}}
\newcommand{\bsP}{{\boldsymbol{\sf P}}}
\newcommand{\bsQ}{{\boldsymbol{\sf Q}}}
\newcommand{\bsR}{{\boldsymbol{\sf R}}}
\newcommand{\bsS}{{\boldsymbol{\sf S}}}
\newcommand{\bsT}{{\boldsymbol{\sf T}}}
\newcommand{\bsU}{{\boldsymbol{\sf U}}}
\newcommand{\bsZ}{{\boldsymbol{\sf Z}}}

\newcommand{\fA}{\mathfrak{A}}

\newcommand{\fR}{\mathfrak{R}}

\newcommand{\fF}{\mathfrak{F}}

\newcommand{\fP}{\mathfrak{P}}

\newcommand{\fg}{\mathfrak{g}}

\newcommand{\cl}{C\kern -0.2em \ell}

\newcommand{\e}{\mbox{\bf e}}

\newcommand{\ld}{\left[}
\newcommand{\rd}{\right]}

\begin{document}
\title{Group Theoretical Description of Mendeleev Periodic System}
\author{V.~V. Varlamov\thanks{Siberian State Industrial University,
Kirova 42, Novokuznetsk 654007, Russia, e-mail:
varlamov@sibsiu.ru}}
\date{}
\maketitle
\begin{abstract}
The group theoretical description of the periodic system of elements in the framework of the Rumer-Fet model is considered. We introduce the concept of a single quantum system, the generating core of which is an abstract $C^\ast$-algebra. It is shown that various concrete implementations of the operator algebra depend on the structure of generators of the fundamental symmetry group attached to the energy operator. In the case of generators of the complex shell of a group algebra of a conformal group, the spectrum of states of a single quantum system is given in the framework of the basic representation of the Rumer-Fet group, which leads to a group-theoretic interpretation of the Mendeleev's periodic system of elements. A mass formula is introduced that allows to give the termwise mass splitting for the main multiplet of the Rumer-Fet group. The masses of elements of the Seaborg table (eight-periodic extension of the Mendeleev table) are calculated starting from the atomic number $Z=121$ to $Z=220$. The continuation of Seaborg homology between lanthanides and actinides is established to the group of superactinides. A 10-periodic extension of the periodic table is introduced in the framework of the group-theoretic approach. The multiplet structure of the extended table periods is considered in detail. It is shown that the period lengths of the system of elements are determined by the structure of the basic representation of the Rumer-Fet group. The theoretical masses of the elements of 10th and 11th periods are calculated starting from $Z=221$ to $Z=364$. The concept of hypertwistor is introduced.
\end{abstract}

\section{Introduction}
In 2019 marks the 150th anniversary of the discovery of the periodic law of chemical elements by Dmitry Ivanovich Mendeleev. Mendeleev's periodic table shed light on a huge number of experimental facts and allowed to predict the existence and basic properties of new, previously unknown, elements. However, the reasons (more precisely, root causes) for periodicity, in particular, the reasons for the periodic recurrence of similar electronic configurations of atoms, are still not clear to the end. Also, the limits of applicability of the periodic law have not yet been delineated -- the controversy regarding the specifics of nuclear and electronic properties of atoms of heavy elements continues.

The now generally accepted structure of the periodic system, based on the Bohr model, proceeds from the fact that the arrangement of elements in the system with increasing their atomic numbers is uniquely determined by the individual features of the electronic structure of atoms described in the framework of one-electronic approximation (Hartree method), and directly reflects the energy sequence of atomic orbitals of $s$-, $p$-, $d$-, $f$-shells populated by electrons with increasing their total number as the charge of the nucleus of the atom increases in accordance with the principle of minimum energy. However, this is only possible in the simplest version of Hartree approximation, but in the variant of the Hartree-Fock approximation total energy of an atom is not equal to the sum of orbital energies, and electron configuration of an atom is determined by the minimum of its full energy. As noted in the book \cite{KK05}, the traditional interpretation of the structure of the periodic system on the basis of the sequence of filling of electronic atomic orbitals in accordance with their relative energies $\varepsilon_{nl}$ is very approximate, has, of course, a number of drawbacks and has narrow limits of applicability. There is no universal sequence of orbital energies $\varepsilon_{nl}$, moreover, such a sequence does not completely determine the order of the atomic orbitals settling by electrons, since it is necessary to take into account the configuration interactions (superposition of configurations in the multi-configuration approximation). And, of course, periodicity is not only and not completely orbital-energy effects. The reason for the repetition of similar electronic configurations of atoms in their ground states escapes us, and within one-electronic approximation can hardly be revealed at all. Moreover, it is possible that the theory of periodicity in general awaits a fate somewhat reminiscent of the fate of the theory of planetary retrogressions in the Ptolemaic system after the creation of the Copernican system. It is quite possible that what we call the periodicity principle is the result of the nonspatial symmetries of the atom (permutation and dynamical symmetries).

In 1971 academician V.A. Fock in his work \cite{Fock71} put the main question for the doctrine of the principle of periodicity and the theory of the periodic system: ``Do the properties of atoms and their constituent parts fit into the framework of purely spatial representations, or do we need to somehow expand the concepts of space and spatial symmetry to accommodate the inherent degrees of freedom of atoms and their constituent parts?'' \cite[p.~108]{Fock71}. As is known, Bohr's model in its original formulation uses quantum numbers relating to electrons in a field with spherical symmetry, which allowed Bohr to introduce the concept of closed electron shells and bring this concept closer to the periods of the Mendeleev's table. Despite this success, the problem of explaining the periodic system was far from solved. Moreover, for all the depth and radicality of these new ideas, they still fit into the framework of conventional spatial representations. A further important step was associated with the discovery of the internal, not spatial, degree of freedom of the electron -- spin, which is not a mechanical concept. The discovery of spin is closely related to the discovery of Pauli principle, which was formulated before quantum mechanics as requiring that each orbit, characterized by certain quantum numbers, contains no more than two electrons. At the end of the article \cite{Fock71} Fock himself answers his own question: ``Purely spatial degrees of freedom of the electron is not enough to describe the properties of the electron shell of the atom and need to go beyond purely spatial concepts to express the laws that underlie the periodic system. The new degree of freedom of the electron -- its spin -- allows us to describe the properties of physical systems that are alien to classical concepts. This internal degree of electron freedom is essential for the formulation of the properties of multi-electron systems, thus for the theoretical justification of the Mendeleev's periodic system'' \cite[c.~116]{Fock71}.

In this paper, we consider the group theoretical description of the periodic system in the framework of the Rumer-Fet model. Unlike Bohr's model, in which spatial and internal (spin) symmetries are combined on the basis of a classical composite system borrowed from celestial mechanics\footnote{It is obvious that the visual spatial image used in Bohr's model is a vestige of classical representations. So, in the middle of 19th century, numerous attempts were made to build mechanical models of electromagnetic phenomena, even Maxwell's treatise contains a large number of mechanical analogies. As time has shown, all mechanical models of electromagnetism turned out to be nothing more than auxiliary scaffolding, which were later discarded as unnecessary.}, the Rumer-Fet group $G$ describes non-spatial symmetries\footnote{The group $G$ also contains the Lorentz group (rotation group of the Minkowski space-time) as a subgroup.}. Moreover, the Rumer-Fet model is entirely based on the mathematical apparatus of quantum mechanics and group theory without involving any classical analogies, such as the concept of a composite system. The concept of a composite system, which directly follows from the principle of separability (the basic principle of reductionism), is known to have limited application in quantum mechanics, since in the microcosm, in contrast to the composite structure of the macrocosm, the superposition structure preveals. Heisenberg argued that the concept of ``consists of'' does not work in particle physics. On the other hand, the problem of ``critical'' elements of Bohr's model is also a consequence of visual spatial representations. Feynmann's solution, representing the atomic nucleus as a point, leads to the Klein paradox for the element \textbf{Uts} (Untriseptium) with atomic number $Z=137$. Another spatial image, used in the Greiner-Reinhardt solution, represents the atomic nucleus as a charged ball, resulting in a loss of electroneutrality for atoms above the value $Z=173$ (see section 6).

The most important characteristic feature of the Rumer-Fet model is the representation of the periodic table of elements as a single quantum system. While Bohr's model considers an atom of one element as a separate quantum system (and the atomic number is included in the theory as a parameter, so that there are as many quantum systems as there are elements), in the Rumer-Fet model atoms of various elements are considered as states of a single quantum system, connected to each other by the action of the symmetry group. A peculiar feature of the Rumer-Fet model is that it ``ignores'' the atomic structure underlying Bohr's model. In contrast to Bohr's model, which represents each atom as a composite aggregate of protons, neutrons and electrons, the Rumer-Fet model is distracted from the internal structure of each single atom, presenting the entire set of elements of the periodic table as a single quantum system of structureless states\footnote{The notion of the atom as a ``structureless'' state does not mean that there is no structure at all behind the concept. This only means that this structure is of a different order, not imported from the outside, from the ``repertoire of classical physics'', but a structure that naturally follows from the mathematical apparatus of quantum mechanics (state vectors, symmetry group, Hilbert space, tensor products of Hilbert ($\K$-Hilbert) spaces, and so on).}. In this paper, the single quantum system $\bsU$ is defined by a $C^\ast$-algebra consisting of the \textit{energy operator} $H$ and the generators of the \textit{fundamental symmetry group} $G_f$ attached to $H$. The states of the system $\bsU$ are formed within the framework of the Gelfand-Naimark-Segal construction (GNS) \cite{GN43,Seg47}, that is, as cyclic representations of the operator algebra. Due to the generality of the task of the system $\bsU$ and the flexibility of the GNS-construction for the each particular implementation of the operator algebra (the so-called ``dressing'' of the $C^\ast$-algebra), we obtain our (corresponding to this implementation) spectrum of states of the system $\bsU$\footnote{Thus, in the case when the generators of the fundamental symmetry group ($G_f=\SO_0(1,3)$ is the Lorentz group) attached to $H$ are generators of the complex shell of the group algebra $\mathfrak{sl}(2,\C)$ (see Appendix A), we obtain a linearly growing spectrum of state masses (``elementary particles'') \cite{Var17}. In this case, the ``dressing'' of the operator algebra and the construction of the cyclic representations of the GNS-construction are carried out in the framework of spinor structure (charged, neutral, truly neutral (Majorana) states and their discrete symmetries set through morphisms of the spinor structure, see \cite{Var99,Var01,Var05,Var11,Var15}). In \cite{Var17} it is shown that the masses of ``elementary particles'' are multiples of the mass of the electron with an accuracy of 0.41\%. Here there is a direct analogy with the electric charge. Any electric charge is a multiple of the charge of the electron, and a multiple of exactly. If any electric charge is absolutely a multiple of the electron charge, then in the case of masses this multiplicity takes place with an accuracy of 0.41\% (on average).}. In section 2 of this article, a conformal group is considered as a fundamental symmetry group ($G_f=\SO(2,4)$). In this case, the concrete implementation of the operator $C^\ast$-algebra is given by means of the generators of the complex shell of the group algebra $\mathfrak{so}(2,4)$ attached to $H$ and the twistor structure associated with the group $\SU(2,2)$ (the double covering of the conformal group). The complex shell of the algebra $\mathfrak{so}(2,4)$ leads to a representation $F^+_{ss^\prime}$ of the Rumer-Fet group, within which a group theoretical description of the periodic system of elements is given (section 3--4). At this point, atoms are considered as states (discrete stationary states) of the \textit{matter spectrum}\footnote{A term introduced by Heisenberg in the book \cite{Hesen4} with reference to particle physics.}, each atom is given by a state vector of the physical Hilbert space, in which a symmetry group acts, translating some state vectors into others (that is, a group that specifies quantum transitions between elements of the periodic system). In the section 6, Seaborg table (eight-periodic extension of the Mendeleev table) is formulated in the framework of the basic representation of the Rumer-Fet group for two different group chains, which specify the split of the main multiplet into smaller multiplets. It also calculates the average mass of the multiplets included in the Seaborg table (in addition to those multiplets that belong to the Mendeleev table with exception of elements \textbf{Uue} and \textbf{Ubn}). In subsection 6.1 the mass formula is introduced to allow a termwise mass splitting for the basic representation of the Rumer-Fet group. The masses of elements are calculated starting from the atomic number $Z=121$ to $Z=220$. In section 7 the 10-periodic extension of the Mendeleev table is studied. The multiplet structure of the extended table is considered in detail. It is shown that the period lengths of the system of elements are determined by the structure of the basic representation of the Rumer-Fet group. The theoretical masses of the elements of the 10th and 11th periods are calculated. In section 8 quantum transitions between state vectors of the physical Hilbert space, formed by the set of elements of periodic system, are considered.

It is possible to imagine an electron in any way: whether as a point (particle or wave), or a charged ball, or as an electron claud on an atomic orbital, all these mental images only obscure the essence of the matter, because they remain within the framework of visual spatial representations. However, there is a mathematical structure that is far from visualization, and yet accurately describes the electron: it is a two-component spinor, the vector of the fundamental representation of the double covering $\SL(2,\C)\simeq\spin_+(1,3)$ of the Lorentz group. Similarly, apart from any visual representations of the atom, it can be argued that the meaning is only the mathematical structure, which is directly derived from the symmetry group of the periodic system. In the section 9 it is shown that such structure is a hypertwistor acting in the $\K$-Hilbert space $\bsH_8\otimes\bsH_\infty$.

Bohr's model does not explain the periodicity, but only approximates it within the framework of one-electronic Hartree approximation. Apparently, the explanation of the periodic law lies on the path indicated by academician Fock, that is, it is necessary to go beyond the classical (space-time) representations in the description of the periodic system of elements. It is obvious that the most suitable scheme of description in this way is the group-theoretic approach.

\section{Single Quantum System}
As already noted in the introduction, the starting point of the construction of the group theoretical description of the periodic system of elements is the concept of a single quantum system $\bsU$. Following Heisenberg, we assume that at the fundamental level, the definition of the system $\bsU$ is based on two concepts: \textit{energy} and \textit{symmetry}. Let us define a single quantum system $\bsU$ by means of the following axiom system:

\textbf{A.I} (Energy and fundamental symmetry) \textit{A single quantum system $\bsU$ at the fundamental level is characterized by a $C^\ast$-algebra $\fA$ with a unit consisting of the energy operator $H$ and the generators of the fundamental symmetry group $G_f$ attached to $H$, forming a common system of eigenfunctions with $H$}.

\textbf{A.II} (States and GNS construction) \textit{The physical state of a $C^\ast$-algebra $\fA$ is determined by the cyclic vector $\left|\Phi\right\rangle$ of the representation $\pi$ of a $C^\ast$-algebra in a separable Hilbert space $\sH_\infty$:
\[
\omega_\Phi(H)=\frac{\langle\Phi\mid\pi(H)\Phi\rangle}{\langle\Phi\mid\Phi\rangle}.
\]
The set $PS(\fA)$ of all pure states of a $C^\ast$-algebra $\fA$ coincides with the set of all states $\omega_\Phi(H)$ associated with all irreducible cyclic representations $\pi$ of an algebra $\fA$, $\left|\Phi\right\rangle\in\sH_\infty$ (Gelfand-Naimark-Segal construction)}.

\textbf{A.III} (Physical Hilbert space) \textit{The set of all pure states $\omega_\Phi(H)$ under the condition $\omega_\Phi(H)\geq 0$ forms a physical Hilbert space $\bsH_{\rm phys}$ (in general, the space $\bsH_{\rm phys}$ is nonseparable). For each state vector $\left|\Psi\right\rangle\in\bsH_{\rm phys}$ there is a unit ray $\boldsymbol{\Psi}=e^{i\alpha}\left|\Psi\right\rangle$, where $\alpha$ runs through all real numbers and $\sqrt{\left\langle\Psi\right.\!\left|\Psi\right\rangle}=1$. The ray space is a quotient-space $\hat{H}=\bsH_{\rm phys}/S^1$, that is, the projective space of one-dimensional subspaces of $\bsH_{\rm phys}$. All states of a single quantum system $\bsU$ are described by the unit rays}.

\textbf{A.IV} (Axiom of spectrality) \textit{In $\hat{H}$ there is a complete system of states with non-negative energy}.

\textbf{A.V} (Superposition principle) \textit{The basic correspondence between physical states and elements of space $\hat{H}$ involves the superposition principle of quantum theory, that is, there is a set of basic states such that arbitrary states can be constructed from them using linear superpositions}.


We choose a conformal group as the fundamental symmetry. The conformal group occurs in modern physics in a wide variety of situations and is essentially as universal as the Lorentz group, there are many relativistic theories and, similarly, conformal ones\footnote{Moreover, as Segal showed \cite{Seg51}, the Lie algebra of an inhomogeneous Lorentz group (that is, a Poincar\'{e} group) can be obtained by deformation from a conformal Lie algebra. In turn, the conformal Lie algebra is ``rigid'', that is, cannot be obtained by deforming another Lie algebra. Because of this property, the conformal algebra (the algebra of a non-compact real pseudo-orthogonal group in a six-dimensional space with signature $(-,-,-,-,+,+)$) has a unique (complete) character and occupies a special place among other algebras.}.

\section{Rumer-Fet group}
In the previous section we define conformal group as the fundamental symmetry. The next logical step is to construct a concrete implementation of the operator algebra. We begin this construction by defining a complex shell of a group algebra $\mathfrak{so}(2,4)$ that leads to a basic representation $F^+_{ss^\prime}$ of the Rumer-Fet group\footnote{The first work in this direction is \cite{RF71}, where it was noted a striking similarity between the structure of the system of chemical elements and the structure of the energy spectrum of the hydrogen atom. This similarity is explained in \cite{RF71} within the framework of the Fock representation $F$ \cite{Fock35} for the group $\spin(4)$ (the double covering of the group $\SO(4)$). However, the main drawback of the description in \cite{RF71} is the reducibility of the representation $F$, which did not allow to consider the system as ``elementary'' in the sense of group mechanics. In 1972, Konopelchenko \cite{Kon72} extended the Fock representation $F$ to the representation $F^+$ of the conformal group, thus eliminating the above drawback. Further, based on the connection with the Madelung numbering, Fet defines the $F^+_{s}$ and $F^+_{ss^\prime}$ representations (to define the $F^+_{ss^\prime}$ representation, Madelung's ``lexicographic rule'' had to be changed). After a rather long period of oblivion (totally undeserved), interest in the Rumer-Fet model will resume (see Kibler \cite{Cib}).}.

As is known \cite{Fet}, a system of fifteen generators of the conformal group $\SO_0(2,4)$ satisfies the following commutativity relations:
\[
\left[\bsL_{\alpha\beta},\bsL_{\gamma\delta}\right]=i\left(g_{\alpha\delta}\bsL_{\beta\gamma}+g_{\beta\gamma}\bsL_{\alpha\delta}
-g_{\alpha\gamma}\bsL_{\beta\delta}-g_{\beta\delta}\bsL_{\alpha\gamma}\right),
\]
\[
(\alpha,\beta,\gamma,\delta=1,\ldots,6,\alpha\neq\beta,\;\gamma\neq\delta).
\]
Generators $\bsL_{\alpha\beta}$ form a basis of the group algebra $\mathfrak{so}(2,4)$. In order to move to the complex shell of the algebra $\mathfrak{so}(2,4)$, consider another system of generators proposed by Tsu Yao \cite{Yao,Fet}. Let
\[
\bsJ_1=1/2\left(\bsL_{23}-\bsL_{14}\right),\quad\bsJ_2=1/2\left(\bsL_{31}-\bsL_{24}\right),\quad
\bsJ_3=1/2\left(\bsL_{12}-\bsL_{34}\right),
\]
\[
\bsK_1=1/2\left(\bsL_{23}+\bsL_{14}\right),\quad\bsK_2=1/2\left(\bsL_{31}+\bsL_{24}\right),\quad
\bsK_3=1/2\left(\bsL_{12}+\bsL_{34}\right),
\]
\[
\bsP_1=1/2\left(-\bsL_{35}-\bsL_{16}\right),\quad\bsP_2=1/2\left(\bsL_{45}-\bsL_{36}\right),\quad
\bsP_0=1/2\left(-\bsL_{34}-\bsL_{56}\right),
\]
\[
\bsQ_1=1/2\left(\bsL_{35}-\bsL_{46}\right),\quad\bsQ_2=1/2\left(\bsL_{45}+\bsL_{36}\right),\quad
\bsQ_0=1/2\left(\bsL_{34}-\bsL_{56}\right),
\]
\[
\bsS_1=1/2\left(-\bsL_{15}+\bsL_{26}\right),\quad\bsS_2=1/2\left(-\bsL_{25}-\bsL_{16}\right),\quad
\bsS_0=1/2\left(\bsL_{12}-\bsL_{56}\right),
\]
\begin{equation}\label{Yao}
\bsT_1=1/2\left(-\bsL_{15}-\bsL_{26}\right),\quad\bsT_2=1/2\left(\bsL_{25}-\bsL_{16}\right),\quad
\bsT_0=1/2\left(-\bsL_{12}-\bsL_{56}\right).
\end{equation}
This system of eighteen generators is tied by the three relations
\begin{equation}\label{Yao2}
\bsJ_3-\bsK_3=\bsP_0-\bsQ_0,\quad\bsJ_3+\bsK_3=\bsS_0-\bsT_0,\quad\bsP_0+\bsQ_0=\bsS_0+\bsT_0.
\end{equation}
In virtue of independence of the generators $\bsL_{\alpha\beta}$ ($\alpha<\beta$), the system (\ref{Yao}) defines a surplus system of generators of $\SO_0(2,4)$, from which we can obtain the basis of $\mathfrak{so}(2,4)$ excluding three generators by means of (\ref{Yao2}).

Introducing the generators
\[
\bsJ_\pm=\bsJ_1\pm i\bsJ_2,\quad\bsP_\pm=\bsP_1\pm i\bsP_2,\quad\bsS_\pm=\bsS_1\pm i\bsS_2,
\]
\begin{equation}\label{Envelope}
\bsK_\pm=\bsK_1\pm i\bsK_2,\quad\bsQ_\pm=\bsQ_1\pm i\bsQ_2,\quad\bsT_\pm=\bsT_1\pm i\bsT_2,
\end{equation}
we come to the complex shell of the algebra $\mathfrak{so}(2,4)$.

Then
\[
\left[\bsJ_3,\bsJ_+\right]=\bsJ_+,\quad\left[\bsJ_3,\bsJ_-\right]=-\bsJ_-,\quad\left[\bsJ_+,\bsJ_-\right]=2\bsJ_3,
\]
\[
\left[\bsK_3,\bsK_+\right]=\bsK_+,\quad\left[\bsK_3,\bsK_-\right]=-\bsK_-,\quad\left[\bsK_+,\bsK_-\right]=2\bsK_3,
\]
\[
\left[\bsP_0,\bsP_+\right]=\bsP_+,\quad\left[\bsP_0,\bsP_-\right]=-\bsP_-,\quad\left[\bsP_+,\bsP_-\right]=-2\bsP_0,
\]
\[
\left[\bsQ_0,\bsQ_+\right]=\bsQ_+,\quad\left[\bsQ_0,\bsQ_-\right]=-\bsQ_-,\quad\left[\bsQ_+,\bsQ_-\right]=-2\bsQ_0,
\]
\[
\left[\bsS_0,\bsS_+\right]=\bsS_+,\quad\left[\bsS_0,\bsS_-\right]=-\bsS_-,\quad\left[\bsS_+,\bsS_-\right]=-2\bsS_0,
\]
\[
\left[\bsT_0,\bsT_+\right]=\bsT_+,\quad\left[\bsT_0,\bsT_-\right]=-\bsT_-,\quad\left[\bsT_+,\bsT_-\right]=-2\bsT_0,
\]
\[
\left[\bsJ_i,\bsK_j\right]=0\quad(i,j=+,-,3),
\]
\[
\left[\bsJ_+,\bsP_+\right]=0,\quad\left[\bsJ_+,\bsP_-\right]=-\bsT_-,\quad\left[\bsJ_+,\bsP_0\right]=-1/2\bsJ_+,
\]
\[
\left[\bsJ_-,\bsP_+\right]=\bsT_+,\quad\left[\bsJ_-,\bsP_-\right]=0,\quad\left[\bsJ_-,\bsP_0\right]=1/2\bsJ_-,
\]
\[
\left[\bsJ_3,\bsP_+\right]=1/2\bsP_+,\quad\left[\bsJ_3,\bsP_-\right]=-1/2\bsP_-,\quad
\left[\bsJ_3,\bsP_0\right]=0,
\]
\[
\left[\bsJ_+,\bsQ_+\right]=\bsS_+,\quad\left[\bsJ_+,\bsQ_-\right]=0,\quad\left[\bsJ_+,\bsQ_0\right]=1/2\bsJ_+,
\]
\[
\left[\bsJ_-,\bsQ_+\right]=0,\quad\left[\bsJ_-,\bsQ_-\right]=-\bsS_-,\quad\left[\bsJ_-,\bsQ_0\right]=-1/2\bsJ_-,
\]
\[
\left[\bsJ_3,\bsQ_+\right]=-1/2\bsQ_+,\quad\left[\bsJ_3,\bsQ_-\right]=1/2\bsQ_-,
\quad\left[\bsJ_3,\bsQ_0\right]=0,
\]
\[
\left[\bsJ_+,\bsS_+\right]=0,\quad\left[\bsJ_+,\bsS_-\right]=-\bsQ_-,\quad\left[\bsJ_+,\bsS_0\right]=-1/2\bsJ_+,
\]
\[
\left[\bsJ_-,\bsS_+\right]=\bsQ_+,\quad\left[\bsJ_-,\bsS_-\right]=0,\quad\left[\bsJ_-,\bsS_0\right]=1/2\bsJ_-,
\]
\[
\left[\bsJ_3,\bsS_+\right]=1/2\bsS_+,\quad\left[\bsJ_3,\bsQ_-\right]=-1/2\bsS_-,
\quad\left[\bsJ_3,\bsS_0\right]=0,
\]
\[
\left[\bsJ_+,\bsT_+\right]=\bsP_+,\quad\left[\bsJ_+,\bsT_-\right]=0,\quad\left[\bsJ_+,\bsT_0\right]=1/2\bsJ_+,
\]
\[
\left[\bsJ_-,\bsT_+\right]=0,\quad\left[\bsJ_-,\bsT_-\right]=-\bsP_-,\quad\left[\bsJ_-,\bsT_0\right]=-1/2\bsJ_-,
\]
\[
\left[\bsJ_3,\bsT_+\right]=-1/2\bsT_+,\quad\left[\bsJ_3,\bsT_-\right]=1/2\bsT_-,
\quad\left[\bsJ_3,\bsT_0\right]=0,
\]
\[
\left[\bsK_+,\bsP_+\right]=-\bsS_+,\quad\left[\bsK_+,\bsP_-\right]=0,\quad\left[\bsK_+,\bsP_0\right]=1/2\bsK_+,
\]
\[
\left[\bsK_-,\bsP_+\right]=0,\quad\left[\bsK_-,\bsP_-\right]=\bsS_-,\quad\left[\bsK_-,\bsP_0\right]=-1/2\bsK_-,
\]
\[
\left[\bsK_3,\bsP_+\right]=-1/2\bsP_+,\quad\left[\bsK_3,\bsP_-\right]=1/2\bsP_-,
\quad\left[\bsK_3,\bsP_0\right]=0,
\]
\[
\left[\bsK_+,\bsQ_+\right]=0,\quad\left[\bsK_+,\bsQ_-\right]=\bsT_-,\quad\left[\bsK_+,\bsQ_0\right]=-1/2\bsK_+,
\]
\[
\left[\bsK_-,\bsQ_+\right]=-\bsT_+,\quad\left[\bsK_-,\bsQ_-\right]=0,\quad\left[\bsK_-,\bsQ_0\right]=1/2\bsK_-,
\]
\[
\left[\bsK_3,\bsQ_+\right]=1/2\bsQ_+,\quad\left[\bsK_3,\bsQ_-\right]=-1/2\bsQ_-,
\quad\left[\bsK_3,\bsQ_0\right]=0,
\]
\[
\left[\bsK_+,\bsS_+\right]=0,\quad\left[\bsK_+,\bsS_-\right]=\bsP_-,\quad\left[\bsK_+,\bsS_0\right]=-1/2\bsK_+,
\]
\[
\left[\bsK_-,\bsS_+\right]=-\bsP_+,\quad\left[\bsK_-,\bsS_-\right]=0,\quad\left[\bsK_-,\bsS_0\right]=1/2\bsK_-,
\]
\[
\left[\bsK_3,\bsS_+\right]=1/2\bsS_+,\quad\left[\bsK_3,\bsS_-\right]=-1/2\bsS_-,
\quad\left[\bsK_3,\bsS_0\right]=0,
\]
\[
\left[\bsK_+,\bsT_+\right]=-\bsQ_+,\quad\left[\bsK_+,\bsT_-\right]=0,\quad\left[\bsK_+,\bsT_0\right]=1/2\bsK_+,
\]
\[
\left[\bsK_-,\bsT_+\right]=0,\quad\left[\bsK_-,\bsT_-\right]=\bsQ_-,\quad\left[\bsK_-,\bsT_0\right]=-1/2\bsK_-,
\]
\[
\left[\bsK_3,\bsT_+\right]=-1/2\bsT_+,\quad\left[\bsK_3,\bsT_-\right]=1/2\bsT_-,
\quad\left[\bsK_3,\bsT_0\right]=0,
\]
\[
\left[\bsP_i,\bsQ_j\right]=0\quad(i,j=+,-,0),
\]
\[
\left[\bsP_+,\bsS_+\right]=0,\quad\left[\bsP_+,\bsS_-\right]=\bsK_-,\quad\left[\bsP_+,\bsS_0\right]=-1/2\bsP_+,
\]
\[
\left[\bsP_-,\bsS_+\right]=-\bsK_+,\quad\left[\bsP_-,\bsS_-\right]=0,\quad\left[\bsP_-,\bsS_0\right]=1/2\bsP_-,
\]
\[
\left[\bsP_0,\bsS_+\right]=1/2\bsS_+,\quad\left[\bsP_0,\bsS_-\right]=-1/2\bsS_-,
\quad\left[\bsP_0,\bsS_0\right]=0,
\]
\[
\left[\bsP_+,\bsT_+\right]=0,\quad\left[\bsP_+,\bsT_-\right]=-\bsJ_+,\quad\left[\bsP_+,\bsT_0\right]=-1/2\bsP_+,
\]
\[
\left[\bsP_-,\bsT_+\right]=\bsJ_-,\quad\left[\bsP_-,\bsT_-\right]=0,\quad\left[\bsP_-,\bsT_0\right]=1/2\bsP_-,
\]
\[
\left[\bsP_0,\bsT_+\right]=1/2\bsT_+,\quad\left[\bsP_0,\bsT_-\right]=-1/2\bsT_-,
\quad\left[\bsP_0,\bsT_0\right]=0,
\]
\[
\left[\bsQ_+,\bsS_+\right]=0,\quad\left[\bsQ_+,\bsS_-\right]=-\bsJ_-,\quad\left[\bsQ_+,\bsS_0\right]=-1/2\bsQ_+,
\]
\[
\left[\bsQ_-,\bsS_+\right]=\bsJ_+,\quad\left[\bsQ_-,\bsS_-\right]=0,\quad\left[\bsQ_-,\bsS_0\right]=1/2\bsQ_-,
\]
\[
\left[\bsQ_0,\bsS_+\right]=1/2\bsS_+,\quad\left[\bsQ_0,\bsS_-\right]=-1/2\bsS_-,
\quad\left[\bsQ_0,\bsS_0\right]=0,
\]
\[
\left[\bsQ_+,\bsT_+\right]=0,\quad\left[\bsQ_+,\bsT_-\right]=\bsK_+,\quad\left[\bsQ_+,\bsT_0\right]=-1/2\bsQ_+,
\]
\[
\left[\bsQ_-,\bsT_+\right]=-\bsK_-,\quad\left[\bsQ_-,\bsT_-\right]=0,\quad\left[\bsQ_-,\bsT_0\right]=1/2\bsQ_-,
\]
\[
\left[\bsQ_0,\bsT_+\right]=1/2\bsT_+,\quad\left[\bsQ_0,\bsT_-\right]=-1/2\bsT_-,
\quad\left[\bsQ_0,\bsT_0\right]=0,
\]
\[
\left[\bsS_i,\bsT_j\right]=0\quad(i,j=+,-,0).
\]

Let us consider a special representation of the conformal group $\SO_0(2,4)$, which is analogous to the van der Waerden representation (\ref{Waerden}) for the Lorentz group $\SO_0(1,3)$ (see Appendix A). This \textit{local} representation of $\SO_0(2,4)$ is related immediately with the Fock representation for the group $\SO(4)$ (see Appendix B). In essence this representation is an extension of the Fock representation for $\SO(4)$ to the unitary representation of the conformal group $\SO_0(2,4)$ in the \textit{Fock space} $\fF$ by the basis (\ref{Fock2}). Using generators (\ref{Envelope}) of the complex shell of the algebra $\mathfrak{so}(2,4)$, we obtain
\begin{eqnarray}
&&\bsJ_-|j,\sigma,\tau\rangle= \sqrt{(j+\sigma)(j-\sigma+1)}|j,\sigma-1,\tau\rangle,\nonumber\\
&&\bsJ_+|j,\sigma,\tau\rangle= \sqrt{(j-\sigma)(j+\sigma+1)}|j,\sigma+1,\tau\rangle,\nonumber\\
&&\bsJ_3|j,\sigma,\tau\rangle=\sigma|j,\sigma,\tau\rangle,\nonumber\\
&&\bsK_-|j,\sigma,\tau\rangle= \sqrt{(j+\tau)(j-\tau+1)}|j,\sigma,\tau-1\rangle,\nonumber\\
&&\bsK_+|j,\sigma,\tau\rangle= \sqrt{(j-\tau)(j+\tau+1)}|j,\sigma,\tau+1\rangle,\nonumber\\
&&\bsK_3|j,\sigma,\tau\rangle=\tau|j,\sigma,\tau\rangle,\nonumber
\end{eqnarray}
\begin{eqnarray}
&&\bsP_-|j,\sigma,\tau\rangle= -i\sqrt{(j+\sigma)(j-\tau)}\left|j-\frac{1}{2},\sigma-\frac{1}{2},\tau+\frac{1}{2}\right\rangle,\nonumber\\
&&\bsP_+|j,\sigma,\tau\rangle= i\sqrt{(j-\tau+1)(j+\sigma+1)}\left|j+\frac{1}{2},\sigma+\frac{1}{2},\tau-\frac{1}{2}\right\rangle,\nonumber\\
&&\bsP_0|j,\sigma,\tau\rangle=\left(j+\frac{\sigma-\tau+1}{2}\right)\sigma\left|j,\sigma,\tau\right\rangle,\nonumber\\
&&\bsQ_-|j,\sigma,\tau\rangle= i\sqrt{(j+\tau)(j-\sigma)}\left|j-\frac{1}{2},\sigma+\frac{1}{2},\tau-\frac{1}{2}\right\rangle,\label{Fock3}\\
&&\bsQ_+|j,\sigma,\tau\rangle= -i\sqrt{(j-\sigma+1)(j+\tau+1)}\left|j+\frac{1}{2},\sigma-\frac{1}{2},\tau+\frac{1}{2}\right\rangle,\nonumber\\
&&\bsQ_0|j,\sigma,\tau\rangle=\left(j-\frac{\sigma-\tau-1}{2}\right)\left|j,\sigma,\tau\right\rangle,\nonumber
\end{eqnarray}
\begin{eqnarray}
&&\bsS_-|j,\sigma,\tau\rangle= i\sqrt{(j+\sigma)(j+\tau)}\left|j-\frac{1}{2},\sigma-\frac{1}{2},\tau-\frac{1}{2}\right\rangle,\nonumber\\
&&\bsS_+|j,\sigma,\tau\rangle= -i\sqrt{(j+\tau+1)(j+\sigma+1)}\left|j+\frac{1}{2},\sigma+\frac{1}{2},\tau+\frac{1}{2}\right\rangle,\nonumber\\
&&\bsS_0|j,\sigma,\tau\rangle=\left(j+\frac{\sigma+\tau+1}{2}\right)\sigma\left|j,\sigma,\tau\right\rangle,\nonumber\\
&&\bsT_-|j,\sigma,\tau\rangle= -i\sqrt{(j-\tau)(j-\sigma)}\left|j-\frac{1}{2},\sigma+\frac{1}{2},\tau+\frac{1}{2}\right\rangle,\nonumber\\
&&\bsT_+|j,\sigma,\tau\rangle= i\sqrt{(j-\sigma+1)(j-\tau+1)}\left|j+\frac{1}{2},\sigma-\frac{1}{2},\tau-\frac{1}{2}\right\rangle,\nonumber\\
&&\bsT_0|j,\sigma,\tau\rangle=\left(j-\frac{\sigma+\tau-1}{2}\right)\left|j,\sigma,\tau\right\rangle.\nonumber
\end{eqnarray}

The formulas (\ref{Fock3}) define an unitary representation of the conformal group $\SO_0(2,4)$ in the Fock space $\fF$. The formulas (\ref{Fock3}) include the formulas for $\sJ_k$, $\sK_k$, giving representations $\Phi_n$ in subspaces $\fF_n$ (see (\ref{Fock}), where $j_1=j_2=j$) and thus the Fock representation $\Phi$ on the subgroup $\SO(4)$. Moreover, if we restrict $\SO_0(2,4)$ to a subgroup $\SO_0(1,3)$ (Lorentz group), we obtain the van der Waerden representation (\ref{Waerden}) given by the generators $\sX_k$, $\sY_k$, which proves the similarity of the complex shells of the group algebras $\mathfrak{so}(2,4)$ and $\mathfrak{sl}(2,\C)$. The representation, defined by (\ref{Fock3}), is called \textit{an extension $F^+$ of the Fock representation on the conformal group} \cite{Fet}. The representation $F^+$ is insufficient yet for the description of periodic system of elements. With that end in view it is necessary to include a fourth Madelung number $s$ (which is analogous to a spin) that leads to a group (the first ``doubling'')
\begin{equation}\label{RF}
\SO(2,4)\otimes\SU(2).
\end{equation}
A representation $F^+_s=\varphi_2\otimes F^+$ of the group (\ref{RF}), where $\varphi_2$ is an unitary representation of the group $\SU(2)$ in the space $C(2)$, already satisfies to this requirement (inclusion of the Madelung number $s$). A basis of the space $\fF^2=C(2)\otimes\fF$ of the representation $F^+_s$ has the form
\begin{equation}
|n,l,m,s\rangle,\quad n=1,2,\ldots;\; l=0,1,\ldots, n-1;\;
m=-l,-l+1,\ldots,l-1,l;\;s=-1/2,1/2.\label{RF2}
\end{equation}
Here $n$, $l$, $m$ are quantum numbers of the conformal group.

Let $\boldsymbol{\tau}_k$ be generators of the Lie algebra of $\SU(2)$, then a generator $\boldsymbol{\tau}_3$ commutes with the all generators of the subgroup $\SO(2,4)\otimes\boldsymbol{1}$. For that reason generators $\bsR_0$, $\bsL^2$, $\bsJ_3+\bsK_3$, $\boldsymbol{\tau}_3$ commute with the each other. Eigenvectors of the operators, representing these generators in the space $\fF^2$, have the form
\[
\left|n,l,m,\frac{1}{2}\right\rangle=\begin{bmatrix}
\Psi^1_{nlm}\\
0
\end{bmatrix},\quad
\left|n,l,m,-\frac{1}{2}\right\rangle=\begin{bmatrix}
0\\
\Psi^2_{nlm}
\end{bmatrix}
\]
with eigenvalues $n$, $l(l+1)$, $m$, $\frac{1}{2}$ and $n$, $l(l+1)$, $m$, $-\frac{1}{2}$. An action of the operators, representing generators $\boldsymbol{\tau}_+$, $\boldsymbol{\tau}_-$, $\boldsymbol{\tau}_3$ in the space $\fF^2$, is defined by the following formulas:
\[
\boldsymbol{\tau}_+\left|n,l,m,-\frac{1}{2}\right\rangle=\left|n,l,m,\frac{1}{2}\right\rangle,\quad
\boldsymbol{\tau}_-\left|n,l,m,\frac{1}{2}\right\rangle=\left|n,l,m,-\frac{1}{2}\right\rangle,
\]
\[
\boldsymbol{\tau}_3\left|n,l,m,s\right\rangle= s\left|n,l,m,s\right\rangle.
\]
In virtue of the Madelung numbering the basis $\left|n,l,m,s\right\rangle$ stands in one-to-one correspondence with the elements of periodic system. A relation between the arrangement of the element in the Mendeleev table and a number collection $(n,l,m,s)$ is defined by a so-called Madelung ``lexicographic rule'' $Z\leftrightarrow(n,l,m,s)$\footnote{Erwin Madelung was the first to apply ``hydrogen'' quantum numbers $n$, $l$, $m$, $s$ to the numbering of elements of the periodic table. It should be noted that the numbers $n$, $l$, $m$, $s$ are not quantum numbers in Bohr's model, because in this model there is no single quantum-mechanical description of the system of elements, the latter is assigned an atomic number $Z$, distinguishing rather than combining individual quantum systems. The resulting classification of elements Madelung called ``empirical'', because he cold not connect it with the Bohr model. Apparently, it was because of the lack of theoretical justification at the time (20-ies of the last century), he published it as a reference material in \cite{Mad68}. Theoretical justification (understanding of the nature of Madelung numbers) was later given by Fet \cite{Fet} from the position of group-theoretic vision. The history of Madelung numbers was also developed in a slightly different direction in the works of Kleczkowski \cite{Clech}, where the filling of the electronic levels of the atom was considered according to the rule of sequential filling of $(n+l)$-groups (the so-called Madelung-Kleczkowski groups).}:

1) elements are arranged in increasing order of atomic number $Z$;

2) collections $(n,l,m,s)$ are arranged in increasing order of $n+l$; at the given $n+l$ in increasing order of $n$; at the given $n+l$, $n$ in increasing order of $m$; at the given $n+l$, $n$, $m$ in increasing order of $s$;

3) $Z$-th element corresponds to $Z$-th collection.\\
In Madelung numbering the sum $n+l$ does not have a group sense: it is a sum of quantum number $n$ (an eigenvalue of the operator $\bsR_0=-\bsL_{56}$) and the number $l$ which is not quantum number. In this case, a quantum number is $l(l+1)$ (an eigenvalue of the operator $\bsL^2=\bsL^2_{12}+\bsL^2_{23}+\bsL^2_{31}$), and $l$ only defines this quantum number. Therefore, in accordance with group theoretical viewpoint the number $n+l$ should be excluded from the formulation of ``lexicographic rule''. In \cite{Fet} Fet introduced a new quantum number $\nu$, which equal to $\nu=1/2(n+l+1)$ for the odd value of $n+l$ and $\nu=1/2(n+l)$ for the even value of $n+l$. Introduction of the quantum number $\nu$ allows us to change Madelung numbering, that leads to the following ``lexicographic rule'' (Fet rule):

1) elements are arranged in increasing order of atomic number $Z$;

2) collections $(\nu,\lambda,\mu,s,s^\prime)$ are arranged in increasing order of $\nu$; at the given $\nu$ in increasing order of $s^\prime$; at the given $\nu$, $s^\prime$ in decreasing order of $\lambda$; at the given $\nu$, $s^\prime$, $\lambda$ in increasing order of $\mu$; at the given $\nu$, $s^\prime$, $\lambda$, $\mu$ in increasing order of $s$;

3) $Z$-th element corresponds to $Z$-th collection.\\
The introduction of the fifth quantum number leads to another (second) ``doubling'' of the representation space. In this space we have the Rumer-Fet group
\begin{equation}\label{RF3}
\SO(2,4)\otimes\SU(2)\otimes\SU(2)^\prime.
\end{equation}
A representation $F^+_{ss^\prime}=\varphi^\prime_2\otimes F^+_s$ of the group (\ref{RF3}), where $\varphi^\prime_2$ is an unitary representation of the group $\SU(2)^\prime$ in the space $C(2)$, satisfies to the requirement of inclusion of the fifth quantum number. A basis of the space $\fF^4=C(2)\otimes\fF^2$ of the representation $F^+_{ss^\prime}$ has the form
\begin{multline}
|\nu,\lambda,\mu,s,s^\prime\rangle,\quad \nu=1,2,\ldots;\; \lambda=0,1,\ldots, \nu-1;\\
\mu=-\lambda,-\lambda+1,\ldots,\lambda-1,\lambda;\;s=-1/2,1/2,\;s^\prime=-1/2,1/2.\label{RF4}
\end{multline}

Let $\boldsymbol{\tau}^\prime_k$ be generators of the Lie algebra $\SU(2)^\prime$, then the generator $\boldsymbol{\tau}^\prime_3$ commutes with the all generators of the subgroup $\SO(2,4)\otimes\SU(2)\otimes\boldsymbol{1}$. Therefore, generators $\bsR_0=-\bsL_{56}=\bsP_0+\bsQ_0=\bsS_0+\bsT_0$, $\bsL^2$, $\bsJ_3+\bsK_3$, $\boldsymbol{\tau}_3$, $\boldsymbol{\tau}^\prime_3$ commute with the each other. Common eigenvectors of the operators, which represent these generators  in the space $\fF^4$, have the form
\[
\left|\nu,\lambda,\mu,s,+\frac{1}{2}\right\rangle=\begin{bmatrix}
\Psi^1_{\nu\lambda,\mu}\\
\Psi^2_{\nu\lambda,\mu}\\
0\\
0
\end{bmatrix},\quad
\left|\nu,\lambda,\mu,s,-\frac{1}{2}\right\rangle=\begin{bmatrix}
0\\
0\\
\Psi^3_{\nu\lambda,\mu}\\
\Psi^4_{\nu\lambda,\mu}
\end{bmatrix}
\]
with eigenvalues $\nu$, $\lambda(\lambda+1)$, $\mu$, $s$, $\frac{1}{2}$ and $\nu$, $\lambda(\lambda+1)$, $\mu$, $s$, $-\frac{1}{2}$.

An action of the operators, representing the generators $\boldsymbol{\tau}_+$, $\boldsymbol{\tau}_-$, $\boldsymbol{\tau}_3$ and $\boldsymbol{\tau}^\prime_+$, $\boldsymbol{\tau}^\prime_-$, $\boldsymbol{\tau}^\prime_3$ in the space $\fF^4$, is defined by the following formulas:
\[
\boldsymbol{\tau}_+\left|\nu,\lambda,\mu,-\frac{1}{2},s^\prime\right\rangle=\left|\nu,\lambda,\mu,\frac{1}{2},
s^\prime\right\rangle,\quad
\boldsymbol{\tau}_-\left|\nu,\lambda,\mu,\frac{1}{2},s^\prime\right\rangle=\left|\nu,\lambda,\mu,-\frac{1}{2},
s^\prime\right\rangle,
\]
\[
\boldsymbol{\tau}_3\left|\nu,\lambda,\mu,s,s^\prime\right\rangle= s\left|\nu,\lambda,\mu,s,s^\prime\right\rangle.
\]
\[
\boldsymbol{\tau}^\prime_+\left|\nu,\lambda,\mu,s,-\frac{1}{2}\right\rangle=\left|\nu,\lambda,\mu,s,\frac{1}{2}
\right\rangle,\quad
\boldsymbol{\tau}^\prime_-\left|\nu,\lambda,\mu,s,\frac{1}{2}\right\rangle=\left|\nu,\lambda,\mu,s,-\frac{1}{2}
\right\rangle,
\]
\[
\boldsymbol{\tau}^\prime_3\left|\nu,\lambda,\mu,s,s^\prime\right\rangle= s^\prime\left|\nu,\lambda,\mu,s,s^\prime\right\rangle.
\]
``Addresses'' of the elements (see Fig.\,1) are defined by a collection of quantum numbers of the Rumer-Fet group (\ref{RF3}), which numbering basis vectors $\left|\nu,\lambda,\mu,s,s^\prime\right\rangle$ of the space $\fF^4$. Generators of the Lie algebra of $\SO(2,4)$ act on the quantum numbers $\nu$, $\lambda$, $\mu$ by means of the formulas (\ref{Fock3}) with a replacement of $n$, $l$, $m$ via $\nu$, $\lambda$, $\mu$.

There is the following chain of groups:
\begin{equation}\label{Chain1}
G\supset G_1\supset G_2\longmapsto\SO(2,4)\otimes\SU(2)\otimes\SU(2)\supset\SO(4)\otimes\SU(2)\supset\SO(3)\otimes\SU(2).
\end{equation}

\unitlength=1mm
\begin{center}
\begin{picture}(120,220)(7,0)
\put(-2,205){$\lambda=0$}
\put(10,212){$\overbrace{\scr s^\prime=-1/2\;\; s^\prime=1/2}^{\nu=1}$} \put(10,200){\framebox(10,10){}}\put(13,206){H}\put(13,202){He}
\put(23,200){\framebox(10,10){}}\put(26,206){Li}\put(26,202){Be}
\put(36,212){$\overbrace{\scr s^\prime=-1/2\;\; s^\prime=1/2}^{\nu=2}$}
\put(36,200){\framebox(10,10){}}\put(39,206){Na}\put(39,202){Mg}
\put(49,200){\framebox(10,10){}}\put(51,206){K}\put(51,202){Ca}
\put(62,212){$\overbrace{\scr s^\prime=-1/2\;\; s^\prime=1/2}^{\nu=3}$}
\put(62,200){\framebox(10,10){}}\put(65,206){Rb}\put(65,202){Sr}
\put(75,200){\framebox(10,10){}}\put(78,206){Cs}\put(78,202){Ba}
\put(88,212){$\overbrace{\scr s^\prime=-1/2\;\; s^\prime=1/2}^{\nu=4}$}
\put(88,200){\framebox(10,10){}}\put(91,206){Fr}\put(91,202){Ra}
\put(101,200){\framebox(10,10){}}
\put(102.5,206){{\bf Uue}}\put(102.5,202){{\bf Ubn}}
\put(112,206){$\scr s=-1/2$}
\put(112,202){$\scr s=1/2$}
\put(121,204){$\left.\phantom{\framebox(1,6){}}\right\}$}
\put(125,204){$\mu=0$}
\put(-2,178){$\lambda=1$}
\put(12,178){$\left\{\phantom{\framebox(1,16){}}\right.$}
\put(36,184){\framebox(10,10){}}\put(39,190){B}\put(39,186){C}
\put(49,184){\framebox(10,10){}}\put(51,190){Al}\put(51,186){Si}
\put(62,184){\framebox(10,10){}}\put(65,190){Ga}\put(65,186){Ge}
\put(75,184){\framebox(10,10){}}\put(78,190){In}\put(78,186){Sn}
\put(88,184){\framebox(10,10){}}\put(91,190){Tl}\put(91,186){Pb}
\put(101,184){\framebox(10,10){}}
\put(104,190){{\bf Nh}}\put(104,186){{\bf Fl}}
\put(112,190){$\scr s=-1/2$}
\put(112,186){$\scr s=1/2$}
\put(121,188){$\left.\phantom{\framebox(1,6){}}\right\}$}
\put(125,188){$\mu=-1$}
\put(36,174){\framebox(10,10){}}\put(39,180){N}\put(39,176){O}
\put(49,174){\framebox(10,10){}}\put(51,180){P}\put(51,176){S}
\put(62,174){\framebox(10,10){}}\put(65,180){As}\put(65,176){Se}
\put(75,174){\framebox(10,10){}}\put(78,180){Sb}\put(78,176){Te}
\put(88,174){\framebox(10,10){}}\put(91,180){Bi}\put(91,176){Po}
\put(101,174){\framebox(10,10){}}
\put(104,180){{\bf Mc}}\put(104,176){{\bf Lv}}
\put(112,180){$\scr s=-1/2$}
\put(112,176){$\scr s=1/2$}
\put(121,178){$\left.\phantom{\framebox(1,6){}}\right\}$}
\put(125,178){$\mu=0$}
\put(36,164){\framebox(10,10){}}\put(39,170){F}\put(39,166){Ne}
\put(49,164){\framebox(10,10){}}\put(51,170){Cl}\put(51,166){Ar}
\put(62,164){\framebox(10,10){}}\put(65,170){Br}\put(65,166){Kr}
\put(75,164){\framebox(10,10){}}\put(78,170){I}\put(78,166){Xe}
\put(88,164){\framebox(10,10){}}\put(91,170){At}\put(91,166){Rn}
\put(101,164){\framebox(10,10){}}
\put(104,170){{\bf Ts}}\put(104,166){{\bf Og}}
\put(112,170){$\scr s=-1/2$}
\put(112,166){$\scr s=1/2$}
\put(121,168){$\left.\phantom{\framebox(1,6){}}\right\}$}
\put(125,168){$\mu=1$}
\put(-2,135){$\lambda=2$}
\put(12,135){$\left\{\phantom{\framebox(1,25){}}\right.$}
\put(62,151){\framebox(10,10){}}\put(65,157){Sc}\put(65,153){Ti}
\put(75,151){\framebox(10,10){}}\put(78,157){Y}\put(78,153){Zr}
\put(88,151){\framebox(10,10){}}\put(91,157){Lu}\put(91,153){Hf}
\put(101,151){\framebox(10,10){}}\put(104,157){Lr}\put(104,153){Rf}
\put(112,157){$\scr s=-1/2$}
\put(112,153){$\scr s=1/2$}
\put(121,155){$\left.\phantom{\framebox(1,6){}}\right\}$}
\put(125,155){$\mu=-2$}
\put(62,141){\framebox(10,10){}}\put(65,147){V}\put(65,143){Cr}
\put(75,141){\framebox(10,10){}}\put(78,147){Nb}\put(78,143){Mo}
\put(88,141){\framebox(10,10){}}\put(91,147){Ta}\put(91,143){W}
\put(101,141){\framebox(10,10){}}
\put(104,147){{\bf Db}}\put(104,143){{\bf Sg}}
\put(112,147){$\scr s=-1/2$}
\put(112,143){$\scr s=1/2$}
\put(121,145){$\left.\phantom{\framebox(1,6){}}\right\}$}
\put(125,145){$\mu=-1$}
\put(62,131){\framebox(10,10){}}\put(65,137){Mn}\put(65,133){Fe}
\put(75,131){\framebox(10,10){}}\put(78,137){Tc}\put(78,133){Ru}
\put(88,131){\framebox(10,10){}}\put(91,137){Re}\put(91,133){Os}
\put(101,131){\framebox(10,10){}}
\put(104,137){{\bf Bh}}\put(104,133){{\bf Hs}}
\put(112,137){$\scr s=-1/2$}
\put(112,133){$\scr s=1/2$}
\put(121,135){$\left.\phantom{\framebox(1,6){}}\right\}$}
\put(125,135){$\mu=0$}
\put(62,121){\framebox(10,10){}}\put(65,127){Co}\put(65,123){Ni}
\put(75,121){\framebox(10,10){}}\put(78,127){Rh}\put(78,123){Pd}
\put(88,121){\framebox(10,10){}}\put(91,127){Ir}\put(91,123){Pt}
\put(101,121){\framebox(10,10){}}
\put(104,127){{\bf Mt}}\put(104,123){{\bf Ds}}
\put(112,127){$\scr s=-1/2$}
\put(112,123){$\scr s=1/2$}
\put(121,125){$\left.\phantom{\framebox(1,6){}}\right\}$}
\put(125,125){$\mu=1$}
\put(62,111){\framebox(10,10){}}\put(65,117){Cu}\put(65,113){Zn}
\put(75,111){\framebox(10,10){}}\put(78,117){Ag}\put(78,113){Cd}
\put(88,111){\framebox(10,10){}}\put(91,117){Au}\put(91,113){Hg}
\put(101,111){\framebox(10,10){}}
\put(104,117){{\bf Rg}}\put(104,113){{\bf Cn}}
\put(112,117){$\scr s=-1/2$}
\put(112,113){$\scr s=1/2$}
\put(121,115){$\left.\phantom{\framebox(1,6){}}\right\}$}
\put(125,115){$\mu=2$}
\put(-2,72){$\lambda=3$}
\put(12,72){$\left\{\phantom{\framebox(1,36){}}\right.$}
\put(88,98){\framebox(10,10){}}\put(91,104){La}\put(91,100){Ce}
\put(101,98){\framebox(10,10){}}\put(104,104){Ac}\put(104,100){Th}
\put(112,104){$\scr s=-1/2$}
\put(112,100){$\scr s=1/2$}
\put(121,102){$\left.\phantom{\framebox(1,6){}}\right\}$}
\put(125,102){$\mu=-3$}
\put(88,88){\framebox(10,10){}}\put(91,94){Pr}\put(91,90){Nd}
\put(101,88){\framebox(10,10){}}\put(104,94){Pa}\put(104,90){U}
\put(112,94){$\scr s=-1/2$}
\put(112,90){$\scr s=1/2$}
\put(121,92){$\left.\phantom{\framebox(1,6){}}\right\}$}
\put(125,92){$\mu=-2$}
\put(88,78){\framebox(10,10){}}\put(91,84){Pm}\put(91,80){Sm}
\put(101,78){\framebox(10,10){}}\put(104,84){Np}\put(104,80){Pu}
\put(112,84){$\scr s=-1/2$}
\put(112,80){$\scr s=1/2$}
\put(121,82){$\left.\phantom{\framebox(1,6){}}\right\}$}
\put(125,82){$\mu=-1$}
\put(88,68){\framebox(10,10){}}\put(91,74){Eu}\put(91,70){Gd}
\put(101,68){\framebox(10,10){}}\put(104,74){Am}\put(104,70){Cm}
\put(112,74){$\scr s=-1/2$}
\put(112,70){$\scr s=1/2$}
\put(121,72){$\left.\phantom{\framebox(1,6){}}\right\}$}
\put(125,72){$\mu=0$}
\put(88,58){\framebox(10,10){}}\put(91,64){Tb}\put(91,60){Dy}
\put(101,58){\framebox(10,10){}}\put(104,64){Bk}\put(104,60){Cf}
\put(112,64){$\scr s=-1/2$}
\put(112,60){$\scr s=1/2$}
\put(121,62){$\left.\phantom{\framebox(1,6){}}\right\}$}
\put(125,62){$\mu=1$}
\put(88,48){\framebox(10,10){}}\put(91,54){Ho}\put(91,50){Er}
\put(101,48){\framebox(10,10){}}\put(104,54){Es}\put(104,50){Fm}
\put(112,54){$\scr s=-1/2$}
\put(112,50){$\scr s=1/2$}
\put(121,52){$\left.\phantom{\framebox(1,6){}}\right\}$}
\put(125,52){$\mu=2$}
\put(88,38){\framebox(10,10){}}\put(91,44){Tm}\put(91,40){Yb}
\put(101,38){\framebox(10,10){}}\put(104,44){Md}\put(104,40){No}
\put(112,44){$\scr s=-1/2$}
\put(112,40){$\scr s=1/2$}
\put(121,42){$\left.\phantom{\framebox(1,6){}}\right\}$}
\put(125,42){$\mu=3$}
\put(10,20){\begin{minipage}{25pc}{\small {\bf Fig.\,1.} Mendeleev table in the form of the basic representation $F^+_{ss^\prime}$ of the Rumer-Fet group $\SO(2,4)\otimes\SU(2)\otimes\SU(2)^\prime$.}\end{minipage}}
\end{picture}
\end{center}

The reduction of the basic representation $F^+_{ss^\prime}$ of the Rumer-Fet group on the subgroups is realized in accordance with the chain (\ref{Chain1}). So, multiplets of a subgroup $\SO(3)\otimes\SU(2)$ are represented by vertical rectangles of Fig.\,1, and their elements compose well-known $s$-, $p$-, $d$- and $f$-families (in particular, lanthanides and actinides are selected as multiplets of the subgroup $\SO(3)\otimes\SU(2)$). The each element occupies a quite definite place, which is defined by its ``address'' in the table $(\nu,\lambda,\mu,s,s^\prime)$, that is, by corresponding quantum numbers of the symmetry group. Thus, atoms of all possible elements stand in one-to-one correspondence with the vectors of the basis (\ref{RF4}).
\section{Mendeleev Table}
In Fig.\,1 we mark by bold script the elements that had not yet been discovered or received their official names during lifetime of Rumer and Fet. These elements belong the last column of the Fig.\,1 with the quantum numbers $\nu=4$ and $s^\prime=1/2$: \textbf{Db} -- Dubnium, \textbf{Sg} -- Seaborgium, \textbf{Bh} -- Bohrium, \textbf{Hs} -- Hassium (eka-osmium), \textbf{Mt} -- Meitnerium, \textbf{Ds} -- Darmstadtium, \textbf{Rg} -- Roentgenium, \textbf{Cn} -- Copernicium (eka-mercury). All these elements belong to a multiplet with a quantum number $\lambda=2$. A multiplet with $\lambda=1$ ($\nu=4$, $s^\prime=1/2$) consists of recently discovered elements: \textbf{Nh} -- Nihonium (eka-thallium), \textbf{Fl} -- Flerovium (eka-lead), \textbf{Mc} -- Moscovium (eka-bismuth), \textbf{Lv} -- Livermorium (eka-polonium), \textbf{Ts} -- Tennessine (eka-astatine), \textbf{Og} -- Oganesson (eka-radon). Further, a multiplet with a quantum number $\lambda=0$ ($\nu=4$, $s^\prime=1/2$) is formed by undetected yet hypothetical elements \textbf{Uue} -- Ununennium (eka-francium) with supposed atomic mass 316 a.u. and \textbf{Ubn} -- Unbinillium (eka-radium). All the enumerated elements accomplish the filling of the Mendeleev table (from 1-th to 120-th number) inclusive the value of quantum number $\nu=4$.

\begin{center}
{\textbf{Tab.\,1.} Average masses of ``heavy'' multiplets.}
\vspace{0.1cm}
{\renewcommand{\arraystretch}{1.0}
\begin{tabular}{|c||l|l|c|l|}\hline
 & Multiplet & Mass (exp.) & Mass (theor.) & Approx. \%\\ \hline\hline
1. & $(\nu=3,s^\prime=-1/2,\lambda=2)$ & $55,31$ & $53$ & $-4,36$\\
2. & $(\nu=3,s^\prime=-1/2,\lambda=1)$ & $76,65$ & $75$ & $-2,2$\\
3. & $(\nu=3,s^\prime=-1/2,\lambda=0)$ & $86,54$ & $86$ & $-0,63$\\
4. & $(\nu=3,s^\prime=1/2,\lambda=2)$ & $99,76$ & $104$ & $4,07$\\
5. & $(\nu=3,s^\prime=1/2,\lambda=1)$ & $123,51$ & $126$ & $1,98$\\
6. & $(\nu=3,s^\prime=1/2,\lambda=0)$ & $135,12$ & $137$ & $1,37$\\
7. & $(\nu=4,s^\prime=-1/2,\lambda=3)$ & $154,59$ & $156$ & $0,90$\\
8. & $(\nu=4,s^\prime=-1/2,\lambda=2)$ & $187,96$ & $189$ & $0,55$\\
9. & $(\nu=4,s^\prime=-1/2,\lambda=1)$ & $210,21$ & $211$ & $0,37$\\
10. & $(\nu=4,s^\prime=-1/2,\lambda=0)$ & $224,52$ & $222$ & $-1,13$\\
11. & $(\nu=4,s^\prime=1/2,\lambda=3)$ & $244,56$ & $241$ & $-1,48$\\
12. & $(\nu=4,s^\prime=1/2,\lambda=2)$ & $273,10$ & $274$ & $0,33$\\
13. & $(\nu=4,s^\prime=1/2,\lambda=1)$ & $290,83$ & $296$ & $1,75$\\
14. & $(\nu=4,s^\prime=1/2,\lambda=0)$ & $318^\ast$ & $307$ & $-3,58^\ast$\\
\hline
\end{tabular}
}
\end{center}

Let us calculate masses of the hypothetical elements \textbf{Uue} and \textbf{Ubn}. Before to proceed this, we calculate average masses of the multiplets belonging to the Mendeleev table. With this aim in view we use a mass formula\footnote{This formula is analogous to a Gell-Mann--Okubo formula for the hadrons in $\SU(3)$-theory \cite{Gel61,OR64}, and also to a B\'{e}g-Singh formula in $\SU(6)$-theory \cite{BS}.} proposed in \cite{Fet}:
\begin{equation}\label{Mass1}
m=m_0+a\left[s^\prime(2\nu-3)-5\nu+\frac{11}{2}+2(\nu^2-1)\right]-b\cdot\lambda(\lambda+1),
\end{equation}
where $m_0$, $a$, $b$ are the coefficients which are underivable from the theory. The formula (\ref{Mass1}) is analogous to the ``first perturbation'' in $\SU(3)$- and $\SU(6)$-theories which allows to calculate an average mass of the elements of the multiplet (an analogue of the ``second perturbation'' for the Rumer-Fet group, which leads to the mass splitting inside the multiplet, we will give in section 6). The Tab.\,1 contains average masses of ``heavy'' multiplets ($\nu=3,\,4$) calculated according the formula (\ref{Mass1}) at $m_0=1$, $a=17$, $b=5,5$. From the Tab.\,1 we see that an accuracy between experimental and theoretical masses rises with growth of the ``weight'' of the multiplet, therefore, the formula (\ref{Mass1}) is asymptotic. An exception is the last multiplet $(\nu=4,s^\prime=1/2,\lambda=0)$, consisting of the hypothetical elements \textbf{Uue} and \textbf{Ubn}, which have masses unconfirmed by experiment.

\section{Implementation of the Operator Algebra}
As is known, in the foundation of an algebraic formulation of quantum theory we have a Gelfand-Naimark-Segal (GNS) construction, which is defined by a canonical correspondence $\omega\leftrightarrow\pi_\omega$ between states and cyclic representations of $C^\ast$-algebra \cite{Emh,BLOT,Hor86}.

Let us suppose that according to the axiom \textbf{A.I} (section \,2) generators of the conformal group (a fundamental symmetry $G_f=\SO(2,4)$ in this context) are attached to the energy operator $H$. Therefore, the each eigensubspace $\sH_E$ of the energy operator is invariant with respect to the operators of the representation $F^+$ of the conformal group\footnote{It follows from a similarity of the complex shells of the group algebras $\mathfrak{sl}(2,\C)$, $\mathfrak{so}(4)$ and $\mathfrak{so}(2,4)$.}. It allows us to obtain a concrete implementation (``dressing'') of the operator algebra $\pi(\fA)\rightarrow\pi(H)$, where $\pi\equiv F^+$. Thus, the each possible value of energy (an energy level) is a vector state of the form (axiom \textbf{A.II}):
\[
\omega_\Phi(H)=\frac{\langle\Phi\mid\pi(H)\Phi\rangle}{\langle\Phi\mid\Phi\rangle}=
\frac{\langle\Phi\mid F^+(H)\Phi\rangle}{\langle\Phi\mid\Phi\rangle},
\]
where $\left|\Phi\right\rangle$ is a cyclic vector of the Hilbert space $\sH_\infty$.

Further, in virtue of an isomorphism $\SU(2,2)\simeq\spin_+(2,4)$ (see Appendix C) we will consider an universal covering $\widetilde{G}_f$ as a \textit{spinor group}. It allows us to associate in addition a \textit{twistor structure} with the each cyclic vector $\left|\Phi\right\rangle\in\sH_\infty$. Spintensor representations of the group $\widetilde{G}_f=\spin_+(2,4)$ form a substrate of finite-dimensional representations $\boldsymbol{\tau}_{k/2,r/2}$, $\overline{\boldsymbol{\tau}}_{k/2,r/2}$ of the conformal group realized in the spaces $\Sym_{(k,r)}\subset\dS_{2^{k+r}}$ and $\overline{\Sym}_{(k,r)}\subset\overline{\dS}_{2^{k+r}}$, where $\dS_{2^{k+r}}$ is a spinspace. Indeed, a twistor $\bsZ^\alpha=\left(\boldsymbol{s}^\alpha,\boldsymbol{s}_{\dot{\alpha}}\right)$ is a vector of the fundamental representation of the group $\spin_+(2,4)$, where $\alpha,\dot{\alpha}=0,1$. A vector of the general spintensor representation of the group $\spin_+(2,4)$ is
\begin{equation}\label{TT}
\bsZ=\begin{bmatrix}
\boldsymbol{S}\\
\overline{\boldsymbol{S}}
\end{bmatrix},
\end{equation}
where $\boldsymbol{S}$ is a spintensor of the form
\begin{equation}\label{Spintensor}
\boldsymbol{S}=\boldsymbol{s}^{\alpha_1\alpha_2\ldots\alpha_k}_{\dot{\alpha}_1\dot{\alpha}_2\ldots
\dot{\alpha}_r}=\sum \boldsymbol{s}^{\alpha_1}\otimes
\boldsymbol{s}^{\alpha_2}\otimes\cdots\otimes
\boldsymbol{s}^{\alpha_k}\otimes
\boldsymbol{s}_{\dot{\alpha}_1}\otimes
\boldsymbol{s}_{\dot{\alpha}_2}\otimes\cdots\otimes
\boldsymbol{s}_{\dot{\alpha}_r},\quad\alpha_i,\dot{\alpha}_i=0,1;
\end{equation}
that is, the vector of the spinspace $\dS_{2^{k+r}}=\dS_{2^k}\otimes\dot{\dS}_{2^r}$, where $\dot{\dS}_{2^r}$ is a dual spinspace. $\overline{\boldsymbol{S}}$ is a spintensor from the conjugated spinspace $\overline{\dS}_{2^{k+r}}$. Symmetrizing the each spintensor $\boldsymbol{S}$ and $\overline{\boldsymbol{S}}$ in (\ref{TT}), we obtain symmetric \textit{twisttensor} $\bsZ$. In turn, as is known \cite{Lou91}, spinspace is a minimal left ideal of the Clifford algebra $\cl_{p,q}$, that is, there is an isomorphism $\dS_{2^m}(\K)\simeq I_{p,q}=\cl_{p,q}f$, where $f$ is a primitive idempotent of the algebra $\cl_{p,q}$, $\K=f\cl_{p,q}f$ is a division ring for $\cl_{p,q}$, $m=(p+q)/2$. A complex spinspace $\dS_{2^m}(\C)$ is a complexification $\C\otimes I_{p,q}$ of the minimal left ideal $I_{p,q}$ of a real subalgebra $\cl_{p,q}$. Hence, $\dS_{2^{k+r}}$ is a minimal left ideal of the complex algebra $\C_{2k}\otimes\dot{\C}_{2r}\simeq\C_{2(k+r)}$ (for more details see \cite{Var00,Var15c}).

Now we are in a position that allows us to define a system of \textit{basic cyclic vectors} endowed with a complex twistor structure (these vectors correspond to the system of finite-dimensional representations of the conformal group). Let
\begin{eqnarray}
&&\mid\C_0,\boldsymbol{\tau}_{0,0}(H)\Phi\rangle;\nonumber\\
&&\mid\C_2,\boldsymbol{\tau}_{1/2,0}(H)\Phi\rangle,\quad\mid\dot{\C}_2,\boldsymbol{\tau}_{0,1/2}(H)\Phi\rangle;
\nonumber\\
&&\mid\C_2\otimes\C_2,\boldsymbol{\tau}_{1,0}(H)\Phi\rangle,\quad
\mid\C_2\otimes\dot{\C}_2,\boldsymbol{\tau}_{1/2,1/2}(H)\Phi\rangle,\quad
\mid\dot{\C}_2\otimes\dot{\C}_2,\boldsymbol{\tau}_{0,1}(H)\Phi\rangle;\nonumber\\
&&\mid\C_2\otimes\C_2\otimes\C_2,\boldsymbol{\tau}_{3/2,0}(H)\Phi\rangle,\quad
\mid\C_2\otimes\C_2\otimes\dot{\C}_2,\boldsymbol{\tau}_{1,1/2}(H)\Phi\rangle,\nonumber\\
&&\phantom{\C_2\otimes\dot{\C}_2\otimes\dot{\C}_2}\mid\C_2\otimes\dot{\C}_2\otimes\dot{\C}_2,\boldsymbol{\tau}_{1/2,1}(H)\Phi\rangle,\quad
\mid\dot{\C}_2\otimes\dot{\C}_2\otimes\dot{\C}_2,\boldsymbol{\tau}_{0,3/2}(H)\Phi\rangle;
\nonumber\\
&&\ldots\ldots\ldots\ldots\ldots\ldots\ldots\ldots\ldots\ldots\ldots\ldots\ldots\ldots\ldots\ldots\ldots\ldots\ldots
\ldots\ldots\ldots\ldots
\nonumber
\end{eqnarray}
\begin{eqnarray}
&&\mid\overline{\C_0,\boldsymbol{\tau}_{0,0}(H)}\Phi\rangle;\nonumber\\
&&\mid\overline{\C_2,\boldsymbol{\tau}_{1/2,0}(H)}\Phi\rangle,\quad
\mid\overline{\dot{\C}_2,\boldsymbol{\tau}_{0,1/2}(H)}\Phi\rangle;
\nonumber\\
&&\mid\overline{\C_2\otimes\C_2,\boldsymbol{\tau}_{1,0}(H)}\Phi\rangle,\quad
\mid\overline{\C_2\otimes\dot{\C}_2,\boldsymbol{\tau}_{1/2,1/2}(H)}\Phi\rangle,\quad
\mid\overline{\dot{\C}_2\otimes\dot{\C}_2,\boldsymbol{\tau}_{0,1}(H)}\Phi\rangle;\nonumber\\
&&\mid\overline{\C_2\otimes\C_2\otimes\C_2,\boldsymbol{\tau}_{3/2,0}(H)}\Phi\rangle,\quad
\mid\overline{\C_2\otimes\C_2\otimes\dot{\C}_2,\boldsymbol{\tau}_{1,1/2}(H)}\Phi\rangle,\nonumber\\
&&\phantom{\C_2\otimes\dot{\C}_2\otimes\dot{\C}_2}\mid
\overline{\C_2\otimes\dot{\C}_2\otimes\dot{\C}_2,\boldsymbol{\tau}_{1/2,1}(H)}\Phi\rangle,\quad
\mid\overline{\dot{\C}_2\otimes\dot{\C}_2\otimes\dot{\C}_2,\boldsymbol{\tau}_{0,3/2}(H)}\Phi\rangle;
\nonumber\\
&&\ldots\ldots\ldots\ldots\ldots\ldots\ldots\ldots\ldots\ldots\ldots\ldots\ldots\ldots\ldots\ldots\ldots\ldots\ldots
\ldots\ldots\ldots\ldots
\nonumber
\end{eqnarray}
Therefore, in accordance with GNS-construction (axiom \textbf{A.II}) we have complex vector states of the form
\[
\omega^c_\Phi(H)=
\frac{\langle\Phi\mid\C_{2(k+r)},\boldsymbol{\tau}_{k/2,r/2}(H)\Phi\rangle}{\langle\Phi\mid\Phi\rangle},
\]
\[
\overline{\omega}^c_\Phi(H)=
\frac{\langle\Phi\mid\overline{\C_{2(k+r)},\boldsymbol{\tau}_{k/2,r/2}(H)}\Phi\rangle}{\langle\Phi\mid\Phi\rangle}.
\]
According to (\ref{TT}), the pairs $(\omega^c_\Phi(H),\overline{\omega}^c_\Phi(H))$ form \textit{neutral states}. Further, at execution of the condition $\omega^c_\Phi(H)\geq 0$ ($\overline{\omega}^c_\Phi(H)\geq 0$) a set of all pure states $(\omega^c_\Phi(H),\overline{\omega}^c_\Phi(H))$ forms a physical Hilbert space $\bsH_{\rm phys}$ (axiom \textbf{A.III}) and, correspondingly, a space of rays $\hat{H}=\bsH_{\rm phys}/S^1$. All the pure states of the physical quantum system $\bsU$ are described by the unit rays and at this realization of the operator algebra these states correspond to atoms of the periodic system of elements. At this point, there is superposition principle (axiom \textbf{A.V}).

Following to Heisenberg's classification \cite{Heisen}, the all set of symmetry groups $G$ should be divided on the two classes: 1) \textit{groups of fundamental (primary) symmetries} $G_f$, which participate in construction of state vectors of the quantum system $\bsU$; 2) \textit{groups of dynamic (secondary) symmetries} $G_d$, which describe approximate symmetries between state vectors of $\bsU$. Dynamic symmetries $G_d$ relate different states (state vectors $\left|\Psi\right\rangle\in\bsH_{\rm phys}$) among the quantum system $\bsU$. The symmetry $G_d$ of the system $\bsU$ can be represented as a \textit{quantum transition} between its states (levels of state spectrum of $\bsU$).

We now show that Rumer-Fet group is a dynamic symmetry. Indeed, the group (\ref{RF3}) is equivalent to $\widetilde{\SO}(2,4)\otimes\SU(2)=\SU(2,2)\otimes\SU(2)$ (see \cite{Fet75}), since one ``doubling'' in (\ref{RF3}) already actually described by the two-sheeted covering $\SU(2,2)$ of the conformal group\footnote{Throughout the article, the term ``doubling'' occurs many times. ``Doubling'' (or Pauli's ``doubling and decreasing symmetry'' \cite{Heisen}) is one of the leading principles of group-theoretic description. Heisenberg notes \cite{Heisen2} that all the real symmetries of nature arose as a consequence of such doubling. ``Symmetry decreasing'' should be understood as group reduction, that is, if there is a chain of nested groups $G\supset G_1\supset G_2\supset\ldots\supset G_k$ and an irreducible unitary representation $\fP$ of group $G$ in space $\bsH_{\rm phys}$ is given, then the reduction $G/G_1$ of the representation $\fP$ of group $G$ by subgroup $G_1$ leads to the decomposition of $\fP$ into orthogonal sum of irreducible representations $\fP^{(1)}_i$ of subgroup $G_1$. In turn, the reduction $G_1/G_2$ of the representation of the group $G_1$ over the subgroup $G_2$ leads to the decomposition of the representations $\fP^{(1)}_i$ into irreducible representations $\fP^{(2)}_{ij}$ of the group $G_2$ and so on (see \cite{Var15b,Var15d}). Thus there is a reduction (``symmetry decreasing'' of Pauli) of the group $G$ with high symmetry to lower symmetries of the subgroups.}. At this point, atoms of different elements stand in one-to-one correspondence with the vectors belonging the basis (\ref{RF4}) of the space of the representation $F^+_{ss^\prime}$. Here we have a direct analogue with the physics of ``elementary particles''. According to \cite{Wig39}, a quantum system, described by an irreducible unitary \textit{representation} of the Poincar\'{e} group $\cP$, is called elementary particle. On the other hand, in accordance with $\SU(3)$- and $\SU(6)$-theories elementary particle is described by a \emph{vector} of an irreducible representation of the group $\SU(3)$ (or $\SU(6)$). Therefore, we have two group theoretical interpretations of the elementary particle: as a \textit{representation} of the group $\cP$ (group of fundamental symmetry) and as a  \textit{vector} of the representation of the group of dynamic symmetry $\SU(3)$ (or $\SU(6)$). Besides, the structure of the mass formula (\ref{Mass1}) for the Rumer-Fet group is analogous to Gell-Mann-Okubo and B\'{e}g-Singh mass formulas for the groups $\SU(3)$ and $\SU(6)$. An action of the group $G_d=\SU(2,2)\otimes\SU(2)$, which lifted into $\bsH_{\rm phys}$ via a central extension (see, for example, \cite{Var15}), moves state vectors $\left|\Psi\right\rangle\in\bsH_{\rm phys}$, corresponding to different atoms of the periodic system, into each other.

\section{Seaborg Table}
As is known, Mendeleev table includes 118 elements, from which at present time 118 elements have been discovered (the last detected element \textbf{Og} -- Oganesson (eka-radon) with the atomic number $Z=118$). Two as yet undiscovered hypothetical elements \textbf{Uue} -- Ununennium (eka-francium) with $Z=119$ and \textbf{Ubn} -- Unbinillium (eka-radium) with $Z=120$ begin to fill the eight period. According to Bohr model, both elements belong to $s$-shell. Mendeleev table contains seven periods (rows), including $s$-, $p$-, $d$- and $f$-families (shells). The next (eight) period involves the construction of the $g$-shell. In 1969, Glenn Seaborg \cite{Sea69} proposed an eight-periodic table containing $g$-shell. The first element of $g$-shell is \textbf{Ubu} (Unbiunium) with the atomic number $Z=121$ (superactinide group starts also with this element). The full number of elements of Seaborg table is equal to 218.

No one knows how many elements can be in the periodic system. The Reserford-Bohr structural model leads to the following restriction (so-called ``Bohr model breakdown'') on the number of physically possible elements. So, for elements with atomic number greater than 137 a ``speed'' of an electron in $1s$ orbital is given by
\[
v=Z\alpha c\approx\frac{Zc}{137,036},
\]
where $\alpha$ is the fine structure constant. Under this approximation, any element with $Z>137$ would require $1s$ electrons to be travelling faster than $c$. On the other hand, Feynman pointed out that a relativistic Dirac equation leads also into problems with $Z>137$, since a ground state energy for the electron on $1s$-subshell is given by an expression $E=m_0c^2\sqrt{1-Z^2\alpha^2}$, where $m_0$ is the rest mass of the electron. In the case of $Z>137$, an energy value becomes an imaginary number, and, therefore, the wave function of the ground state is oscillatory, that is, there is no gap between the positive and negative energy spectra, as in the Klein paradox. For that reason 137-th element \textbf{Uts} (Untriseptium) was proclaimed as the ``end'' of the periodic system, and also in honour of Feynman this element was called Feynmanium (symbol: \textbf{Fy}). As is known, Feynman derived this result with the assimption that the atomic nucleus is point-like.
\unitlength=1mm
\begin{center}
\begin{picture}(120,220)(7,0)
\put(-4,203){$\lambda=0$}
\put(10,209){$\overbrace{\scriptscriptstyle s^\prime=-1/2\; s^\prime=1/2}^{\nu=1}$} \put(10,200){\framebox(7,7){}}\put(12,204){\footnotesize H}\put(12,201){\footnotesize He}
\put(21,200){\framebox(7,7){}}\put(23,204){\footnotesize Li}\put(23,201){\footnotesize Be}
\put(32,209){$\overbrace{\scriptscriptstyle s^\prime=-1/2\; s^\prime=1/2}^{\nu=2}$}
\put(32,200){\framebox(7,7){}}\put(34,204){\footnotesize Na}\put(34,201){\footnotesize Mg}
\put(43,200){\framebox(7,7){}}\put(45,204){\footnotesize K}\put(45,201){\footnotesize Ca}
\put(54,209){$\overbrace{\scriptscriptstyle s^\prime=-1/2\;\; s^\prime=1/2}^{\nu=3}$}
\put(54,200){\framebox(7,7){}}\put(56,204){\footnotesize Rb}\put(56,201){\footnotesize Sr}
\put(65,200){\framebox(7,7){}}\put(67,204){\footnotesize Cs}\put(67,201){\footnotesize Ba}
\put(76,209){$\overbrace{\scriptscriptstyle s^\prime=-1/2\;\; s^\prime=1/2}^{\nu=4}$}
\put(76,200){\framebox(7,7){}}\put(78,204){\footnotesize Fr}\put(78,201){\footnotesize Ra}
\put(87,200){\framebox(7,7){}}
\put(87.25,204){{\footnotesize Uue}}\put(87.25,201){{\footnotesize Ubn}}
\put(98,209){$\overbrace{\scriptscriptstyle s^\prime=-1/2\;\; s^\prime=1/2}^{\nu=5}$}
\put(98,200){\framebox(7,7){}}\put(98.25,204){\footnotesize Uhe}\put(98.25,201){\footnotesize Usn}
\put(109,200){\framebox(7,7){}}
\put(109.25,204){{\footnotesize Bue}}\put(109.25,201){{\footnotesize Bbn}}
\put(118,204){$\scriptscriptstyle s=-1/2$}
\put(118,201){$\scriptscriptstyle s=1/2$}
\put(125,203){$\left.\phantom{\framebox(1,4){}}\right\}$}
\put(129,203){$\scr\mu=0$}
\put(-4,182){$\lambda=1$}
\put(12,182){$\left\{\phantom{\framebox(1,12){}}\right.$}
\put(32,187){\framebox(7,7){}}\put(34,191){\footnotesize B}\put(34,188){\footnotesize C}
\put(43,187){\framebox(7,7){}}\put(45,191){\footnotesize Al}\put(45,188){\footnotesize Si}
\put(54,187){\framebox(7,7){}}\put(56,191){\footnotesize Ga}\put(56,188){\footnotesize Ge}
\put(65,187){\framebox(7,7){}}\put(67,191){\footnotesize In}\put(67,188){\footnotesize Sn}
\put(76,187){\framebox(7,7){}}\put(78,191){\footnotesize Tl}\put(78,188){\footnotesize Pb}
\put(87,187){\framebox(7,7){}}\put(89,191){{\footnotesize Nh}}\put(89,188){{\footnotesize Fl}}
\put(98,187){\framebox(7,7){}}\put(98.25,191){\footnotesize Uht}\put(98.25,188){\footnotesize Uhq}
\put(109,187){\framebox(7,7){}}\put(109.25,191){{\footnotesize But}}\put(109.25,188){{\footnotesize Buq}}
\put(118,191){$\scriptscriptstyle s=-1/2$}
\put(118,188){$\scriptscriptstyle s=1/2$}
\put(125,190){$\left.\phantom{\framebox(1,4){}}\right\}$}
\put(129,190){$\scr\mu=-1$}
\put(32,180){\framebox(7,7){}}\put(34,184){\footnotesize N}\put(34,181){\footnotesize O}
\put(43,180){\framebox(7,7){}}\put(45,184){\footnotesize P}\put(45,181){\footnotesize S}
\put(54,180){\framebox(7,7){}}\put(56,184){\footnotesize As}\put(56,181){\footnotesize Se}
\put(65,180){\framebox(7,7){}}\put(67,184){\footnotesize Sb}\put(67,181){\footnotesize Te}
\put(76,180){\framebox(7,7){}}\put(78,184){\footnotesize Bi}\put(78,181){\footnotesize Po}
\put(87,180){\framebox(7,7){}}\put(89,184){{\footnotesize Mc}}\put(89,181){{\footnotesize Lv}}
\put(98,180){\framebox(7,7){}}\put(98.25,184){\footnotesize Uhp}\put(98.25,181){\footnotesize Uhn}
\put(109,180){\framebox(7,7){}}\put(109.25,184){{\footnotesize Bup}}\put(109.25,181){{\footnotesize Buh}}
\put(118,184){$\scriptscriptstyle s=-1/2$}
\put(118,181){$\scriptscriptstyle s=1/2$}
\put(125,183){$\left.\phantom{\framebox(1,4){}}\right\}$}
\put(129,183){$\scr\mu=0$}
\put(32,173){\framebox(7,7){}}\put(34,177){\footnotesize F}\put(34,174){\footnotesize Ne}
\put(43,173){\framebox(7,7){}}\put(45,177){\footnotesize Cl}\put(45,174){\footnotesize Ar}
\put(54,173){\framebox(7,7){}}\put(56,177){\footnotesize Br}\put(56,174){\footnotesize Kr}
\put(65,173){\framebox(7,7){}}\put(67,177){\footnotesize I}\put(67,174){\footnotesize Xe}
\put(76,173){\framebox(7,7){}}\put(78,177){\footnotesize At}\put(78,174){\footnotesize Rn}
\put(87,173){\framebox(7,7){}}\put(89,177){{\footnotesize Ts}}\put(89,174){{\footnotesize Og}}
\put(98,173){\framebox(7,7){}}\put(98.25,177){\footnotesize Uhs}\put(98.25,174){\footnotesize Uho}
\put(109,173){\framebox(7,7){}}\put(109.25,177){{\footnotesize Bus}}\put(109.25,174){{\footnotesize Buo}}
\put(118,177){$\scriptscriptstyle s=-1/2$}
\put(118,174){$\scriptscriptstyle s=1/2$}
\put(125,176){$\left.\phantom{\framebox(1,4){}}\right\}$}
\put(129,176){$\scr\mu=1$}
\put(-4,151){$\lambda=2$}
\put(12,151){$\left\{\phantom{\framebox(1,18){}}\right.$}
\put(54,163){\framebox(7,7){}}\put(56,167){\footnotesize Sc}\put(56,164){\footnotesize Ti}
\put(65,163){\framebox(7,7){}}\put(67,167){\footnotesize Y}\put(67,164){\footnotesize Zr}
\put(76,163){\framebox(7,7){}}\put(78,167){\footnotesize Lu}\put(78,164){\footnotesize Hf}
\put(87,163){\framebox(7,7){}}\put(89,167){\footnotesize Lr}\put(89,164){\footnotesize Rf}
\put(98,163){\framebox(7,7){}}\put(98.25,167){\footnotesize Upt}\put(98.25,164){\footnotesize Upq}
\put(109,163){\framebox(7,7){}}\put(109.25,167){{\footnotesize Bnt}}\put(109.25,164){{\footnotesize Bnq}}
\put(118,167){$\scriptscriptstyle s=-1/2$}
\put(118,164){$\scriptscriptstyle s=1/2$}
\put(125,166){$\left.\phantom{\framebox(1,4){}}\right\}$}
\put(129,166){$\scr\mu=-2$}
\put(54,156){\framebox(7,7){}}\put(56,160){\footnotesize V}\put(56,157){\footnotesize Cr}
\put(65,156){\framebox(7,7){}}\put(67,160){\footnotesize Nb}\put(67,157){\footnotesize Mo}
\put(76,156){\framebox(7,7){}}\put(78,160){\footnotesize Ta}\put(78,157){\footnotesize W}
\put(87,156){\framebox(7,7){}}\put(89,160){{\footnotesize Db}}\put(89,157){{\footnotesize Sg}}
\put(98,156){\framebox(7,7){}}\put(98.25,160){\footnotesize Upp}\put(98.25,157){\footnotesize Uph}
\put(109,156){\framebox(7,7){}}\put(109.25,160){{\footnotesize Bnp}}\put(109.25,157){{\footnotesize Bnh}}
\put(118,160){$\scriptscriptstyle s=-1/2$}
\put(118,157){$\scriptscriptstyle s=1/2$}
\put(125,159){$\left.\phantom{\framebox(1,4){}}\right\}$}
\put(129,159){$\scr\mu=-1$}
\put(54,149){\framebox(7,7){}}\put(56,153){\footnotesize Mn}\put(56,150){\footnotesize Fe}
\put(65,149){\framebox(7,7){}}\put(67,153){\footnotesize Tc}\put(67,150){\footnotesize Ru}
\put(76,149){\framebox(7,7){}}\put(78,153){\footnotesize Re}\put(78,150){\footnotesize Os}
\put(87,149){\framebox(7,7){}}\put(89,153){{\footnotesize Bh}}\put(89,150){{\footnotesize Hs}}
\put(98,149){\framebox(7,7){}}\put(98.25,153){\footnotesize Ups}\put(98.25,150){\footnotesize Upo}
\put(109,149){\framebox(7,7){}}\put(109.25,153){{\footnotesize Bns}}\put(109.25,150){{\footnotesize Bno}}
\put(118,153){$\scriptscriptstyle s=-1/2$}
\put(118,150){$\scriptscriptstyle s=1/2$}
\put(125,152){$\left.\phantom{\framebox(1,4){}}\right\}$}
\put(129,152){$\scr\mu=0$}
\put(54,142){\framebox(7,7){}}\put(56,146){\footnotesize Co}\put(56,143){\footnotesize Ni}
\put(65,142){\framebox(7,7){}}\put(67,146){\footnotesize Rh}\put(67,143){\footnotesize Pd}
\put(76,142){\framebox(7,7){}}\put(78,146){\footnotesize Ir}\put(78,143){\footnotesize Pt}
\put(87,142){\framebox(7,7){}}\put(89,146){{\footnotesize Mt}}\put(89,143){{\footnotesize Ds}}
\put(98,142){\framebox(7,7){}}\put(98.25,146){\footnotesize Upe}\put(98.25,143){\footnotesize Uhn}
\put(109,142){\framebox(7,7){}}\put(109.25,146){{\footnotesize Bne}}\put(109.25,143){{\footnotesize Bun}}
\put(118,146){$\scriptscriptstyle s=-1/2$}
\put(118,143){$\scriptscriptstyle s=1/2$}
\put(125,145){$\left.\phantom{\framebox(1,4){}}\right\}$}
\put(129,145){$\scr\mu=1$}
\put(54,135){\framebox(7,7){}}\put(56,139){\footnotesize Cu}\put(56,136){\footnotesize Zn}
\put(65,135){\framebox(7,7){}}\put(67,139){\footnotesize Ag}\put(67,136){\footnotesize Cd}
\put(76,135){\framebox(7,7){}}\put(78,139){\footnotesize Au}\put(78,136){\footnotesize Hg}
\put(87,135){\framebox(7,7){}}\put(89,139){{\footnotesize Rg}}\put(89,136){{\footnotesize Cn}}
\put(98,135){\framebox(7,7){}}\put(98.25,139){\footnotesize Uhu}\put(98.25,136){\footnotesize Uhb}
\put(109,135){\framebox(7,7){}}\put(109.25,139){{\footnotesize Buu}}\put(109.25,136){{\footnotesize Bub}}
\put(118,139){$\scriptscriptstyle s=-1/2$}
\put(118,136){$\scriptscriptstyle s=1/2$}
\put(125,138){$\left.\phantom{\framebox(1,4){}}\right\}$}
\put(129,138){$\scr\mu=2$}
\put(-4,106){$\lambda=3$}
\put(12,106){$\left\{\phantom{\framebox(1,25){}}\right.$}
\put(76,125){\framebox(7,7){}}\put(78,129){\footnotesize La}\put(78,126){\footnotesize Ce}
\put(87,125){\framebox(7,7){}}\put(89,129){\footnotesize Ac}\put(89,126){\footnotesize Th}
\put(98,125){\framebox(7,7){}}\put(98.25,129){\footnotesize Ute}\put(98.25,126){\footnotesize Uqn}
\put(109,125){\framebox(7,7){}}\put(109.25,129){{\footnotesize Uoe}}\put(109.25,126){{\footnotesize Uen}}
\put(118,129){$\scriptscriptstyle s=-1/2$}
\put(118,126){$\scriptscriptstyle s=1/2$}
\put(125,128){$\left.\phantom{\framebox(1,4){}}\right\}$}
\put(129,128){$\scr\mu=-3$}
\put(76,118){\framebox(7,7){}}\put(78,122){\footnotesize Pr}\put(78,119){\footnotesize Nd}
\put(87,118){\framebox(7,7){}}\put(89,122){\footnotesize Pa}\put(89,119){\footnotesize U}
\put(98,118){\framebox(7,7){}}\put(98.25,122){\footnotesize Uqu}\put(98.25,119){\footnotesize Uqb}
\put(109,118){\framebox(7,7){}}\put(109.25,122){{\footnotesize Ueu}}\put(109.25,119){{\footnotesize Ueb}}
\put(118,122){$\scriptscriptstyle s=-1/2$}
\put(118,119){$\scriptscriptstyle s=1/2$}
\put(125,121){$\left.\phantom{\framebox(1,4){}}\right\}$}
\put(129,121){$\scr\mu=-2$}
\put(76,111){\framebox(7,7){}}\put(78,115){\footnotesize Pm}\put(78,112){\footnotesize Sm}
\put(87,111){\framebox(7,7){}}\put(89,115){\footnotesize Np}\put(89,112){\footnotesize Pu}
\put(98,111){\framebox(7,7){}}\put(98.25,115){\footnotesize Uqt}\put(98.25,112){\footnotesize Uqq}
\put(109,111){\framebox(7,7){}}\put(109.25,115){{\footnotesize Uet}}\put(109.25,112){{\footnotesize Ueq}}
\put(118,115){$\scriptscriptstyle s=-1/2$}
\put(118,112){$\scriptscriptstyle s=1/2$}
\put(125,114){$\left.\phantom{\framebox(1,4){}}\right\}$}
\put(129,114){$\scr\mu=-1$}
\put(76,104){\framebox(7,7){}}\put(78,108){\footnotesize Eu}\put(78,105){\footnotesize Gd}
\put(87,104){\framebox(7,7){}}\put(89,108){\footnotesize Am}\put(89,105){\footnotesize Cm}
\put(98,104){\framebox(7,7){}}\put(98.25,108){\footnotesize Uqp}\put(98.25,105){\footnotesize Uqh}
\put(109,104){\framebox(7,7){}}\put(109.25,108){{\footnotesize Uep}}\put(109.25,105){{\footnotesize Ueh}}
\put(118,108){$\scriptscriptstyle s=-1/2$}
\put(118,105){$\scriptscriptstyle s=1/2$}
\put(125,107){$\left.\phantom{\framebox(1,4){}}\right\}$}
\put(129,107){$\scr\mu=0$}
\put(76,97){\framebox(7,7){}}\put(78,101){\footnotesize Tb}\put(78,98){\footnotesize Dy}
\put(87,97){\framebox(7,7){}}\put(89,101){\footnotesize Bk}\put(89,98){\footnotesize Cf}
\put(98,97){\framebox(7,7){}}\put(98.25,101){\footnotesize Uqs}\put(98.25,98){\footnotesize Uqo}
\put(109,97){\framebox(7,7){}}\put(109.25,101){{\footnotesize Ues}}\put(109.25,98){{\footnotesize Ueo}}
\put(118,101){$\scriptscriptstyle s=-1/2$}
\put(118,98){$\scriptscriptstyle s=1/2$}
\put(125,100){$\left.\phantom{\framebox(1,4){}}\right\}$}
\put(129,100){$\scr\mu=1$}
\put(76,90){\framebox(7,7){}}\put(78,94){\footnotesize Ho}\put(78,91){\footnotesize Er}
\put(87,90){\framebox(7,7){}}\put(89,94){\footnotesize Es}\put(89,91){\footnotesize Fm}
\put(98,90){\framebox(7,7){}}\put(98.25,94){\footnotesize Uqe}\put(98.25,91){\footnotesize Upn}
\put(109,90){\framebox(7,7){}}\put(109.25,94){{\footnotesize Uee}}\put(109.25,91){{\footnotesize Bnn}}
\put(118,94){$\scriptscriptstyle s=-1/2$}
\put(118,91){$\scriptscriptstyle s=1/2$}
\put(125,93){$\left.\phantom{\framebox(1,4){}}\right\}$}
\put(129,93){$\scr\mu=2$}
\put(76,83){\framebox(7,7){}}\put(78,87){\footnotesize Tm}\put(78,84){\footnotesize Yb}
\put(87,83){\framebox(7,7){}}\put(89,87){\footnotesize Md}\put(89,84){\footnotesize No}
\put(98,83){\framebox(7,7){}}\put(98.25,87){\footnotesize Upu}\put(98.25,84){\footnotesize Upb}
\put(109,83){\framebox(7,7){}}\put(109.25,87){{\footnotesize Bnu}}\put(109.25,84){{\footnotesize Bnb}}
\put(8,82){\dashbox{1}(88,135)[b]{\bf M}}
\put(118,87){$\scriptscriptstyle s=-1/2$}
\put(118,84){$\scriptscriptstyle s=1/2$}
\put(125,86){$\left.\phantom{\framebox(1,4){}}\right\}$}
\put(129,86){$\scr\mu=3$}
\put(-4,48){$\lambda=4$}
\put(12,48){$\left\{\phantom{\framebox(1,33){}}\right.$}
\put(98,73){\framebox(7,7){}}\put(98.25,77){\footnotesize Ubu}\put(98.25,74){\footnotesize Ubb}
\put(109,73){\framebox(7,7){}}\put(109.25,77){{\footnotesize Usu}}\put(109.25,74){{\footnotesize Usb}}
\put(118,77){$\scriptscriptstyle s=-1/2$}
\put(118,74){$\scriptscriptstyle s=1/2$}
\put(125,76){$\left.\phantom{\framebox(1,4){}}\right\}$}
\put(129,76){$\scr\mu=-4$}
\put(98,66){\framebox(7,7){}}\put(98.25,70){\footnotesize Ubt}\put(98.25,67){\footnotesize Ubq}
\put(109,66){\framebox(7,7){}}\put(109.25,70){{\footnotesize Ust}}\put(109.25,67){{\footnotesize Usq}}
\put(118,70){$\scriptscriptstyle s=-1/2$}
\put(118,67){$\scriptscriptstyle s=1/2$}
\put(125,69){$\left.\phantom{\framebox(1,4){}}\right\}$}
\put(129,69){$\scr\mu=-3$}
\put(98,59){\framebox(7,7){}}\put(98.25,63){\footnotesize Ubp}\put(98.25,60){\footnotesize Ubn}
\put(109,59){\framebox(7,7){}}\put(109.25,63){{\footnotesize Usp}}\put(109.25,60){{\footnotesize Ush}}
\put(118,63){$\scriptscriptstyle s=-1/2$}
\put(118,60){$\scriptscriptstyle s=1/2$}
\put(125,62){$\left.\phantom{\framebox(1,4){}}\right\}$}
\put(129,62){$\scr\mu=-2$}
\put(98,52){\framebox(7,7){}}\put(98.25,56){\footnotesize Ubs}\put(98.25,53){\footnotesize Ubo}
\put(109,52){\framebox(7,7){}}\put(109.25,56){{\footnotesize Uss}}\put(109.25,53){{\footnotesize Uso}}
\put(118,56){$\scriptscriptstyle s=-1/2$}
\put(118,53){$\scriptscriptstyle s=1/2$}
\put(125,55){$\left.\phantom{\framebox(1,4){}}\right\}$}
\put(129,55){$\scr\mu=-1$}
\put(98,45){\framebox(7,7){}}\put(98.25,49){\footnotesize Ube}\put(98.25,46){\footnotesize Utn}
\put(109,45){\framebox(7,7){}}\put(109.25,49){{\footnotesize Use}}\put(109.25,46){{\footnotesize Uon}}
\put(118,49){$\scriptscriptstyle s=-1/2$}
\put(118,46){$\scriptscriptstyle s=1/2$}
\put(125,48){$\left.\phantom{\framebox(1,4){}}\right\}$}
\put(129,48){$\scr\mu=0$}
\put(98,38){\framebox(7,7){}}\put(98.25,42){\footnotesize Utu}\put(98.25,39){\footnotesize Utb}
\put(109,38){\framebox(7,7){}}\put(109.25,42){{\footnotesize Uou}}\put(109.25,39){{\footnotesize Uob}}
\put(118,42){$\scriptscriptstyle s=-1/2$}
\put(118,39){$\scriptscriptstyle s=1/2$}
\put(125,41){$\left.\phantom{\framebox(1,4){}}\right\}$}
\put(129,41){$\scr\mu=1$}
\put(98,31){\framebox(7,7){}}\put(98.25,35){\footnotesize Utt}\put(98.25,32){\footnotesize Utq}
\put(109,31){\framebox(7,7){}}\put(109.25,35){{\footnotesize Uot}}\put(109.25,32){{\footnotesize Uoq}}
\put(118,35){$\scriptscriptstyle s=-1/2$}
\put(118,32){$\scriptscriptstyle s=1/2$}
\put(125,34){$\left.\phantom{\framebox(1,4){}}\right\}$}
\put(129,34){$\scr\mu=2$}
\put(98,24){\framebox(7,7){}}\put(98.25,28){\footnotesize Utp}\put(98.25,25){\footnotesize Uth}
\put(109,24){\framebox(7,7){}}\put(109.25,28){{\footnotesize Uop}}\put(109.25,25){{\footnotesize Uoh}}
\put(118,28){$\scriptscriptstyle s=-1/2$}
\put(118,25){$\scriptscriptstyle s=1/2$}
\put(125,27){$\left.\phantom{\framebox(1,4){}}\right\}$}
\put(129,27){$\scr\mu=3$}
\put(98,17){\framebox(7,7){}}\put(98.25,21){\footnotesize Uts}\put(98.25,18){\footnotesize Uto}
\put(109,17){\framebox(7,7){}}\put(109.25,21){{\footnotesize Uos}}\put(109.25,18){{\footnotesize Uoo}}
\put(118,21){$\scriptscriptstyle s=-1/2$}
\put(118,18){$\scriptscriptstyle s=1/2$}
\put(125,20){$\left.\phantom{\framebox(1,4){}}\right\}$}
\put(129,20){$\scr\mu=4$}
\put(10,5){\begin{minipage}{25pc}{\small {\bf Fig.\,2.} Seaborg table in the form of the basic representation $F^+_{ss^\prime}$ of the Rumer-Fet group (basis $|\nu,\lambda,\mu,s,s^\prime\rangle$).}\end{minipage}}
\end{picture}
\end{center}

Further, Greiner-Reinhardt solution \cite{GR09}, representing the atomic nucleus by a charged ball of the radius $R=1,2A^{1/3}fm$, where $A$ is the atomic mass, moves aside the Feynman limit to the value $Z=173$. For $Z\approx 173$ under action of the electric field of the nucleus $1s$-subshell ``dives'' into the negative continuum (Dirac sea), that leads to spontaneous emission of electron-positron pairs and, as a consequence, to the absence of neutral atoms above the element \textbf{Ust} (Unsepttrium) with $Z=173$. Atoms with $Z>Z_{cr}\approx 173$ are called \textit{supercritical} atoms. It is supposed that elements with $Z>Z_{cr}$ could only exist as ions.

As shown earlier, the Seaborg table is an eight-periodic extension of the Mendeleev table (from 119-th to the 218-th element). The Seaborg table contains both ``critical'' elements of the Bohr model: \textbf{Uts} (Untriseptium, $Z=137$) and \textbf{Ust} (Unsepttrium, $Z=173$). According to Bohr model, the filling of the $g$-shell (formation of $g$-family) begins with the 121-th element. In the Rumer-Fet model \cite{Fet} $g$-shell corresponds to quantum numbers $\nu=5$ and $\lambda=4$ of the symmetry group $\SO(2,4)\otimes\SU(2)\otimes\SU(2)^\prime$. The Seaborg table is presented on the Fig.\,2 in the form of the basic representation $F^+_{ss^\prime}$ of the Rumer-Fet group. The Mendeleev table (as part of the Seaborg table) is highlighted by a dotted border. Within eight-periodic extension (quantum numbers $\nu=5$, $\lambda=4$) in addition to 20 multiplets of the Mendeleev table we have 10 multiplets.

Let us calculate average masses of these multiplets. With this aim in view we use the mass formula (\ref{Mass1}). The formula (\ref{Mass1}) corresponds to the chain of groups (\ref{Chain1}), according to which we have a reduction of the basic representation $F^+_{ss^\prime}$ on the subgroups of this chain, that is, a partition of the basic multiplet into the lesser multiplets. As noted above, the formula (\ref{Mass1}) is analogous to the ``first perturbation'' in $\SU(3)$- and $\SU(6)$-theories, which allows us to calculate an average mass of the elements belonging to a given multiplet\footnote{So, in $\SU(3)$-theory we have a Gell-Mann--Okubo mass formula
\[
m=m_0+\alpha+\beta Y+\gamma\left[I(I+1)-\frac{1}{4}Y^2\right]
+\alpha^\prime-\beta^\prime
Q+\gamma^\prime\left[U(U+1)-\frac{1}{4}Q^2\right],
\]
in which, according to $\SU(3)/\SU(2)$-reduction, quantum numbers (isospin $I$, hypercharge $Y$), standing in the first square bracket, define the ``first perturbation'' that leads to a so-called \textit{hypercharge} mass splitting, that is, a partition of the multiplet of $\SU(3)$ into the lesser multiplets of the subgroup $\SU(2)$. A ``second perturbation'' is defined by the quantum numbers, standing in the second square bracket (charge $Q$ and isospin $U$ which in difference from $I$ corresponds to other choice of the basis in the subgroup $\SU(2)$), that leads to a \textit{charge} mass splitting inside the multiplets of $\SU(2)$.
}. At $m_0=1$, $a=17$, $b=5,5$ from (\ref{Mass1}) we obtain the average masses of multiplets (see Tab.\,2).
\begin{center}
{\textbf{Tab.\,2.} Average masses of multiplets of Seaborg table.}
\vspace{0.1cm}
{\renewcommand{\arraystretch}{1.0}
\begin{tabular}{|c||l|c|}\hline
 & Multiplet & Mass (theor.) \\ \hline\hline
1. & $(\nu=5,s^\prime=-1/2,\lambda=4)$ & $316$ \\
2. & $(\nu=5,s^\prime=-1/2,\lambda=3)$ & $360$ \\
3. & $(\nu=5,s^\prime=-1/2,\lambda=2)$ & $393$ \\
4. & $(\nu=5,s^\prime=-1/2,\lambda=1)$ & $415$ \\
5. & $(\nu=5,s^\prime=-1/2,\lambda=0)$ & $426$ \\
6. & $(\nu=5,s^\prime=1/2,\lambda=4)$ & $435$ \\
7. & $(\nu=5,s^\prime=1/2,\lambda=3)$ & $479$ \\
8. & $(\nu=5,s^\prime=1/2,\lambda=2)$ & $512$ \\
9. & $(\nu=5,s^\prime=1/2,\lambda=1)$ & $534$ \\
10. & $(\nu=5,s^\prime=1/2,\lambda=0)$& $545$ \\
\hline
\end{tabular}
}
\end{center}

With the aim to obtain an analogue of the ``second perturbation'', which leads to a mass splitting inside the multiplets of the group $G_2=\SO(3)\otimes\SU(2)$, it needs to find a subsequent lengthening of the group chain $G\supset G_1\supset G_2$ (\ref{Chain1}). Therefore, we need to find another subgroup $G_3$. Then $G_2/G_3$-reduction gives a termwise mass splitting. As is known, a representation $\{u_O\}$ of the group $\SU(2)$ compares the each rotation $O$ from $\SO(3)$ with the matrix $u_O$ from $\SU(2)$ and thereby the pair $(O,u_O)$, that is, the element of $G_2$. At the multiplication $(O,u_O)$ give the pairs of the same form: $(O_1,u_{O_1})(O_2,u_{O_2})=(O_1O_2,u_{O_1}u_{O_2})=(O_1O_2,u_{O_1O_2})$, reverse pairs are analogous: $(O,u_O)^{-1}=(O^{-1},u_{O^{-1}})$. Therefore, such pairs form a subgroup $G_3$ in $G_2$. The subgroup $G_3$ is locally isomorphic to $\SO(3)$. Following to Fet, we will denote it via $\SO(3)_c$. Further, one-parameter subgroups of the group $\SO(3)$ have the form $\left\{e^{-i\alpha A_k}\right\}$ ($k=1,2,3$); since the representation $\{u_O\}$ converts them into one-parameter subgroups $\left\{e^{-i\alpha\boldsymbol{\tau}_k}\right\}$ of the group $\SU(2)$, then corresponding one-parameter subgroups in $\SO(3)_c$ have the form
\[
\left(e^{-i\alpha A_k},e^{-i\alpha\boldsymbol{\tau}_k}\right)=\left(e^{-i\alpha A_k},1\right)
\left(1,e^{-i\alpha\boldsymbol{\tau}_k}\right).
\]
Since the matrices $A_1$, $A_2$, $A_3$ correspond to rotations $\bsL_{23}=\bsJ_1+\bsK_1$, $\bsL_{31}=\bsJ_2+\bsK_2$, $\bsL_{12}=\bsJ_3+\bsK_3$ in the group $G$ (in the basic representation of the group $G$, see section 3), the pair $\left(e^{-i\alpha A_k},1\right)$ is represented by the operator $e^{-i\alpha(\bsJ_k+\bsK_k)}$; the pair $\left(1,e^{-i\alpha\boldsymbol{\tau}_k}\right)$ is represented by $e^{-i\alpha\boldsymbol{\tau}_k}$. Thus, one-parameter subgroups of $G_3=\SO(3)_c$ correspond to subgroups of operators $e^{-i\alpha(\bsJ_k+\bsK_k)}e^{-i\alpha\boldsymbol{\tau}_k}$ ($k=1,2,3$). Irreducible representations of the group $G_2=\SO(3)\otimes\SU(2)$ are numbered by the collections of quantum numbers $(\nu,s^\prime,\lambda)$. These representations form vertical rectangles in Fig.\,2. Each of them is defined by a fundamental representation of $\SU(2)$ and $(2\lambda+1)$-dimensional irreducible representation of the group $\SO(3)$. At $G_2/G_3$-reduction from such representation we obtain an irreducible representation of the subgroup $G_3=\SO(3)_c$, for which a Clebsch-Gordan sequence $\left|j_1-j_2\right|$, $\ldots$, $j_1+j_2$ with the values $j_1=1/2$ and $j_2=\lambda$ is reduced to two terms $\lambda-1/2$, $\lambda+1/2$ at $\lambda>0$ and to one term $1/2$ at $\lambda=0$. Therefore, at $\lambda>0$ the representation $(\nu,s^\prime,\lambda)$ of the group $G_2$ is reduced to two irreducible representations of the subgroup $G_3$ with dimensionality $2(\lambda-1/2)+1=2\lambda$ and $2(\lambda+1/2)+1=2\lambda+2$, and at $\lambda=0$ to one two-dimensional irreducible representation. Thus, at $\lambda>0$ multiplets of the subgroup $G_2$ are reduced to two multiplets of the subgroup $G_3$. $G_2/G_3$-reduction leads to the following (lengthened) chain of groups:
\begin{multline}
G\supset G_1\supset G_2\supset G_3\longmapsto\\
\SO(2,4)\otimes\SU(2)\otimes\SU(2)\supset\SO(4)\otimes\SU(2)\supset\SO(3)\otimes\SU(2)\supset\SO(3)_c,\label{Chain2}
\end{multline}
The lengthening of the group chain requires the introduction of a new basis whose vectors belong to the smallest multiplets of symmetry, that is, multiplets of the subgroup $G_3$. The vectors $|\nu,\lambda,\mu,s,s^\prime\rangle$ of the basis (\ref{RF4}), corresponding to the group chain (\ref{Chain1}), do not compose already a chosen (well-defined) basis, since $\mu$, $s$ do not belong to quantum numbers of the symmetry group, that is, these vectors do not belong to irreducible spaces of the group $G_3$. The new basis is defined as follows. Since $\nu$, $s^\prime$, $\lambda$ are related with the groups $G$, $G_1$, $G_2$, they remain quantum numbers of the chain (\ref{Chain1}), and instead $\mu$, $s$ we have new quantum numbers related with $G_3$. First, quantum number $\iota_\lambda$ relates with the Casimir operator of the subgroup $G_3$, which equals to $\sum^3_{k=1}(\boldsymbol{\tau}_k+\bsJ_k+\bsK_k)^2$. At this point, two multiplets of $G_3$, obtained at $G_2/G_3$-reduction, correspond to $\iota_\lambda=\lambda-1/2$ and $\iota_\lambda=\lambda+1/2$, whence $2\lambda=2\iota_\lambda+1$, $2\lambda+2=2\iota_\lambda+1$. Other quantum number $\kappa$ is an eigenvalue of the operator $q_3=\boldsymbol{\tau}_3+\bsJ_3+\bsK_3$, which belongs to the Lie algebra of the group $G_3=\SO(3)_c$. Thus, the new basis, corresponding to the group chain (\ref{Chain2}), has the form
\begin{multline}
|\nu,s^\prime,\lambda,\iota_\lambda,\kappa\rangle,\quad \nu=1,2,\ldots;\;s^\prime=-1/2,1/2\;\lambda=0,1,\ldots, \nu-1;\\
\iota_\lambda=\lambda-1/2,\lambda+1/2\;\kappa=-\iota_\lambda, -\iota_\lambda+1,\ldots,\iota_\lambda-1,\iota_\lambda.\label{Basis2}
\end{multline}
The Seaborg table recorded in basis (\ref{Basis2}), is shown in Fig.\,3.
\unitlength=1mm
\begin{center}
\begin{picture}(120,220)(7,0)
\put(-4,203){$\lambda=0$}
\put(10,209){$\overbrace{\scriptscriptstyle s^\prime=-1/2\; s^\prime=1/2}^{\nu=1}$} \put(10,200){\framebox(7,7){}}\put(12,204){\footnotesize H}\put(12,201){\footnotesize He}
\put(21,200){\framebox(7,7){}}\put(23,204){\footnotesize Li}\put(23,201){\footnotesize Be}
\put(32,209){$\overbrace{\scriptscriptstyle s^\prime=-1/2\; s^\prime=1/2}^{\nu=2}$}
\put(32,200){\framebox(7,7){}}\put(34,204){\footnotesize Na}\put(34,201){\footnotesize Mg}
\put(43,200){\framebox(7,7){}}\put(45,204){\footnotesize K}\put(45,201){\footnotesize Ca}
\put(54,209){$\overbrace{\scriptscriptstyle s^\prime=-1/2\;\; s^\prime=1/2}^{\nu=3}$}
\put(54,200){\framebox(7,7){}}\put(56,204){\footnotesize Rb}\put(56,201){\footnotesize Sr}
\put(65,200){\framebox(7,7){}}\put(67,204){\footnotesize Cs}\put(67,201){\footnotesize Ba}
\put(76,209){$\overbrace{\scriptscriptstyle s^\prime=-1/2\;\; s^\prime=1/2}^{\nu=4}$}
\put(76,200){\framebox(7,7){}}\put(78,204){\footnotesize Fr}\put(78,201){\footnotesize Ra}
\put(87,200){\framebox(7,7){}}
\put(87.25,204){{\footnotesize Uue}}\put(87.25,201){{\footnotesize Ubn}}
\put(98,209){$\overbrace{\scriptscriptstyle s^\prime=-1/2\;\; s^\prime=1/2}^{\nu=5}$}
\put(98,200){\framebox(7,7){}}\put(98.25,204){\footnotesize Uhe}\put(98.25,201){\footnotesize Usn}
\put(109,200){\framebox(7,7){}}
\put(109.25,204){{\footnotesize Bue}}\put(109.25,201){{\footnotesize Bbn}}
\put(118,204){$\scriptscriptstyle \kappa=-1/2$}
\put(118,201){$\scriptscriptstyle \kappa=1/2$}
\put(125,203){$\left.\phantom{\framebox(1,4){}}\right\}$}
\put(130,203){$\scr\iota_\lambda=1/2$}
\put(-4,182){$\lambda=1$}
\put(12,182){$\left\{\phantom{\framebox(1,12){}}\right.$}
\put(32,187){\framebox(7,7){}}\put(34,191){\footnotesize B}\put(34,188){\footnotesize C}
\put(43,187){\framebox(7,7){}}\put(45,191){\footnotesize Al}\put(45,188){\footnotesize Si}
\put(54,187){\framebox(7,7){}}\put(56,191){\footnotesize Ga}\put(56,188){\footnotesize Ge}
\put(65,187){\framebox(7,7){}}\put(67,191){\footnotesize In}\put(67,188){\footnotesize Sn}
\put(76,187){\framebox(7,7){}}\put(78,191){\footnotesize Tl}\put(78,188){\footnotesize Pb}
\put(87,187){\framebox(7,7){}}\put(89,191){{\footnotesize Nh}}\put(89,188){{\footnotesize Fl}}
\put(98,187){\framebox(7,7){}}\put(98.25,191){\footnotesize Uht}\put(98.25,188){\footnotesize Uhq}
\put(109,187){\framebox(7,7){}}\put(109.25,191){{\footnotesize But}}\put(109.25,188){{\footnotesize Buq}}
\put(118,191){$\scriptscriptstyle \kappa=-1/2$}
\put(118,188){$\scriptscriptstyle \kappa=1/2$}
\put(125,190){$\left.\phantom{\framebox(1,4){}}\right\}$}
\put(130,190){$\scr\iota_\lambda=1/2$}
\put(32,180){\framebox(7,7){}}\put(34,184){\footnotesize N}\put(34,181){\footnotesize O}
\put(43,180){\framebox(7,7){}}\put(45,184){\footnotesize P}\put(45,181){\footnotesize S}
\put(54,180){\framebox(7,7){}}\put(56,184){\footnotesize As}\put(56,181){\footnotesize Se}
\put(65,180){\framebox(7,7){}}\put(67,184){\footnotesize Sb}\put(67,181){\footnotesize Te}
\put(76,180){\framebox(7,7){}}\put(78,184){\footnotesize Bi}\put(78,181){\footnotesize Po}
\put(87,180){\framebox(7,7){}}\put(89,184){{\footnotesize Mc}}\put(89,181){{\footnotesize Lv}}
\put(98,180){\framebox(7,7){}}\put(98.25,184){\footnotesize Uhp}\put(98.25,181){\footnotesize Uhn}
\put(109,180){\framebox(7,7){}}\put(109.25,184){{\footnotesize Bup}}\put(109.25,181){{\footnotesize Buh}}
\put(118,184){$\scriptscriptstyle \kappa=-3/2$}
\put(118,181){$\scriptscriptstyle \kappa=-1/2$}
\put(32,173){\framebox(7,7){}}\put(34,177){\footnotesize F}\put(34,174){\footnotesize Ne}
\put(43,173){\framebox(7,7){}}\put(45,177){\footnotesize Cl}\put(45,174){\footnotesize Ar}
\put(54,173){\framebox(7,7){}}\put(56,177){\footnotesize Br}\put(56,174){\footnotesize Kr}
\put(65,173){\framebox(7,7){}}\put(67,177){\footnotesize I}\put(67,174){\footnotesize Xe}
\put(76,173){\framebox(7,7){}}\put(78,177){\footnotesize At}\put(78,174){\footnotesize Rn}
\put(87,173){\framebox(7,7){}}\put(89,177){{\footnotesize Ts}}\put(89,174){{\footnotesize Og}}
\put(98,173){\framebox(7,7){}}\put(98.25,177){\footnotesize Uhs}\put(98.25,174){\footnotesize Uho}
\put(109,173){\framebox(7,7){}}\put(109.25,177){{\footnotesize Bus}}\put(109.25,174){{\footnotesize Buo}}
\put(118,177){$\scriptscriptstyle \kappa=1/2$}
\put(118,174){$\scriptscriptstyle \kappa=3/2$}
\put(125,179){$\left.\phantom{\framebox(1,8){}}\right\}$}
\put(130,179){$\scr\iota_\lambda=3/2$}
\put(-4,151){$\lambda=2$}
\put(12,151){$\left\{\phantom{\framebox(1,18){}}\right.$}
\put(54,163){\framebox(7,7){}}\put(56,167){\footnotesize Sc}\put(56,164){\footnotesize Ti}
\put(65,163){\framebox(7,7){}}\put(67,167){\footnotesize Y}\put(67,164){\footnotesize Zr}
\put(76,163){\framebox(7,7){}}\put(78,167){\footnotesize Lu}\put(78,164){\footnotesize Hf}
\put(87,163){\framebox(7,7){}}\put(89,167){\footnotesize Lr}\put(89,164){\footnotesize Rf}
\put(98,163){\framebox(7,7){}}\put(98.25,167){\footnotesize Upt}\put(98.25,164){\footnotesize Upq}
\put(109,163){\framebox(7,7){}}\put(109.25,167){{\footnotesize Bnt}}\put(109.25,164){{\footnotesize Bnq}}
\put(118,167){$\scriptscriptstyle \kappa=-3/2$}
\put(118,164){$\scriptscriptstyle \kappa=-1/2$}
\put(54,156){\framebox(7,7){}}\put(56,160){\footnotesize V}\put(56,157){\footnotesize Cr}
\put(65,156){\framebox(7,7){}}\put(67,160){\footnotesize Nb}\put(67,157){\footnotesize Mo}
\put(76,156){\framebox(7,7){}}\put(78,160){\footnotesize Ta}\put(78,157){\footnotesize W}
\put(87,156){\framebox(7,7){}}\put(89,160){{\footnotesize Db}}\put(89,157){{\footnotesize Sg}}
\put(98,156){\framebox(7,7){}}\put(98.25,160){\footnotesize Upp}\put(98.25,157){\footnotesize Uph}
\put(109,156){\framebox(7,7){}}\put(109.25,160){{\footnotesize Bnp}}\put(109.25,157){{\footnotesize Bnh}}
\put(118,160){$\scriptscriptstyle \kappa=1/2$}
\put(118,157){$\scriptscriptstyle \kappa=3/2$}
\put(125,162){$\left.\phantom{\framebox(1,8){}}\right\}$}
\put(130,162){$\scr\iota_\lambda=3/2$}
\put(54,149){\framebox(7,7){}}\put(56,153){\footnotesize Mn}\put(56,150){\footnotesize Fe}
\put(65,149){\framebox(7,7){}}\put(67,153){\footnotesize Tc}\put(67,150){\footnotesize Ru}
\put(76,149){\framebox(7,7){}}\put(78,153){\footnotesize Re}\put(78,150){\footnotesize Os}
\put(87,149){\framebox(7,7){}}\put(89,153){{\footnotesize Bh}}\put(89,150){{\footnotesize Hs}}
\put(98,149){\framebox(7,7){}}\put(98.25,153){\footnotesize Ups}\put(98.25,150){\footnotesize Upo}
\put(109,149){\framebox(7,7){}}\put(109.25,153){{\footnotesize Bns}}\put(109.25,150){{\footnotesize Bno}}
\put(118,153){$\scriptscriptstyle \kappa=-5/2$}
\put(118,150){$\scriptscriptstyle \kappa=-3/2$}
\put(54,142){\framebox(7,7){}}\put(56,146){\footnotesize Co}\put(56,143){\footnotesize Ni}
\put(65,142){\framebox(7,7){}}\put(67,146){\footnotesize Rh}\put(67,143){\footnotesize Pd}
\put(76,142){\framebox(7,7){}}\put(78,146){\footnotesize Ir}\put(78,143){\footnotesize Pt}
\put(87,142){\framebox(7,7){}}\put(89,146){{\footnotesize Mt}}\put(89,143){{\footnotesize Ds}}
\put(98,142){\framebox(7,7){}}\put(98.25,146){\footnotesize Upe}\put(98.25,143){\footnotesize Uhn}
\put(109,142){\framebox(7,7){}}\put(109.25,146){{\footnotesize Bne}}\put(109.25,143){{\footnotesize Bun}}
\put(118,146){$\scriptscriptstyle \kappa=-1/2$}
\put(118,143){$\scriptscriptstyle \kappa=1/2$}
\put(125,144){$\left.\phantom{\framebox(1,12){}}\right\}$}
\put(130,144){$\scr\iota_\lambda=5/2$}
\put(54,135){\framebox(7,7){}}\put(56,139){\footnotesize Cu}\put(56,136){\footnotesize Zn}
\put(65,135){\framebox(7,7){}}\put(67,139){\footnotesize Ag}\put(67,136){\footnotesize Cd}
\put(76,135){\framebox(7,7){}}\put(78,139){\footnotesize Au}\put(78,136){\footnotesize Hg}
\put(87,135){\framebox(7,7){}}\put(89,139){{\footnotesize Rg}}\put(89,136){{\footnotesize Cn}}
\put(98,135){\framebox(7,7){}}\put(98.25,139){\footnotesize Uhu}\put(98.25,136){\footnotesize Uhb}
\put(109,135){\framebox(7,7){}}\put(109.25,139){{\footnotesize Buu}}\put(109.25,136){{\footnotesize Bub}}
\put(118,139){$\scriptscriptstyle \kappa=3/2$}
\put(118,136){$\scriptscriptstyle \kappa=5/2$}
\put(-4,106){$\lambda=3$}
\put(12,106){$\left\{\phantom{\framebox(1,25){}}\right.$}
\put(76,125){\framebox(7,7){}}\put(78,129){\footnotesize La}\put(78,126){\footnotesize Ce}
\put(87,125){\framebox(7,7){}}\put(89,129){\footnotesize Ac}\put(89,126){\footnotesize Th}
\put(98,125){\framebox(7,7){}}\put(98.25,129){\footnotesize Ute}\put(98.25,126){\footnotesize Uqn}
\put(109,125){\framebox(7,7){}}\put(109.25,129){{\footnotesize Uoe}}\put(109.25,126){{\footnotesize Uen}}
\put(118,129){$\scriptscriptstyle \kappa=-5/2$}
\put(118,126){$\scriptscriptstyle \kappa=-3/2$}
\put(76,118){\framebox(7,7){}}\put(78,122){\footnotesize Pr}\put(78,119){\footnotesize Nd}
\put(87,118){\framebox(7,7){}}\put(89,122){\footnotesize Pa}\put(89,119){\footnotesize U}
\put(98,118){\framebox(7,7){}}\put(98.25,122){\footnotesize Uqu}\put(98.25,119){\footnotesize Uqb}
\put(109,118){\framebox(7,7){}}\put(109.25,122){{\footnotesize Ueu}}\put(109.25,119){{\footnotesize Ueb}}
\put(118,122){$\scriptscriptstyle \kappa=-1/2$}
\put(118,119){$\scriptscriptstyle \kappa=1/2$}
\put(76,111){\framebox(7,7){}}\put(78,115){\footnotesize Pm}\put(78,112){\footnotesize Sm}
\put(87,111){\framebox(7,7){}}\put(89,115){\footnotesize Np}\put(89,112){\footnotesize Pu}
\put(98,111){\framebox(7,7){}}\put(98.25,115){\footnotesize Uqt}\put(98.25,112){\footnotesize Uqq}
\put(109,111){\framebox(7,7){}}\put(109.25,115){{\footnotesize Uet}}\put(109.25,112){{\footnotesize Ueq}}
\put(118,115){$\scriptscriptstyle \kappa=3/2$}
\put(118,112){$\scriptscriptstyle \kappa=5/2$}
\put(125,121){$\left.\phantom{\framebox(1,12){}}\right\}$}
\put(130,121){$\scr\iota_\lambda=5/2$}
\put(76,104){\framebox(7,7){}}\put(78,108){\footnotesize Eu}\put(78,105){\footnotesize Gd}
\put(87,104){\framebox(7,7){}}\put(89,108){\footnotesize Am}\put(89,105){\footnotesize Cm}
\put(98,104){\framebox(7,7){}}\put(98.25,108){\footnotesize Uqp}\put(98.25,105){\footnotesize Uqh}
\put(109,104){\framebox(7,7){}}\put(109.25,108){{\footnotesize Uep}}\put(109.25,105){{\footnotesize Ueh}}
\put(118,108){$\scriptscriptstyle \kappa=-7/2$}
\put(118,105){$\scriptscriptstyle \kappa=-5/2$}
\put(76,97){\framebox(7,7){}}\put(78,101){\footnotesize Tb}\put(78,98){\footnotesize Dy}
\put(87,97){\framebox(7,7){}}\put(89,101){\footnotesize Bk}\put(89,98){\footnotesize Cf}
\put(98,97){\framebox(7,7){}}\put(98.25,101){\footnotesize Uqs}\put(98.25,98){\footnotesize Uqo}
\put(109,97){\framebox(7,7){}}\put(109.25,101){{\footnotesize Ues}}\put(109.25,98){{\footnotesize Ueo}}
\put(118,101){$\scriptscriptstyle \kappa=-3/2$}
\put(118,98){$\scriptscriptstyle \kappa=-1/2$}
\put(76,90){\framebox(7,7){}}\put(78,94){\footnotesize Ho}\put(78,91){\footnotesize Er}
\put(87,90){\framebox(7,7){}}\put(89,94){\footnotesize Es}\put(89,91){\footnotesize Fm}
\put(98,90){\framebox(7,7){}}\put(98.25,94){\footnotesize Uqe}\put(98.25,91){\footnotesize Upn}
\put(109,90){\framebox(7,7){}}\put(109.25,94){{\footnotesize Uee}}\put(109.25,91){{\footnotesize Bnn}}
\put(118,94){$\scriptscriptstyle \kappa=1/2$}
\put(118,91){$\scriptscriptstyle \kappa=3/2$}
\put(76,83){\framebox(7,7){}}\put(78,87){\footnotesize Tm}\put(78,84){\footnotesize Yb}
\put(87,83){\framebox(7,7){}}\put(89,87){\footnotesize Md}\put(89,84){\footnotesize No}
\put(98,83){\framebox(7,7){}}\put(98.25,87){\footnotesize Upu}\put(98.25,84){\footnotesize Upb}
\put(109,83){\framebox(7,7){}}\put(109.25,87){{\footnotesize Bnu}}\put(109.25,84){{\footnotesize Bnb}}
\put(8,82){\dashbox{1}(88,135)[b]{\bf M}}
\put(118,87){$\scriptscriptstyle \kappa=5/2$}
\put(118,84){$\scriptscriptstyle \kappa=7/2$}
\put(125,96){$\left.\phantom{\framebox(1,15){}}\right\}$}
\put(130,96){$\scr\iota_\lambda=7/2$}
\put(-4,48){$\lambda=4$}
\put(12,48){$\left\{\phantom{\framebox(1,33){}}\right.$}
\put(98,73){\framebox(7,7){}}\put(98.25,77){\footnotesize Ubu}\put(98.25,74){\footnotesize Ubb}
\put(109,73){\framebox(7,7){}}\put(109.25,77){{\footnotesize Usu}}\put(109.25,74){{\footnotesize Usb}}
\put(118,77){$\scriptscriptstyle \kappa=-7/2$}
\put(118,74){$\scriptscriptstyle \kappa=-5/2$}
\put(98,66){\framebox(7,7){}}\put(98.25,70){\footnotesize Ubt}\put(98.25,67){\footnotesize Ubq}
\put(109,66){\framebox(7,7){}}\put(109.25,70){{\footnotesize Ust}}\put(109.25,67){{\footnotesize Usq}}
\put(118,70){$\scriptscriptstyle \kappa=-3/2$}
\put(118,67){$\scriptscriptstyle \kappa=-1/2$}
\put(98,59){\framebox(7,7){}}\put(98.25,63){\footnotesize Ubp}\put(98.25,60){\footnotesize Ubn}
\put(109,59){\framebox(7,7){}}\put(109.25,63){{\footnotesize Usp}}\put(109.25,60){{\footnotesize Ush}}
\put(118,63){$\scriptscriptstyle \kappa=1/2$}
\put(118,60){$\scriptscriptstyle \kappa=3/2$}
\put(98,52){\framebox(7,7){}}\put(98.25,56){\footnotesize Ubs}\put(98.25,53){\footnotesize Ubo}
\put(109,52){\framebox(7,7){}}\put(109.25,56){{\footnotesize Uss}}\put(109.25,53){{\footnotesize Uso}}
\put(118,56){$\scriptscriptstyle \kappa=5/2$}
\put(118,53){$\scriptscriptstyle \kappa=7/2$}
\put(125,65){$\left.\phantom{\framebox(1,15){}}\right\}$}
\put(130,65){$\scr\iota_\lambda=7/2$}
\put(98,45){\framebox(7,7){}}\put(98.25,49){\footnotesize Ube}\put(98.25,46){\footnotesize Utn}
\put(109,45){\framebox(7,7){}}\put(109.25,49){{\footnotesize Use}}\put(109.25,46){{\footnotesize Uon}}
\put(118,49){$\scriptscriptstyle \kappa=-9/2$}
\put(118,46){$\scriptscriptstyle \kappa=-7/2$}
\put(98,38){\framebox(7,7){}}\put(98.25,42){\footnotesize Utu}\put(98.25,39){\footnotesize Utb}
\put(109,38){\framebox(7,7){}}\put(109.25,42){{\footnotesize Uou}}\put(109.25,39){{\footnotesize Uob}}
\put(118,42){$\scriptscriptstyle \kappa=-5/2$}
\put(118,39){$\scriptscriptstyle \kappa=-3/2$}
\put(98,31){\framebox(7,7){}}\put(98.25,35){\footnotesize Utt}\put(98.25,32){\footnotesize Utq}
\put(109,31){\framebox(7,7){}}\put(109.25,35){{\footnotesize Uot}}\put(109.25,32){{\footnotesize Uoq}}
\put(118,35){$\scriptscriptstyle \kappa=-1/2$}
\put(118,32){$\scriptscriptstyle \kappa=1/2$}
\put(98,24){\framebox(7,7){}}\put(98.25,28){\footnotesize Utp}\put(98.25,25){\footnotesize Uth}
\put(109,24){\framebox(7,7){}}\put(109.25,28){{\footnotesize Uop}}\put(109.25,25){{\footnotesize Uoh}}
\put(118,28){$\scriptscriptstyle \kappa=3/2$}
\put(118,25){$\scriptscriptstyle \kappa=5/2$}
\put(98,17){\framebox(7,7){}}\put(98.25,21){\footnotesize Uts}\put(98.25,18){\footnotesize Uto}
\put(109,17){\framebox(7,7){}}\put(109.25,21){{\footnotesize Uos}}\put(109.25,18){{\footnotesize Uoo}}
\put(118,21){$\scriptscriptstyle \kappa=7/2$}
\put(118,18){$\scriptscriptstyle \kappa=9/2$}
\put(125,33){$\left.\phantom{\framebox(1,18){}}\right\}$}
\put(130,33){$\scr\iota_\lambda=9/2$}
\put(10,5){\begin{minipage}{25pc}{\small {\bf Fig.\,3.} Seaborg table in the form of the basic representation $F^+_{ss^\prime}$ of the Rumer-Fet group (basis $|\nu,s^\prime,\lambda,\iota_\lambda,\kappa\rangle$).}\end{minipage}}
\end{picture}
\end{center}
\subsection{Masses of Elements}
The lengthened group chain $G\supset G_1\supset G_2\supset G_3$ (\ref{Chain2}) allows us to provide a termwise mass splitting of the basic representation $F^+_{ss^\prime}$ of the Rumer-Fet group. With this aim in view we introduce the following mass formula:
\begin{multline}
m=m_0+a\left[s^\prime(2\nu-3)-5\nu+\frac{11}{2}+2(\nu^2-1)\right]-b\cdot\lambda(\lambda+1)+\\
+a^\prime\left[2\kappa-0,1666\kappa^3+0,0083\kappa^5-0,0001\kappa^7\right]+\left(b^\prime\iota_\lambda\right)^p-1,
\label{Mass2}
\end{multline}
where
\[
p=\left\{\begin{array}{rl}
0, & \mbox{if $\iota_\lambda=\lambda-1/2$};\\
1, & \mbox{if $\iota_\lambda=\lambda+1/2$}.
\end{array}\right.
\]
As the ``first perturbation'' we have in (\ref{Mass2}) the Fet formula (\ref{Mass1}) corresponding to the group chain (\ref{Chain1}), where the basic representation $F^+_{ss^\prime}$ is divided into the multiplets $(\nu,s^\prime,\lambda)$ with average masses of $(\nu,s^\prime,\lambda)$. An analogue of the ``second perturbation'' in the formula (\ref{Mass2}) is defined by quantum numbers $\iota_\lambda$, $\kappa$, that, according to the chain (\ref{Chain2}), leads to a partition of the multiplets $(\nu,s^\prime,\lambda)$ into the pair of multiplets of the subgroup $G_3$ ($G_2/G_3$-reduction), and thereby we have here a termwise mass splitting. The masses of elements of the Seaborg table, starting from atomic number $Z=121$ to $Z=220$\footnote{As noted above, the Seaborg table is an extension of the Mendeleev table, highlighted in the Fig.\,3 by a dotted border. The Table 2 shows the masses of the elements outside the dotted frame. In turn, the Mendeleev table contains two not yet discovered elements \textbf{Uue} ($Z=119$) and \textbf{Ubn} ($Z=120$) corresponding to the basis vectors $|4,1/2,0,1/2,-1/2\rangle$ and $|4,1/2,0,1/2,1/2\rangle$. The masses of \textbf{Uue} and \textbf{Ubn}, calculated according to (\ref{Mass2}), are equal respectively to $304,8942$ and $309,1057$.}, are calculated according to the mass formula (\ref{Mass2}) at the values $m_0=1$, $a=17$, $b=5,5$, $a^\prime=2,15$, $b^\prime=5,3$ (see Tab.\,3). The first column of Tab.\,3 contains atomic number of the element; in the second column we have a generally accepted (according to IUPAC\footnote{IUPAC -- International Union of Pure and Applied Chemistry.}) designation of the element; the third column contains quantum numbers of the element defining the vector $|\nu,s^\prime,\lambda,\iota_\lambda,\kappa\rangle$ of the basis (\ref{Basis2})\footnote{Recall, that according to group-theoretical description, the each element of periodic system corresponds to the vector $|\nu,s^\prime,\lambda,\iota_\lambda,\kappa\rangle$ of the basis (\ref{Basis2}), forming thereby a single quantum system.}; the fourth column contains the mass of the element calculated via the formula (\ref{Mass2}).
\begin{center}
{\textbf{Tab\,3.} Masses of elements of the Seaborg table.}
\vspace{0.1cm}
{\renewcommand{\arraystretch}{1.0}
\begin{tabular}{|c|c|l|c|}\hline
$Z$ & Element     & Vector $|\nu,s^\prime,\lambda,\iota_\lambda,\kappa\rangle$ & Mass  \\ \hline\hline
121 & \textbf{Ubu}& $|5,-1/2,4,7/2,-7/2\rangle$ & 308,3181 \\
122 & \textbf{Ubb}& $|5,-1/2,4,7/2,-5/2\rangle$ & 309,2352\\
123 & \textbf{Ubt}& $|5,-1/2,4,7/2,-3/2\rangle$ & 310,6271\\
124 & \textbf{Ubq}& $|5,-1/2,4,7/2,-1/2\rangle$ & 313,8942\\
125 & \textbf{Ubp}& $|5,-1/2,4,7/2,1/2\rangle$ & 318,1057\\
126 & \textbf{Ubn}& $|5,-1/2,4,7/2,3/2\rangle$ & 321,3729\\
127 & \textbf{Ubs}& $|5,-1/2,4,7/2,5/2\rangle$ & 322,7647\\
128 & \textbf{Ubo}& $|5,-1/2,4,7/2,7/2\rangle$ & 323,6818\\
\hline
129 & \textbf{Ube}& $|5,-1/2,4,9/2,-9/2\rangle$ & 327,2491\\
130 & \textbf{Utn}& $|5,-1/2,4,9/2,-7/2\rangle$ & 331,1681\\
131 & \textbf{Utu}& $|5,-1/2,4,9/2,-5/2\rangle$ & 332,0852\\
132 & \textbf{Utb}& $|5,-1/2,4,9/2,-3/2\rangle$ & 333,4771\\
133 & \textbf{Utt}& $|5,-1/2,4,9/2,-1/2\rangle$ & 336,7421\\
134 & \textbf{Utq}& $|5,-1/2,4,9/2,1/2\rangle$ & 340,9557\\
135 & \textbf{Utp}& $|5,-1/2,4,9/2,3/2\rangle$ & 344,2229\\
136 & \textbf{Uth}& $|5,-1/2,4,9/2,5/2\rangle$ & 345,6147\\
137 & \textbf{Uts}& $|5,-1/2,4,9/2,7/2\rangle$ & 346,5318\\
138 & \textbf{Uto}& $|5,-1/2,4,9/2,9/2\rangle$ & 350,4551\\
\hline
\end{tabular}
}
\end{center}
\begin{center}
{\renewcommand{\arraystretch}{1.0}
\begin{tabular}{|c|c|l|c|}\hline
$Z$ & Element     & Vector $|\nu,s^\prime,\lambda,\iota_\lambda,\kappa\rangle$ & Mass  \\ \hline\hline
139 & \textbf{Ute}& $|5,-1/2,3,5/2,-5/2\rangle$ & 353,2352\\
140 & \textbf{Uqn}& $|5,-1/2,3,5/2,-3/2\rangle$ & 354,6271\\
141 & \textbf{Uqu}& $|5,-1/2,3,5/2,-1/2\rangle$ & 357,8942\\
142 & \textbf{Uqb}& $|5,-1/2,3,5/2,1/2\rangle$ & 362,1057\\
143 & \textbf{Uqt}& $|5,-1/2,3,5/2,3/2\rangle$ & 365,3729\\
144 & \textbf{Uqq}& $|5,-1/2,3,5/2,5/2\rangle$ & 366,7647\\
\hline
145 & \textbf{Uqp}& $|5,-1/2,3,7/2,-7/2\rangle$ &  369,8681\\
146 & \textbf{Uqh}& $|5,-1/2,3,7/2,-5/2\rangle$ &  370,7852\\
147 & \textbf{Uqs}& $|5,-1/2,3,7/2,-3/2\rangle$ &  372,1771\\
148 & \textbf{Uqo}& $|5,-1/2,3,7/2,-1/2\rangle$ &  375,4442\\
149 & \textbf{Uqe}& $|5,-1/2,3,7/2,1/2\rangle$ &  379,6557\\
150 & \textbf{Upn}& $|5,-1/2,3,7/2,3/2\rangle$ &  382,9229\\
151 & \textbf{Upu}& $|5,-1/2,3,7/2,5/2\rangle$ &  384,3147\\
152 & \textbf{Uqp}& $|5,-1/2,3,7/2,7/2\rangle$ &  385,2318\\
\hline
153 & \textbf{Upt}& $|5,-1/2,2,3/2,-3/2\rangle$ &  387,6271\\
154 & \textbf{Upq}& $|5,-1/2,2,3/2,-1/2\rangle$ &  390,8942\\
155 & \textbf{Upp}& $|5,-1/2,2,3/2,1/2\rangle$ &  395,1057\\
156 & \textbf{Uph}& $|5,-1/2,2,3/2,3/2\rangle$ &  398,3729\\
\hline
157 & \textbf{Ups}& $|5,-1/2,2,5/2,-5/2\rangle$ &  398,4852\\
158 & \textbf{Upo}& $|5,-1/2,2,5/2,-3/2\rangle$ &  399,8771\\
159 & \textbf{Upe}& $|5,-1/2,2,5/2,-1/2\rangle$ &  403,1442\\
160 & \textbf{Uhn}& $|5,-1/2,2,5/2,1/2\rangle$ &  407,3557\\
161 & \textbf{Uhu}& $|5,-1/2,2,5/2,3/2\rangle$ &  410,6229\\
162 & \textbf{Uhb}& $|5,-1/2,2,5/2,5/2\rangle$ &  412,0147\\
\hline
163 & \textbf{Uht}& $|5,-1/2,1,1/2,-1/2\rangle$ &  412,8942\\
164 & \textbf{Uhq}& $|5,-1/2,1,1/2,1/2\rangle$ &  417,1057\\
\hline
165 & \textbf{Uhp}& $|5,-1/2,1,3/2,-3/2\rangle$ &  416,5771\\
166 & \textbf{Uhn}& $|5,-1/2,1,3/2,-1/2\rangle$ &  419,8442\\
167 & \textbf{Uhs}& $|5,-1/2,1,3/2,1/2\rangle$ &  424,0557\\
168 & \textbf{Uho}& $|5,-1/2,1,3/2,3/2\rangle$ &  427,3229\\
\hline
169 & \textbf{Uhe}& $|5,-1/2,0,1/2,-1/2\rangle$ &  425,5452\\
170 & \textbf{Usn}& $|5,-1/2,0,1/2,1/2\rangle$ &  429,7557\\
\hline
171 & \textbf{Usu}& $|5,1/2,4,7/2,-7/2\rangle$ &  427,3181\\
172 & \textbf{Usb}& $|5,1/2,4,7/2,-5/2\rangle$ &  428,2352\\
173 & \textbf{Ust}& $|5,1/2,4,7/2,-3/2\rangle$ &  429,6271\\
174 & \textbf{Usq}& $|5,1/2,4,7/2,-1/2\rangle$ &  432,8942\\
175 & \textbf{Usp}& $|5,1/2,4,7/2,1/2\rangle$ &  437,1057\\
176 & \textbf{Ush}& $|5,1/2,4,7/2,3/2\rangle$ &  440,3729\\
177 & \textbf{Uss}& $|5,1/2,4,7/2,5/2\rangle$ &  441,7647\\
178 & \textbf{Uso}& $|5,1/2,4,7/2,7/2\rangle$ &  442,6818\\
\hline
179 & \textbf{Use}& $|5,1/2,4,9/2,-9/2\rangle$ &  446,2449\\
180 & \textbf{Uon}& $|5,1/2,4,9/2,-7/2\rangle$ &  450,1681\\
181 & \textbf{Uou}& $|5,1/2,4,9/2,-5/2\rangle$ &  451,0852\\
182 & \textbf{Uob}& $|5,1/2,4,9/2,-3/2\rangle$ &  452,4771\\
183 & \textbf{Uot}& $|5,1/2,4,9/2,-1/2\rangle$ &  455,7442\\
\hline
\end{tabular}
}
\end{center}
\begin{center}
{\renewcommand{\arraystretch}{1.0}
\begin{tabular}{|c|c|l|c|}\hline
$Z$ & Element     & Vector $|\nu,s^\prime,\lambda,\iota_\lambda,\kappa\rangle$ & Mass  \\ \hline\hline
184 & \textbf{Uoq}& $|5,1/2,4,9/2,1/2\rangle$ &  459,9557\\
185 & \textbf{Uop}& $|5,1/2,4,9/2,3/2\rangle$ &  463,2229\\
186 & \textbf{Uoh}& $|5,1/2,4,9/2,5/2\rangle$ &  464,6147\\
187 & \textbf{Uos}& $|5,1/2,4,9/2,7/2\rangle$ &  465,5318\\
188 & \textbf{Uoo}& $|5,1/2,4,9/2,9/2\rangle$ &  469,4551\\
\hline
189 & \textbf{Uoe}& $|5,1/2,3,5/2,-5/2\rangle$ &  472,2352\\
190 & \textbf{Uen}& $|5,1/2,3,5/2,-3/2\rangle$ &  473,6271\\
191 & \textbf{Ueu}& $|5,1/2,3,5/2,-1/2\rangle$ &  476,8942\\
192 & \textbf{Ueb}& $|5,1/2,3,5/2,1/2\rangle$ &  481,1057\\
193 & \textbf{Uet}& $|5,1/2,3,5/2,3/2\rangle$ &  484,3729\\
194 & \textbf{Ueq}& $|5,1/2,3,5/2,5/2\rangle$ &  485,7647\\
\hline
195 & \textbf{Uep}& $|5,1/2,3,7/2,-7/2\rangle$ &  488,8681\\
196 & \textbf{Ueh}& $|5,1/2,3,7/2,-5/2\rangle$ &  489,7852\\
197 & \textbf{Ues}& $|5,1/2,3,7/2,-3/2\rangle$ &  491,177\\
198 & \textbf{Ueo}& $|5,1/2,3,7/2,-1/2\rangle$ &  494,4442\\
199 & \textbf{Uee}& $|5,1/2,3,7/2,1/2\rangle$ &  498,6557\\
200 & \textbf{Bnn}& $|5,1/2,3,7/2,3/2\rangle$ &  501,9229\\
201 & \textbf{Bnu}& $|5,1/2,3,7/2,5/2\rangle$ &  503,3147\\
202 & \textbf{Bnb}& $|5,1/2,3,7/2,7/2\rangle$ &  504,2318\\
\hline
203 & \textbf{Bnt}& $|5,1/2,2,3/2,-3/2\rangle$ &  506,6271\\
204 & \textbf{Bnq}& $|5,1/2,2,3/2,-1/2\rangle$ &  509,8942\\
205 & \textbf{Bnp}& $|5,1/2,2,3/2,1/2\rangle$ &  514,1057\\
206 & \textbf{Bnh}& $|5,1/2,2,3/2,3/2\rangle$ &  517,3729\\
\hline
207 & \textbf{Bns}& $|5,1/2,2,5/2,-5/2\rangle$ &  517,4852\\
208 & \textbf{Bno}& $|5,1/2,2,5/2,-3/2\rangle$ &  518,8771\\
209 & \textbf{Bne}& $|5,1/2,2,5/2,-1/2\rangle$ &  522,1442\\
210 & \textbf{Bun}& $|5,1/2,2,5/2,1/2\rangle$ &  526,3557\\
211 & \textbf{Buu}& $|5,1/2,2,5/2,3/2\rangle$ &  529,6229\\
212 & \textbf{Bub}& $|5,1/2,2,5/2,5/2\rangle$ &  531,0147\\
\hline
213 & \textbf{But}& $|5,1/2,1,1/2,-1/2\rangle$ &  531,8942\\
214 & \textbf{Buq}& $|5,1/2,1,1/2,1/2\rangle$ &  536,1057\\
\hline
215 & \textbf{Bup}& $|5,1/2,1,3/2,-3/2\rangle$ &  535,5771\\
216 & \textbf{Buh}& $|5,1/2,1,3/2,-1/2\rangle$ &  538,8442\\
217 & \textbf{Bus}& $|5,1/2,1,3/2,1/2\rangle$ &  543,0557\\
218 & \textbf{Buo}& $|5,1/2,1,3/2,3/2\rangle$ &  546,3229\\
\hline
219 & \textbf{Bue}& $|5,1/2,0,1/2,-1/2\rangle$ &  544,5442\\
220 & \textbf{Bbn}& $|5,1/2,0,1/2,1/2\rangle$ &  548,7557\\
\hline
\end{tabular}
}
\end{center}
\section{10-periodic Extension}
Fig.\,4 shows a 10-periodic extension of the Mendeleev table in the form of the basic representation $F^+_{ss^\prime}$ of the Rumer-Fet group $G$ for the basis (\ref{Basis2}). The Mendeleev and Seaborg tables are separated by dotted frames with the symbols \textbf{M} and \textbf{S}, respectively. The first period of the Mendeleev table, including hydrogen \textbf{H} and helium \textbf{He}, corresponds to the simplest multiplet $(\nu=1,s^\prime=-1/2,\lambda=0,\iota_\lambda=1/2)$ of the group $G$. The second period consists of three multiplets: lithium \textbf{Li} and beryllium \textbf{Be} $(\nu=1,s^\prime=1/2,\lambda=0,\iota_\lambda=1/2)$, boron \textbf{B} and carbon \textbf{C} $(\nu=2,s^\prime=-1/2,\lambda=0,\iota_\lambda=1/2)$, elements \textbf{N}, \textbf{O}, \textbf{F}, \textbf{Ne} form a quadruplet $(\nu=2,s^\prime=-1/2,\lambda=1,\iota_\lambda=3/2)$.

\unitlength=1mm
\begin{center}
\begin{picture}(120,220)(7,-5)
\put(-2,204){$\scr\lambda=0$}
\put(2,209){$\scriptscriptstyle s^\prime:$}
\put(10,209){$\overbrace{\scriptscriptstyle -1/2\;\;\;\; 1/2}^{\nu=1}$}
\put(10,202){\framebox(5,5){}}\put(11,205){\tiny H}\put(11,203){\tiny He}
\put(18,202){\framebox(5,5){}}\put(19,205){\tiny Li}\put(19,203){\tiny Be}
\put(26,209){$\overbrace{\scriptscriptstyle -1/2\;\;\;\; 1/2}^{\nu=2}$}
\put(26,202){\framebox(5,5){}}\put(27,205){\tiny Na}\put(27,203){\tiny Mg}
\put(34,202){\framebox(5,5){}}\put(35,205){\tiny K}\put(35,203){\tiny Ca}
\put(42,209){$\overbrace{\scriptscriptstyle -1/2\;\;\;\; 1/2}^{\nu=3}$}
\put(42,202){\framebox(5,5){}}\put(43,205){\tiny Rb}\put(43,203){\tiny Sr}
\put(50,202){\framebox(5,5){}}\put(51,205){\tiny Cs}\put(51,203){\tiny Ba}
\put(58,209){$\overbrace{\scriptscriptstyle -1/2\;\;\;\; 1/2}^{\nu=4}$}
\put(58,202){\framebox(5,5){}}\put(59,205){\tiny Fr}\put(59,203){\tiny Ra}
\put(66,202){\framebox(5,5){}}\put(66.25,205){{\tiny Uue}}\put(66.25,203){{\tiny Ubn}}
\put(74,209){$\overbrace{\scriptscriptstyle -1/2\;\;\;\; 1/2}^{\nu=5}$}
\put(74,202){\framebox(5,5){}}\put(74.25,205){\tiny Uhe}\put(74.25,203){\tiny Usn}
\put(82,202){\framebox(5,5){}}\put(82.25,205){{\tiny Bue}}\put(82.25,203){{\tiny Bbn}}
\put(90,209){$\overbrace{\scriptscriptstyle -1/2\;\;\;\; 1/2}^{\nu=6}$}
\put(90,202){\framebox(5,5){}}\put(90.25,205){\tiny Beu}\put(90.25,203){\tiny Beb}
\put(98,202){\framebox(5,5){}}\put(98.25,205){{\tiny Tht}}\put(98.25,203){{\tiny Thq}}
\put(106,205.25){$\scriptscriptstyle \kappa=-1/2$}
\put(106,203){$\scriptscriptstyle \kappa=1/2$}
\put(114,203){$\left.\phantom{\framebox(1,3.15){}}\right\}$}
\put(118.5,203){$\scr\iota_\lambda=1/2$}
\put(-2,190.5){$\scr\lambda=1$}
\put(10,190.5){$\left\{\phantom{\framebox(1,8){}}\right.$}
\put(26,194){\framebox(5,5){}}\put(27,197){\tiny B}\put(27,195){\tiny C}
\put(34,194){\framebox(5,5){}}\put(35,197){\tiny Al}\put(35,195){\tiny Si}
\put(42,194){\framebox(5,5){}}\put(43,197){\tiny Ga}\put(43,195){\tiny Ge}
\put(50,194){\framebox(5,5){}}\put(51,197){\tiny In}\put(51,195){\tiny Sn}
\put(58,194){\framebox(5,5){}}\put(59,197){\tiny Tl}\put(59,195){\tiny Pb}
\put(66,194){\framebox(5,5){}}\put(67,197){{\tiny Nh}}\put(67,195){{\tiny Fl}}
\put(74,194){\framebox(5,5){}}\put(74.25,197){\tiny Uht}\put(74.25,195){\tiny Uhq}
\put(82,194){\framebox(5,5){}}\put(82.25,197){{\tiny But}}\put(82.25,195){{\tiny Buq}}
\put(90,194){\framebox(5,5){}}\put(90.25,197){\tiny Bop}\put(90.25,195){\tiny Boh}
\put(98,194){\framebox(5,5){}}\put(98.25,197){{\tiny Tps}}\put(98.25,195){{\tiny Tpo}}
\put(106,197.25){$\scriptscriptstyle \kappa=-1/2$}
\put(106,195){$\scriptscriptstyle \kappa=1/2$}
\put(114,195.75){$\left.\phantom{\framebox(1,3.15){}}\right\}$}
\put(118.5,196){$\scr\iota_\lambda=1/2$}
\put(26,189){\framebox(5,5){}}\put(27,192){\tiny N}\put(27,190){\tiny O}
\put(34,189){\framebox(5,5){}}\put(35,192){\tiny P}\put(35,190){\tiny S}
\put(42,189){\framebox(5,5){}}\put(43,192){\tiny As}\put(43,190){\tiny Se}
\put(50,189){\framebox(5,5){}}\put(51,192){\tiny Sb}\put(51,190){\tiny Te}
\put(58,189){\framebox(5,5){}}\put(59,192){\tiny Bi}\put(59,190){\tiny Po}
\put(66,189){\framebox(5,5){}}\put(67,192){{\tiny Mc}}\put(67,190){{\tiny Lv}}
\put(74,189){\framebox(5,5){}}\put(74.25,192){\tiny Uhp}\put(74.25,190){\tiny Uhn}
\put(82,189){\framebox(5,5){}}\put(82.25,192){{\tiny Bup}}\put(82.25,190){{\tiny Buh}}
\put(90,189){\framebox(5,5){}}\put(90.25,192){\tiny Bos}\put(90.25,190){\tiny Boo}
\put(98,189){\framebox(5,5){}}\put(98.25,192){{\tiny Tpe}}\put(98.25,190){{\tiny Thn}}
\put(106,192.25){$\scriptscriptstyle \kappa=-3/2$}
\put(106,190){$\scriptscriptstyle \kappa=-1/2$}
\put(26,184){\framebox(5,5){}}\put(27,187){\tiny F}\put(27,185){\tiny Ne}
\put(34,184){\framebox(5,5){}}\put(35,187){\tiny Cl}\put(35,185){\tiny Ar}
\put(42,184){\framebox(5,5){}}\put(43,187){\tiny Br}\put(43,185){\tiny Kr}
\put(50,184){\framebox(5,5){}}\put(51,187){\tiny I}\put(51,185){\tiny Xe}
\put(58,184){\framebox(5,5){}}\put(59,187){\tiny At}\put(59,185){\tiny Rn}
\put(66,184){\framebox(5,5){}}\put(67,187){{\tiny Ts}}\put(67,185){{\tiny Og}}
\put(74,184){\framebox(5,5){}}\put(74.25,187){\tiny Uhs}\put(74.25,185){\tiny Uho}
\put(82,184){\framebox(5,5){}}\put(82.25,187){{\tiny Bus}}\put(82.25,185){{\tiny Buo}}
\put(90,184){\framebox(5,5){}}\put(90.25,187){\tiny Boe}\put(90.25,185){\tiny Ben}
\put(98,184){\framebox(5,5){}}\put(98.25,187){{\tiny Thu}}\put(98.25,185){{\tiny Thb}}
\put(106,187.25){$\scriptscriptstyle \kappa=1/2$}
\put(106,185){$\scriptscriptstyle \kappa=3/2$}
\put(114,188){$\left.\phantom{\framebox(1,6.15){}}\right\}$}
\put(118.5,188){$\scr\iota_\lambda=3/2$}
\put(-2,167.5){$\scr\lambda=2$}
\put(10,167.5){$\left\{\phantom{\framebox(1,14){}}\right.$}
\put(42,176){\framebox(5,5){}}\put(43,179){\tiny Sc}\put(43,177){\tiny Ti}
\put(50,176){\framebox(5,5){}}\put(51,179){\tiny Y}\put(51,177){\tiny Zr}
\put(58,176){\framebox(5,5){}}\put(59,179){\tiny Lu}\put(59,177){\tiny Hf}
\put(66,176){\framebox(5,5){}}\put(67,179){\tiny Lr}\put(67,177){\tiny Rf}
\put(74,176){\framebox(5,5){}}\put(74.25,179){\tiny Upt}\put(74.25,177){\tiny Upq}
\put(82,176){\framebox(5,5){}}\put(82.25,179){{\tiny Bnt}}\put(82.25,177){{\tiny Bnq}}
\put(90,176){\framebox(5,5){}}\put(90.25,179){\tiny Bsp}\put(90.25,177){\tiny Bsh}
\put(98,176){\framebox(5,5){}}\put(98.25,179){{\tiny Tqs}}\put(98.25,177){{\tiny Tqo}}
\put(106,179.25){$\scriptscriptstyle \kappa=-3/2$}
\put(106,177){$\scriptscriptstyle \kappa=-1/2$}
\put(42,171){\framebox(5,5){}}\put(43,174){\tiny V}\put(43,172){\tiny Cr}
\put(50,171){\framebox(5,5){}}\put(51,174){\tiny Nb}\put(51,172){\tiny Mo}
\put(58,171){\framebox(5,5){}}\put(59,174){\tiny Ta}\put(59,172){\tiny W}
\put(66,171){\framebox(5,5){}}\put(67,174){{\tiny Db}}\put(67,172){{\tiny Sg}}
\put(74,171){\framebox(5,5){}}\put(74.25,174){\tiny Upp}\put(74.25,172){\tiny Uph}
\put(82,171){\framebox(5,5){}}\put(82.25,174){{\tiny Bnp}}\put(82.25,172){{\tiny Bnh}}
\put(90,171){\framebox(5,5){}}\put(90.25,174){\tiny Bss}\put(90.25,172){\tiny Bso}
\put(98,171){\framebox(5,5){}}\put(98.25,174){{\tiny Tqe}}\put(98.25,172){{\tiny Tpn}}
\put(106,174.25){$\scriptscriptstyle \kappa=1/2$}
\put(106,172){$\scriptscriptstyle \kappa=3/2$}
\put(114,175){$\left.\phantom{\framebox(1,6.15){}}\right\}$}
\put(118.5,175){$\scr\iota_\lambda=3/2$}
\put(42,166){\framebox(5,5){}}\put(43,169){\tiny Mn}\put(43,167){\tiny Fe}
\put(50,166){\framebox(5,5){}}\put(51,169){\tiny Tc}\put(51,167){\tiny Ru}
\put(58,166){\framebox(5,5){}}\put(59,169){\tiny Re}\put(59,167){\tiny Os}
\put(66,166){\framebox(5,5){}}\put(67,169){{\tiny Bh}}\put(67,167){{\tiny Hs}}
\put(74,166){\framebox(5,5){}}\put(74.25,169){\tiny Ups}\put(74.25,167){\tiny Upo}
\put(82,166){\framebox(5,5){}}\put(82.25,169){{\tiny Bns}}\put(82.25,167){{\tiny Bno}}
\put(90,166){\framebox(5,5){}}\put(90.25,169){\tiny Bse}\put(90.25,167){\tiny Bon}
\put(98,166){\framebox(5,5){}}\put(98.25,169){{\tiny Tpu}}\put(98.25,167){{\tiny Tpb}}
\put(106,169.25){$\scriptscriptstyle \kappa=-5/2$}
\put(106,167){$\scriptscriptstyle \kappa=-3/2$}
\put(42,161){\framebox(5,5){}}\put(43,164){\tiny Co}\put(43,162){\tiny Ni}
\put(50,161){\framebox(5,5){}}\put(51,164){\tiny Rh}\put(51,162){\tiny Pd}
\put(58,161){\framebox(5,5){}}\put(59,164){\tiny Ir}\put(59,162){\tiny Pt}
\put(66,161){\framebox(5,5){}}\put(67,164){{\tiny Mt}}\put(67,162){{\tiny Ds}}
\put(74,161){\framebox(5,5){}}\put(74.25,164){\tiny Upe}\put(74.25,162){\tiny Uhn}
\put(82,161){\framebox(5,5){}}\put(82.25,164){{\tiny Bne}}\put(82.25,162){{\tiny Bun}}
\put(90,161){\framebox(5,5){}}\put(90.25,164){\tiny Bou}\put(90.25,162){\tiny Bob}
\put(98,161){\framebox(5,5){}}\put(98.25,164){{\tiny Tpt}}\put(98.25,162){{\tiny Tpq}}
\put(106,164.25){$\scriptscriptstyle \kappa=-1/2$}
\put(106,162){$\scriptscriptstyle \kappa=1/2$}
\put(114,162.75){$\left.\phantom{\framebox(1,8.15){}}\right\}$}
\put(118.5,163){$\scr\iota_\lambda=5/2$}
\put(42,156){\framebox(5,5){}}\put(43,159){\tiny Cu}\put(43,157){\tiny Zn}
\put(50,156){\framebox(5,5){}}\put(51,159){\tiny Ag}\put(51,157){\tiny Cd}
\put(58,156){\framebox(5,5){}}\put(59,159){\tiny Au}\put(59,157){\tiny Hg}
\put(66,156){\framebox(5,5){}}\put(67,159){{\tiny Rg}}\put(67,157){{\tiny Cn}}
\put(74,156){\framebox(5,5){}}\put(74.25,159){\tiny Uhu}\put(74.25,157){\tiny Uhb}
\put(82,156){\framebox(5,5){}}\put(82.25,159){{\tiny Buu}}\put(82.25,157){{\tiny Bub}}
\put(90,156){\framebox(5,5){}}\put(90.25,159){\tiny Bot}\put(90.25,157){\tiny Boq}
\put(98,156){\framebox(5,5){}}\put(98.25,159){{\tiny Tpp}}\put(98.25,157){{\tiny Tph}}
\put(106,159.25){$\scriptscriptstyle \kappa=3/2$}
\put(106,157){$\scriptscriptstyle \kappa=5/2$}
\put(-2,134.5){$\scr\lambda=3$}
\put(10,134.5){$\left\{\phantom{\framebox(1,19){}}\right.$}
\put(58,148){\framebox(5,5){}}\put(59,151){\tiny La}\put(59,149){\tiny Ce}
\put(66,148){\framebox(5,5){}}\put(67,151){\tiny Ac}\put(67,149){\tiny Th}
\put(74,148){\framebox(5,5){}}\put(74.25,151){\tiny Ute}\put(74.25,149){\tiny Uqn}
\put(82,148){\framebox(5,5){}}\put(82.25,151){{\tiny Uoe}}\put(82.25,149){{\tiny Uen}}
\put(90,148){\framebox(5,5){}}\put(90.25,151){\tiny Bhu}\put(90.25,149){\tiny Bhb}
\put(98,148){\framebox(5,5){}}\put(98.25,151){{\tiny Ttt}}\put(98.25,149){{\tiny Ttq}}
\put(106,151.25){$\scriptscriptstyle \kappa=-5/2$}
\put(106,149){$\scriptscriptstyle \kappa=-3/2$}
\put(58,143){\framebox(5,5){}}\put(59,146){\tiny Pr}\put(59,144){\tiny Nd}
\put(66,143){\framebox(5,5){}}\put(67,146){\tiny Pa}\put(67,144){\tiny U}
\put(74,143){\framebox(5,5){}}\put(74.25,146){\tiny Uqu}\put(74.25,144){\tiny Uqb}
\put(82,143){\framebox(5,5){}}\put(82.25,146){{\tiny Ueu}}\put(82.25,144){{\tiny Ueb}}
\put(90,143){\framebox(5,5){}}\put(90.25,146){\tiny Bht}\put(90.25,144){\tiny Bhq}
\put(98,143){\framebox(5,5){}}\put(98.25,146){{\tiny Ttp}}\put(98.25,144){{\tiny Tth}}
\put(106,146.25){$\scriptscriptstyle \kappa=-1/2$}
\put(106,144){$\scriptscriptstyle \kappa=1/2$}
\put(114,144.75){$\left.\phantom{\framebox(1,8.15){}}\right\}$}
\put(118.5,145){$\scr\iota_\lambda=5/2$}
\put(58,138){\framebox(5,5){}}\put(59,141){\tiny Pm}\put(59,139){\tiny Sm}
\put(66,138){\framebox(5,5){}}\put(67,141){\tiny Np}\put(67,139){\tiny Pu}
\put(74,138){\framebox(5,5){}}\put(74.25,141){\tiny Uqt}\put(74.25,139){\tiny Uqq}
\put(82,138){\framebox(5,5){}}\put(82.25,141){{\tiny Uet}}\put(82.25,139){{\tiny Ueq}}
\put(90,138){\framebox(5,5){}}\put(90.25,141){\tiny Bhp}\put(90.25,139){\tiny Bhh}
\put(98,138){\framebox(5,5){}}\put(98.25,141){{\tiny Tts}}\put(98.25,139){{\tiny Tto}}
\put(106,141.25){$\scriptscriptstyle \kappa=3/2$}
\put(106,139){$\scriptscriptstyle \kappa=5/2$}
\put(58,133){\framebox(5,5){}}\put(59,136){\tiny Eu}\put(59,134){\tiny Gd}
\put(66,133){\framebox(5,5){}}\put(67,136){\tiny Am}\put(67,134){\tiny Cm}
\put(74,133){\framebox(5,5){}}\put(74.25,136){\tiny Uqp}\put(74.25,134){\tiny Uqh}
\put(82,133){\framebox(5,5){}}\put(82.25,136){{\tiny Uep}}\put(82.25,134){{\tiny Ueh}}
\put(90,133){\framebox(5,5){}}\put(90.25,136){\tiny Bhs}\put(90.25,134){\tiny Bho}
\put(98,133){\framebox(5,5){}}\put(98.25,136){{\tiny Tte}}\put(98.25,134){{\tiny Tqn}}
\put(106,136.25){$\scriptscriptstyle \kappa=-7/2$}
\put(106,134){$\scriptscriptstyle \kappa=-5/2$}
\put(58,128){\framebox(5,5){}}\put(59,131){\tiny Tb}\put(59,129){\tiny Dy}
\put(66,128){\framebox(5,5){}}\put(67,131){\tiny Bk}\put(67,129){\tiny Cf}
\put(74,128){\framebox(5,5){}}\put(74.25,131){\tiny Uqs}\put(74.25,129){\tiny Uqo}
\put(82,128){\framebox(5,5){}}\put(82.25,131){{\tiny Ues}}\put(82.25,129){{\tiny Ueo}}
\put(90,128){\framebox(5,5){}}\put(90.25,131){\tiny Bhe}\put(90.25,129){\tiny Bsn}
\put(98,128){\framebox(5,5){}}\put(98.25,131){{\tiny Tqu}}\put(98.25,129){{\tiny Tqb}}
\put(106,131.25){$\scriptscriptstyle \kappa=-3/2$}
\put(106,129){$\scriptscriptstyle \kappa=-1/2$}
\put(114,127){$\left.\phantom{\framebox(1,11.20){}}\right\}$}
\put(118.5,127){$\scr\iota_\lambda=7/2$}
\put(58,123){\framebox(5,5){}}\put(59,126){\tiny Ho}\put(59,124){\tiny Er}
\put(66,123){\framebox(5,5){}}\put(67,126){\tiny Es}\put(67,124){\tiny Fm}
\put(74,123){\framebox(5,5){}}\put(74.25,126){\tiny Uqe}\put(74.25,124){\tiny Upn}
\put(82,123){\framebox(5,5){}}\put(82.25,126){{\tiny Uee}}\put(82.25,124){{\tiny Bnn}}
\put(90,123){\framebox(5,5){}}\put(90.25,126){\tiny Bsu}\put(90.25,124){\tiny Bse}
\put(98,123){\framebox(5,5){}}\put(98.25,126){{\tiny Tqt}}\put(98.25,124){{\tiny Tqq}}
\put(106,126.25){$\scriptscriptstyle \kappa=1/2$}
\put(106,124){$\scriptscriptstyle \kappa=3/2$}
\put(58,118){\framebox(5,5){}}\put(59,121){\tiny Tm}\put(59,119){\tiny Yb}
\put(66,118){\framebox(5,5){}}\put(67,121){\tiny Md}\put(67,119){\tiny No}
\put(74,118){\framebox(5,5){}}\put(74.25,121){\tiny Upu}\put(74.25,119){\tiny Upb}
\put(82,118){\framebox(5,5){}}\put(82.25,121){{\tiny Bnu}}\put(82.25,119){{\tiny Bnb}}
\put(90,118){\framebox(5,5){}}\put(90.25,121){\tiny Bst}\put(90.25,119){\tiny Bsq}
\put(98,118){\framebox(5,5){}}\put(98.25,121){{\tiny Tqp}}\put(98.25,119){{\tiny Tqh}}
\put(8,117){\dashbox{1}(64.5,99)[b]{\bf M}}
\put(106,121.25){$\scriptscriptstyle \kappa=5/2$}
\put(106,119){$\scriptscriptstyle \kappa=7/2$}
\put(-2,91.5){$\scr\lambda=4$}
\put(10,91.5){$\left\{\phantom{\framebox(1,23){}}\right.$}
\put(74,110){\framebox(5,5){}}\put(74.25,113){\tiny Ubu}\put(74.25,111){\tiny Ubb}
\put(82,110){\framebox(5,5){}}\put(82.25,113){{\tiny Usu}}\put(82.25,111){{\tiny Usb}}
\put(90,110){\framebox(5,5){}}\put(90.25,113){\tiny Bqt}\put(90.25,111){\tiny Bqq}
\put(98,110){\framebox(5,5){}}\put(98.25,113){{\tiny Tup}}\put(98.25,111){{\tiny Tuh}}
\put(106,113.25){$\scriptscriptstyle \kappa=-7/2$}
\put(106,111){$\scriptscriptstyle \kappa=-5/2$}
\put(74,105){\framebox(5,5){}}\put(74.25,108){\tiny Ubt}\put(74.25,106){\tiny Ubq}
\put(82,105){\framebox(5,5){}}\put(82.25,108){{\tiny Ust}}\put(82.25,106){{\tiny Usq}}
\put(90,105){\framebox(5,5){}}\put(90.25,108){\tiny Bqp}\put(90.25,106){\tiny Bqh}
\put(98,105){\framebox(5,5){}}\put(98.25,108){{\tiny Tus}}\put(98.25,106){{\tiny Tuo}}
\put(106,108.25){$\scriptscriptstyle \kappa=-3/2$}
\put(106,106){$\scriptscriptstyle \kappa=-1/2$}
\put(114,104){$\left.\phantom{\framebox(1,11.20){}}\right\}$}
\put(118.5,104){$\scr\iota_\lambda=7/2$}
\put(74,100){\framebox(5,5){}}\put(74.25,103){\tiny Ubp}\put(74.25,101){\tiny Ubn}
\put(82,100){\framebox(5,5){}}\put(82.25,103){{\tiny Usp}}\put(82.25,101){{\tiny Ush}}
\put(90,100){\framebox(5,5){}}\put(90.25,103){\tiny Bqs}\put(90.25,101){\tiny Bqo}
\put(98,100){\framebox(5,5){}}\put(98.25,103){{\tiny Tue}}\put(98.25,101){{\tiny Tbn}}
\put(106,103.25){$\scriptscriptstyle \kappa=1/2$}
\put(106,101){$\scriptscriptstyle \kappa=3/2$}
\put(74,95){\framebox(5,5){}}\put(74.25,98){\tiny Ubs}\put(74.25,96){\tiny Ubo}
\put(82,95){\framebox(5,5){}}\put(82.25,98){{\tiny Uss}}\put(82.25,96){{\tiny Uso}}
\put(90,95){\framebox(5,5){}}\put(90.25,98){\tiny Bqe}\put(90.25,96){\tiny Bpn}
\put(98,95){\framebox(5,5){}}\put(98.25,98){{\tiny Tbu}}\put(98.25,96){{\tiny Tbb}}
\put(106,98.25){$\scriptscriptstyle \kappa=5/2$}
\put(106,96){$\scriptscriptstyle \kappa=7/2$}
\put(74,90){\framebox(5,5){}}\put(74.25,93){\tiny Ube}\put(74.25,91){\tiny Utn}
\put(82,90){\framebox(5,5){}}\put(82.25,93){{\tiny Use}}\put(82.25,91){{\tiny Uon}}
\put(90,90){\framebox(5,5){}}\put(90.25,93){\tiny Bpu}\put(90.25,91){\tiny Bpb}
\put(98,90){\framebox(5,5){}}\put(98.25,93){{\tiny Tbt}}\put(98.25,91){{\tiny Tbq}}
\put(106,93.25){$\scriptscriptstyle \kappa=-9/2$}
\put(106,91){$\scriptscriptstyle \kappa=-7/2$}
\put(74,85){\framebox(5,5){}}\put(74.25,88){\tiny Utu}\put(74.25,86){\tiny Utb}
\put(82,85){\framebox(5,5){}}\put(82.25,88){{\tiny Uou}}\put(82.25,86){{\tiny Uob}}
\put(90,85){\framebox(5,5){}}\put(90.25,88){\tiny Bpt}\put(90.25,86){\tiny Bpq}
\put(98,85){\framebox(5,5){}}\put(98.25,88){{\tiny Tbp}}\put(98.25,86){{\tiny Tbh}}
\put(106,88.25){$\scriptscriptstyle \kappa=-5/2$}
\put(106,86){$\scriptscriptstyle \kappa=-3/2$}
\put(74,80){\framebox(5,5){}}\put(74.25,83){\tiny Utt}\put(74.25,81){\tiny Utq}
\put(82,80){\framebox(5,5){}}\put(82.25,83){{\tiny Uot}}\put(82.25,81){{\tiny Uoq}}
\put(90,80){\framebox(5,5){}}\put(90.25,83){\tiny Bpp}\put(90.25,81){\tiny Bph}
\put(98,80){\framebox(5,5){}}\put(98.25,83){{\tiny Tbs}}\put(98.25,81){{\tiny Tbo}}
\put(106,83.25){$\scriptscriptstyle \kappa=-1/2$}
\put(106,81){$\scriptscriptstyle \kappa=1/2$}
\put(114,81.75){$\left.\phantom{\framebox(1,14){}}\right\}$}
\put(118.5,82){$\scr\iota_\lambda=9/2$}
\put(74,75){\framebox(5,5){}}\put(74.25,78){\tiny Utp}\put(74.25,76){\tiny Uth}
\put(82,75){\framebox(5,5){}}\put(82.25,78){{\tiny Uop}}\put(82.25,76){{\tiny Uoh}}
\put(90,75){\framebox(5,5){}}\put(90.25,78){\tiny Bps}\put(90.25,76){\tiny Bpo}
\put(98,75){\framebox(5,5){}}\put(98.25,78){{\tiny Tbe}}\put(98.25,76){{\tiny Ttn}}
\put(106,78.25){$\scriptscriptstyle \kappa=3/2$}
\put(106,76){$\scriptscriptstyle \kappa=5/2$}
\put(74,70){\framebox(5,5){}}\put(74.25,73){\tiny Uts}\put(74.25,71){\tiny Uto}
\put(82,70){\framebox(5,5){}}\put(82.25,73){{\tiny Uos}}\put(82.25,71){{\tiny Uoo}}
\put(90,70){\framebox(5,5){}}\put(90.25,73){\tiny Bpe}\put(90.25,71){\tiny Bhn}
\put(98,70){\framebox(5,5){}}\put(98.25,73){{\tiny Ttu}}\put(98.25,71){{\tiny Ttb}}
\put(6,69){\dashbox{1}(82.5,149)[b]{\bf S}}
\put(106,73.25){$\scriptscriptstyle \kappa=7/2$}
\put(106,71){$\scriptscriptstyle \kappa=9/2$}
\put(-2,38){$\scr\lambda=5$}
\put(10,38){$\left\{\phantom{\framebox(1,29){}}\right.$}
\put(90,62){\framebox(5,5){}}\put(90.25,65){\tiny Bbu}\put(90.25,63){\tiny Bbb}
\put(98,62){\framebox(5,5){}}\put(98.25,65){{\tiny Bet}}\put(98.25,63){{\tiny Beq}}
\put(106,65.25){$\scriptscriptstyle \kappa=-9/2$}
\put(106,63){$\scriptscriptstyle \kappa=-7/2$}
\put(90,57){\framebox(5,5){}}\put(90.25,60){\tiny Bbt}\put(90.25,58){\tiny Bbq}
\put(98,57){\framebox(5,5){}}\put(98.25,60){{\tiny Bep}}\put(98.25,58){{\tiny Beh}}
\put(106,60.25){$\scriptscriptstyle \kappa=-5/2$}
\put(106,58){$\scriptscriptstyle \kappa=-3/2$}
\put(90,52){\framebox(5,5){}}\put(90.25,55){\tiny Bbp}\put(90.25,53){\tiny Bbh}
\put(98,52){\framebox(5,5){}}\put(98.25,55){{\tiny Bes}}\put(98.25,53){{\tiny Beo}}
\put(106,55.25){$\scriptscriptstyle \kappa=-1/2$}
\put(106,53){$\scriptscriptstyle \kappa=1/2$}
\put(114,53.75){$\left.\phantom{\framebox(1,14){}}\right\}$}
\put(118.5,54){$\scr\iota_\lambda=9/2$}
\put(90,47){\framebox(5,5){}}\put(90.25,50){\tiny Bbs}\put(90.25,48){\tiny Bbo}
\put(98,47){\framebox(5,5){}}\put(98.25,50){{\tiny Bee}}\put(98.25,48){{\tiny Tnn}}
\put(106,50.25){$\scriptscriptstyle \kappa=3/2$}
\put(106,48){$\scriptscriptstyle \kappa=5/2$}
\put(90,42){\framebox(5,5){}}\put(90.25,45){\tiny Bbe}\put(90.25,43){\tiny Bth}
\put(98,42){\framebox(5,5){}}\put(98.25,45){{\tiny Tnu}}\put(98.25,43){{\tiny Tnb}}
\put(106,45.25){$\scriptscriptstyle \kappa=7/2$}
\put(106,43){$\scriptscriptstyle \kappa=9/2$}
\put(90,37){\framebox(5,5){}}\put(90.25,40){\tiny Btu}\put(90.25,38){\tiny Btb}
\put(98,37){\framebox(5,5){}}\put(98.25,40){{\tiny Tnt}}\put(98.25,38){{\tiny Tnq}}
\put(106,40.25){$\scriptscriptstyle \kappa=-11/2$}
\put(106,38){$\scriptscriptstyle \kappa=-9/2$}
\put(90,32){\framebox(5,5){}}\put(90.25,35){\tiny Btt}\put(90.25,33){\tiny Btq}
\put(98,32){\framebox(5,5){}}\put(98.25,35){{\tiny Tnp}}\put(98.25,33){{\tiny Tnh}}
\put(106,35.25){$\scriptscriptstyle \kappa=-7/2$}
\put(106,33){$\scriptscriptstyle \kappa=-5/2$}
\put(90,27){\framebox(5,5){}}\put(90.25,30){\tiny Btp}\put(90.25,28){\tiny Bth}
\put(98,27){\framebox(5,5){}}\put(98.25,30){{\tiny Tns}}\put(98.25,28){{\tiny Tno}}
\put(106,30.25){$\scriptscriptstyle \kappa=-3/2$}
\put(106,28){$\scriptscriptstyle \kappa=-1/2$}
\put(114,26){$\left.\phantom{\framebox(1,16.15){}}\right\}$}
\put(118.5,26){$\scr\iota_\lambda=11/2$}
\put(90,22){\framebox(5,5){}}\put(90.25,25){\tiny Bts}\put(90.25,23){\tiny Bto}
\put(98,22){\framebox(5,5){}}\put(98.25,25){{\tiny Tne}}\put(98.25,23){{\tiny Tun}}
\put(106,25.25){$\scriptscriptstyle \kappa=1/2$}
\put(106,23){$\scriptscriptstyle \kappa=3/2$}
\put(90,17){\framebox(5,5){}}\put(90.25,20){\tiny Bte}\put(90.25,18){\tiny Bqn}
\put(98,17){\framebox(5,5){}}\put(98.25,20){{\tiny Tuu}}\put(98.25,18){{\tiny Tub}}
\put(106,20.25){$\scriptscriptstyle \kappa=5/2$}
\put(106,18){$\scriptscriptstyle \kappa=7/2$}
\put(90,12){\framebox(5,5){}}\put(90.25,15){\tiny Bqu}\put(90.25,13){\tiny Bqb}
\put(98,12){\framebox(5,5){}}\put(98.25,15){{\tiny Tut}}\put(98.25,13){{\tiny Tuq}}
\put(106,15.25){$\scriptscriptstyle \kappa=9/2$}
\put(106,13){$\scriptscriptstyle \kappa=11/2$}
\put(10,-2){\begin{minipage}{25pc}{\small {\bf Fig.\,4.} 10-periodic extension of Mendeleev table in the form of the basic representation $F^+_{ss^\prime}$ of the Rumer-Fet group (basis $|\nu,s^\prime,\lambda,\iota_\lambda,\kappa\rangle$).}\end{minipage}}
\end{picture}
\end{center}

The third period consists of three multiplets also (two doublets and one quadruplet): doublet \textbf{Na} and \textbf{Mg} $(\nu=2,s^\prime=-1/2,\lambda=0,\iota_\lambda=1/2)$, doublet \textbf{Al} and \textbf{Si} $(\nu=2,s^\prime=1/2,\lambda=1,\iota_\lambda=1/2)$, quadruplet \textbf{P}, \textbf{S}, \textbf{Cl}, \textbf{Ar} $(\nu=2,s^\prime=1/2,\lambda=1,\iota_\lambda=3/2)$. The fourth period includes five multiplets: doublets \textbf{K}, \textbf{Ca} $(\nu=2,s^\prime=1/2,\lambda=0,\iota_\lambda=1/2)$ and \textbf{Ga}, \textbf{Ge} $(\nu=3,s^\prime=-1/2,\lambda=1,\iota_\lambda=1/2)$, quadruplets \textbf{As}, \textbf{Se}, \textbf{Br}, \textbf{Kr} $(\nu=3,s^\prime=-1/2,\lambda=1,\iota_\lambda=3/2)$ and \textbf{Sc}, \textbf{Ti}, \textbf{V}, \textbf{Cr} $(\nu=3,s^\prime=-1/2,\lambda=2,\iota_\lambda=3/2)$, and also sextet $(\nu=3,s^\prime=-1/2,\lambda=2,\iota_\lambda=5/2)$ formed by the elements \textbf{Mn}, $\ldots$, \textbf{Zn}. This sextet and quadruplet $(\nu=3,s^\prime=-1/2,\lambda=2,\iota_\lambda=3/2)$ form the first insertion decade (transitional elements). The fifth period has the analogous structure: doublets \textbf{Rb}, \textbf{Sr} $(\nu=3,s^\prime=-1/2,\lambda=0,\iota_\lambda=1/2)$, \textbf{In}, \textbf{Sn} $(\nu=3,s^\prime=1/2,\lambda=1,\iota_\lambda=1/2)$, quadruplet \textbf{Sb}, \textbf{Te}, \textbf{I}, \textbf{Xe} $(\nu=3,s^\prime=1/2,\lambda=1,\iota_\lambda=3/2)$, quadruplet \textbf{Y}, \textbf{Zr}, \textbf{Nb}, \textbf{Mo} $(\nu=3,s^\prime=1/2,\lambda=2,\iota_\lambda=3/2)$ and sextet \textbf{Tc}, $\ldots$, \textbf{Cd} $(\nu=3,s^\prime=1/2,\lambda=2,\iota_\lambda=5/2)$ (the second insertion decade). The sixth period consists of seven multiplets: doublets \textbf{Cs},\textbf{Ba} $(\nu=3,s^\prime=1/2,\lambda=0,\iota_\lambda=1/2)$ and \textbf{Tl}, \textbf{Pb} $(\nu=4,s^\prime=-1/2,\lambda=1,\iota_\lambda=1/2)$, quadruplets \textbf{Bi}, \textbf{Po}, \textbf{At}, \textbf{Rn} $(\nu=4,s^\prime=-1/2,\lambda=1,\iota_\lambda=3/2)$ and \textbf{Lu}, \textbf{Hf}, \textbf{Ta}, \textbf{W} $(\nu=4,s^\prime=-1/2,\lambda=2,\iota_\lambda=3/2)$,
sextets \textbf{Re}, $\ldots$, \textbf{Hg} $(\nu=4,s^\prime=-1/2,\lambda=2,\iota_\lambda=5/2)$ and \textbf{La}, $\ldots$, \textbf{Sm} $(\nu=4,s^\prime=-1/2,\lambda=3,\iota_\lambda=5/2)$, octet \textbf{Eu}, $\ldots$, \textbf{Yb} $(\nu=4,s^\prime=-1/2,\lambda=3,\iota_\lambda=7/2)$. The seventh period (the last period of the Mendeleev table) duplicates the structure of the sixth period: doublets \textbf{Fr}, \textbf{Ra} $(\nu=4,s^\prime=-1/2,\lambda=0,\iota_\lambda=1/2)$ and \textbf{Nh}, \textbf{Fl} $(\nu=4,s^\prime=1/2,\lambda=1,\iota_\lambda=1/2)$, quadruplets \textbf{Mc}, \textbf{Lv}, \textbf{Ts}, \textbf{Og} $(\nu=4,s^\prime=1/2,\lambda=1,\iota_\lambda=3/2)$ and \textbf{Lr}, \textbf{Rf}, \textbf{Db}, \textbf{Sg} $(\nu=4,s^\prime=1/2,\lambda=2,\iota_\lambda=3/2)$, sextets \textbf{Bh}, $\ldots$, \textbf{Cn} $(\nu=4,s^\prime=1/2,\lambda=2,\iota_\lambda=5/2)$ and \textbf{Ac}, $\ldots$, \textbf{Pu} $(\nu=4,s^\prime=1/2,\lambda=3,\iota_\lambda=5/2)$, octet \textbf{Am}, $\ldots$, \textbf{No} $(\nu=4,s^\prime=1/2,\lambda=3,\iota_\lambda=7/2)$. The eighth period\footnote{The domain of hypothetical (undiscovered) elements of the periodic system begins with the eighth period.}, forming an 8-periodic extension of the Mendeleev table (Seaborg table), consists of nine multiplets: doublets \textbf{Uue}, \textbf{Ubn} $(\nu=4,s^\prime=1/2,\lambda=0,\iota_\lambda=1/2)$ and \textbf{Uht}, \textbf{Uhq} $(\nu=5,s^\prime=-1/2,\lambda=1,\iota_\lambda=1/2)$, quadruplets \textbf{Uhp}, $\ldots$, \textbf{Uho} $(\nu=5,s^\prime=-1/2,\lambda=1,\iota_\lambda=3/2)$ and \textbf{Upt}, $\ldots$, \textbf{Uph} $(\nu=5,s^\prime=-1/2,\lambda=2,\iota_\lambda=3/2)$, sextets \textbf{Ups}, $\ldots$, \textbf{Uhb} $(\nu=5,s^\prime=-1/2,\lambda=2,\iota_\lambda=5/2)$ and \textbf{Ute}, $\ldots$, \textbf{Uqq} $(\nu=5,s^\prime=-1/2,\lambda=3,\iota_\lambda=5/2)$, octets \textbf{Uqp}, $\ldots$, \textbf{Upb} $(\nu=5,s^\prime=-1/2,\lambda=3,\iota_\lambda=7/2)$ and \textbf{Ubu}\footnote{According to Bohr model, filling of $g$-shell is started with the element \textbf{Ubu} (Unbiunium). An analogue of $g$-shell in the Rumer-Fet model is a family of multiplets with quantum number $\lambda=4$ of the group $G$.}, $\ldots$, \textbf{Ubo} $(\nu=5,s^\prime=-1/2,\lambda=4,\iota_\lambda=7/2)$, decuplet \textbf{Ube}, $\ldots$, \textbf{Uto} $(\nu=5,s^\prime=-1/2,\lambda=4,\iota_\lambda=9/2)$. The eighth period contains 50 elements. The ninth period, finishing Seaborg table, contains also nine multiplets: doublets \textbf{Uhe}, \textbf{Usn} $(\nu=5,s^\prime=-1/2,\lambda=0,\iota_\lambda=1/2)$ and \textbf{But}, \textbf{Buq} $(\nu=5,s^\prime=1/2,\lambda=1,\iota_\lambda=1/2)$, quadruplets \textbf{Bup}, $\ldots$, \textbf{Buo} $(\nu=5,s^\prime=1/2,\lambda=1,\iota_\lambda=3/2)$ and \textbf{Bnt}, $\ldots$, \textbf{Bnh} $(\nu=5,s^\prime=1/2,\lambda=2,\iota_\lambda=3/2)$, sextets \textbf{Bns}, $\ldots$, \textbf{Bub} $(\nu=5,s^\prime=1/2,\lambda=2,\iota_\lambda=5/2)$ and \textbf{Uoe}, $\ldots$, \textbf{Ueq} $(\nu=5,s^\prime=1/2,\lambda=3,\iota_\lambda=5/2)$, octets \textbf{Uep}, $\ldots$, \textbf{Bnb} $(\nu=5,s^\prime=1/2,\lambda=3,\iota_\lambda=7/2)$ and \textbf{Usu}, $\ldots$, \textbf{Uso} $(\nu=5,s^\prime=1/2,\lambda=4,\iota_\lambda=7/2)$, decuplet \textbf{Use}, $\ldots$, \textbf{Uoo} $(\nu=5,s^\prime=1/2,\lambda=4,\iota_\lambda=9/2)$. The construction of a family of multiplets with the quantum number $\lambda=5$ of the group $G$ is started with the tenth period (in the Bohr's model it corresponds to the formation of $h$-shell). The tenth period consists of 11 multiplets: doublets \textbf{Bue}, \textbf{Bbn} $(\nu=5,s^\prime=1/2,\lambda=0,\iota_\lambda=1/2)$ and \textbf{Bop}, \textbf{Boh} $(\nu=6,s^\prime=-1/2,\lambda=1,\iota_\lambda=1/2)$, quadruplets \textbf{Bos}, $\ldots$, \textbf{Ben} $(\nu=6,s^\prime=-1/2,\lambda=1,\iota_\lambda=3/2)$ and \textbf{Bsp}, $\ldots$, \textbf{Bso} $(\nu=6,s^\prime=-1/2,\lambda=2,\iota_\lambda=3/2)$, sextets \textbf{Bse}, $\ldots$, \textbf{Boq} $(\nu=6,s^\prime=-1/2,\lambda=2,\iota_\lambda=5/2)$ and \textbf{Bhu}, $\ldots$, \textbf{Bhh} $(\nu=6,s^\prime=-1/2,\lambda=3,\iota_\lambda=5/2)$, octets \textbf{Bhs}, $\ldots$, \textbf{Bsq} $(\nu=6,s^\prime=-1/2,\lambda=3,\iota_\lambda=7/2)$ and \textbf{Bqt}, $\ldots$, \textbf{Bpn} $(\nu=6,s^\prime=-1/2,\lambda=4,\iota_\lambda=7/2)$, decuplets \textbf{Bpu}, $\ldots$, \textbf{Bhn} $(\nu=6,s^\prime=-1/2,\lambda=4,\iota_\lambda=9/2)$ and \textbf{Bbu}, $\ldots$, \textbf{Bth} $(\nu=6,s^\prime=-1/2,\lambda=5,\iota_\lambda=9/2)$, 12-plet \textbf{Btu}, $\ldots$, \textbf{Bqb} $(\nu=6,s^\prime=-1/2,\lambda=5,\iota_\lambda=11/2)$. The eleventh period has the analogous structure: doublets \textbf{Beu}, \textbf{Beb} $(\nu=6,s^\prime=-1/2,\lambda=0,\iota_\lambda=1/2)$ and \textbf{Tps}, \textbf{Tpo} $(\nu=6,s^\prime=1/2,\lambda=1,\iota_\lambda=1/2)$, quadruplets \textbf{Tpe}, $\ldots$, \textbf{Thb} $(\nu=6,s^\prime=1/2,\lambda=1,\iota_\lambda=3/2)$ and \textbf{Tqs}, $\ldots$, \textbf{Tpn} $(\nu=6,s^\prime=1/2,\lambda=2,\iota_\lambda=3/2)$, sextets \textbf{Tpu}, $\ldots$, \textbf{Tph} $(\nu=6,s^\prime=1/2,\lambda=2,\iota_\lambda=5/2)$ and \textbf{Ttt}, $\ldots$, \textbf{Tto} $(\nu=6,s^\prime=1/2,\lambda=3,\iota_\lambda=5/2)$, octets \textbf{Tte}, $\ldots$, \textbf{Tqh} $(\nu=6,s^\prime=1/2,\lambda=3,\iota_\lambda=7/2)$ and \textbf{Tup}, $\ldots$, \textbf{Tbb} $(\nu=6,s^\prime=1/2,\lambda=4,\iota_\lambda=7/2)$, decuplets \textbf{Tbt}, $\ldots$, \textbf{Ttb} $(\nu=6,s^\prime=1/2,\lambda=4,\iota_\lambda=9/2)$ and \textbf{Bet}, $\ldots$, \textbf{Tnb} $(\nu=6,s^\prime=1/2,\lambda=5,\iota_\lambda=9/2)$, 12-plet \textbf{Tnt}, $\ldots$, \textbf{Tuq} $(\nu=6,s^\prime=1/2,\lambda=5,\iota_\lambda=11/2)$. The tenth and eleventh periods each contain 72 elements. The lengths of periods form the following number sequence:
\begin{equation}\label{Length}
2,\,8,\,8,\,18,\,18,\,32,\,32,\,50,\,50,\,72,\,72,\,\ldots
\end{equation}
The numbers of this sequence are defined by the famous Rydberg formula $2p^2$ ($p$ is an integer number). Rydberg series
\[
R=2(1^2+1^2+2^2+2^2+3^2+3^2+4^2+4^2+\ldots)
\]
contains a doubled first period, which is somewhat inconsistent with reality, that is, sequence (\ref{Length}).

Further, 12-th period begins with the elements \textbf{Tht} (Trihexitrium, $Z=363$) and \textbf{Thq} (Trihexiquadium, $Z=364$), forming a doublet $(\nu=6,s^\prime=1/2,\lambda=0,\iota_\lambda=1/2)$. This period, already beyond the table in Fig.\,4, contains 13 multiplets. The length of 12-th period is equal to 98 (in exact correspondence with the sequence (\ref{Length})). A new family of multiplets with quantum number $\lambda=6$ of group $G$ starts from the 12-th period. This family corresponds to the $i$-shell filling. The 13-th period has the analogous structure.

Obviously, as the quantum number $\nu$ increases, we will see new ``steps'' (doubled periods) and corresponding $\lambda$-families of multiplets (shells) in Fig.\,4.

\subsection{Masses of elements of 10-th and 11-th periods}
The table in Fig.\,4 corresponds to the reduction chain (\ref{Chain2}). Theoretical masses of elements of 10-th and 11-th periods, starting from $Z=221$ to $Z=364$, are calculated according to the mass formula (\ref{Mass2}) at the values $m_0=1$, $a=17$, $b=5,5$, $a^\prime=2,15$, $b^\prime=5,3$ (see Tab.\,4).

\begin{center}
{\textbf{Tab.\,4.} Masses of elements of 10-th and 11-th periods.}
\vspace{0.1cm}
{\renewcommand{\arraystretch}{1.0}
\begin{tabular}{|c|c|l|c|}\hline
$Z$ & Element     & Vector $|\nu,s^\prime,\lambda,\iota_\lambda,\kappa\rangle$ & Mass  \\ \hline\hline
221 & \textbf{Bbu}& $|6,-1/2,5,9/2,-9/2\rangle$ & 521,3949 \\
222 & \textbf{Bbb}& $|6,-1/2,5,9/2,-7/2\rangle$ & 525,3181\\
223 & \textbf{Bbt}& $|6,-1/2,5,9/2,-5/2\rangle$ & 526,2352\\
224 & \textbf{Bbq}& $|6,-1/2,5,9/2,-3/2\rangle$ & 527,6270\\
225 & \textbf{Bbp}& $|6,-1/2,5,9/2,-1/2\rangle$ & 530,8342\\
226 & \textbf{Bbh}& $|6,-1/2,5,9/2,1/2\rangle$ & 535,1057\\
227 & \textbf{Bbs}& $|6,-1/2,5,9/2,3/2\rangle$ & 538,3729\\
228 & \textbf{Bbo}& $|6,-1/2,5,9/2,5/2\rangle$ & 539,7647\\
229 & \textbf{Bbe}& $|6,-1/2,5,9/2,7/2\rangle$ & 540,6818\\
230 & \textbf{Btn}& $|6,-1/2,5,9/2,9/2\rangle$ & 544,6050\\
\hline
231 & \textbf{Btu}& $|6,-1/2,5,11/2,-11/2\rangle$ & 545,0151\\
232 & \textbf{Btb}& $|6,-1/2,5,11/2,-9/2\rangle$ & 549,5449\\
233 & \textbf{Btt}& $|6,-1/2,5,11/2,-7/2\rangle$ & 553,4681\\
234 & \textbf{Btq}& $|6,-1/2,5,11/2,-5/2\rangle$ & 554,3852\\
235 & \textbf{Btp}& $|6,-1/2,5,11/2,-3/2\rangle$ & 555,7770\\
236 & \textbf{Bth}& $|6,-1/2,5,11/2,-1/2\rangle$ & 559,0442\\
237 & \textbf{Bts}& $|6,-1/2,5,11/2,1/2\rangle$ & 563,2557\\
238 & \textbf{Bto}& $|6,-1/2,5,11/2,3/2\rangle$ & 566,5229\\
239 & \textbf{Bte}& $|6,-1/2,5,11/2,5/2\rangle$ & 567,9147\\
240 & \textbf{Bqn}& $|6,-1/2,5,11/2,7/2\rangle$ & 568,8318\\
\hline
\end{tabular}
}
\end{center}
\begin{center}
{\renewcommand{\arraystretch}{1.0}
\begin{tabular}{|c|c|l|c|}\hline
$Z$ & Element     & Vector $|\nu,s^\prime,\lambda,\iota_\lambda,\kappa\rangle$ & Mass  \\ \hline\hline
241 & \textbf{Bqu}& $|6,-1/2,5,11/2,9/2\rangle$ & 572,7556\\
242 & \textbf{Bqb}& $|6,-1/2,5,11/2,11/2\rangle$ & 582,2848\\
\hline
243 & \textbf{Bqt}& $|6,-1/2,4,7/2,-7/2\rangle$ & 580,3181\\
244 & \textbf{Bqq}& $|6,-1/2,4,7/2,-5/2\rangle$ & 581,2352\\
245 & \textbf{Bqp}& $|6,-1/2,4,7/2,-3/2\rangle$ & 582,6270\\
246 & \textbf{Bqh}& $|6,-1/2,4,7/2,-1/2\rangle$ & 585,8942\\
247 & \textbf{Bqs}& $|6,-1/2,4,7/2,1/2\rangle$ & 590,1057\\
248 & \textbf{Bqo}& $|6,-1/2,4,7/2,3/2\rangle$ & 593,3729\\
249 & \textbf{Bqe}& $|6,-1/2,4,7/2,5/2\rangle$ & 594,7647\\
250 & \textbf{Bpn}& $|6,-1/2,4,7/2,7/2\rangle$ & 595,6818\\
\hline
251 & \textbf{Bpu}& $|6,-1/2,4,9/2,-9/2\rangle$ & 599,2449\\
252 & \textbf{Bpb}& $|6,-1/2,4,9/2,-7/2\rangle$ & 603,1681\\
253 & \textbf{Bpt}& $|6,-1/2,4,9/2,-5/2\rangle$ & 604,0852\\
254 & \textbf{Bpq}& $|6,-1/2,4,9/2,-3/2\rangle$ & 605,4770\\
255 & \textbf{Bpp}& $|6,-1/2,4,9/2,-1/2\rangle$ & 608,7442\\
256 & \textbf{Bph}& $|6,-1/2,4,9/2,1/2\rangle$ & 612,9557\\
257 & \textbf{Bps}& $|6,-1/2,4,9/2,3/2\rangle$ & 616,2229\\
258 & \textbf{Bpo}& $|6,-1/2,4,9/2,5/2\rangle$ & 617,6147\\
259 & \textbf{Bpe}& $|6,-1/2,4,9/2,7/2\rangle$ & 618,5318\\
260 & \textbf{Bhn}& $|6,-1/2,4,9/2,9/2\rangle$ & 622,4550\\
\hline
261 & \textbf{Bhu}& $|6,-1/2,3,5/2,-5/2\rangle$ & 625,2352\\
262 & \textbf{Bhb}& $|6,-1/2,3,5/2,-3/2\rangle$ & 626,6270\\
263 & \textbf{Bht}& $|6,-1/2,3,5/2,-1/2\rangle$ & 629,8942\\
264 & \textbf{Bhq}& $|6,-1/2,3,5/2,1/2\rangle$ & 634,1057\\
265 & \textbf{Bhp}& $|6,-1/2,3,5/2,3/2\rangle$ & 637,3729\\
266 & \textbf{Bhh}& $|6,-1/2,3,5/2,5/2\rangle$ & 638,7647\\
\hline
267 & \textbf{Bhs}& $|6,-1/2,3,7/2,-7/2\rangle$ & 641,8681\\
268 & \textbf{Bho}& $|6,-1/2,3,7/2,-5/2\rangle$ & 642,7852\\
269 & \textbf{Bhe}& $|6,-1/2,3,7/2,-3/2\rangle$ & 644,1770\\
270 & \textbf{Bsn}& $|6,-1/2,3,7/2,-1/2\rangle$ & 647,4442\\
271 & \textbf{Bsu}& $|6,-1/2,3,7/2,1/2\rangle$ & 651,6557\\
272 & \textbf{Bsb}& $|6,-1/2,3,7/2,3/2\rangle$ & 654,9229\\
273 & \textbf{Bst}& $|6,-1/2,3,7/2,5/2\rangle$ & 656,3147\\
274 & \textbf{Bsq}& $|6,-1/2,3,7/2,7/2\rangle$ & 657,2318\\
\hline
275 & \textbf{Bsp}& $|6,-1/2,2,3/2,-3/2\rangle$ & 659,6270\\
276 & \textbf{Bsh}& $|6,-1/2,2,3/2,-1/2\rangle$ & 662,8942\\
277 & \textbf{Bss}& $|6,-1/2,2,3/2,1/2\rangle$ & 667,1057\\
278 & \textbf{Bso}& $|6,-1/2,2,3/2,3/2\rangle$ & 670,3729\\
\hline
279 & \textbf{Bse}& $|6,-1/2,2,5/2,-5/2\rangle$ & 670,4852\\
280 & \textbf{Bon}& $|6,-1/2,2,5/2,-3/2\rangle$ & 671,8770\\
281 & \textbf{Bou}& $|6,-1/2,2,5/2,-1/2\rangle$ & 675,1442\\
282 & \textbf{Bob}& $|6,-1/2,2,5/2,1/2\rangle$ & 679,3557\\
283 & \textbf{Bot}& $|6,-1/2,2,5/2,3/2\rangle$ & 682,6229\\
284 & \textbf{Boq}& $|6,-1/2,2,5/2,5/2\rangle$ & 684,0147\\
\hline
285 & \textbf{Bop}& $|6,-1/2,1,1/2,-1/2\rangle$ & 684,8942\\
286 & \textbf{Boh}& $|6,-1/2,1,1/2,1/2\rangle$ & 689,1057\\
\hline
\end{tabular}
}
\end{center}
\begin{center}
{\renewcommand{\arraystretch}{1.0}
\begin{tabular}{|c|c|l|c|}\hline
$Z$ & Element     & Vector $|\nu,s^\prime,\lambda,\iota_\lambda,\kappa\rangle$ & Mass  \\ \hline\hline
287 & \textbf{Bos}& $|6,-1/2,1,3/2,-3/2\rangle$ & 689,5770\\
288 & \textbf{Boo}& $|6,-1/2,1,3/2,-1/2\rangle$ & 691,8442\\
289 & \textbf{Boe}& $|6,-1/2,1,3/2,1/2\rangle$ & 696,0557\\
290 & \textbf{Ben}& $|6,-1/2,1,3/2,3/2\rangle$ & 699,3229\\
\hline
291 & \textbf{Beu}& $|6,-1/2,0,1/2,-1/2\rangle$ & 699,8942\\
292 & \textbf{Beb}& $|6,-1/2,0,1/2,1/2\rangle$ & 700,1037\\
\hline
293 & \textbf{Bet}& $|6,1/2,5,9/2,-9/2\rangle$ & 674,3949\\
294 & \textbf{Beq}& $|6,1/2,5,9/2,-7/2\rangle$ & 678,3181\\
295 & \textbf{Bep}& $|6,1/2,5,9/2,-5/2\rangle$ & 679,2352\\
296 & \textbf{Beh}& $|6,1/2,5,9/2,-3/2\rangle$ & 680,6270\\
297 & \textbf{Bes}& $|6,1/2,5,9/2,-1/2\rangle$ & 683,8942\\
298 & \textbf{Beo}& $|6,1/2,5,9/2,1/2\rangle$ & 688,1097\\
299 & \textbf{Bee}& $|6,1/2,5,9/2,3/2\rangle$ & 691,3729\\
300 & \textbf{Tnn}& $|6,1/2,5,9/2,5/2\rangle$ & 692,7647\\
301 & \textbf{Tnu}& $|6,1/2,5,9/2,7/2\rangle$ & 693,6818\\
302 & \textbf{Tnb}& $|6,1/2,5,9/2,9/2\rangle$ & 697,6050\\
303 & \textbf{Tnt}& $|6,1/2,5,11/2,-11/2\rangle$ & 693,0151\\
304 & \textbf{Tnq}& $|6,1/2,5,11/2,-9/2\rangle$ & 702,5449\\
305 & \textbf{Tnp}& $|6,1/2,5,11/2,-7/2\rangle$ & 706,4681\\
306 & \textbf{Tnh}& $|6,1/2,5,11/2,-5/2\rangle$ & 707,3852\\
307 & \textbf{Tns}& $|6,1/2,5,11/2,-3/2\rangle$ & 708,7770\\
308 & \textbf{Tno}& $|6,1/2,5,11/2,-1/2\rangle$ & 712,0442\\
309 & \textbf{Tne}& $|6,1/2,5,11/2,1/2\rangle$ & 716,2557\\
310 & \textbf{Tun}& $|6,1/2,5,11/2,3/2\rangle$ & 719,5229\\
311 & \textbf{Tuu}& $|6,1/2,5,11/2,5/2\rangle$ & 720,9147\\
312 & \textbf{Tub}& $|6,1/2,5,11/2,7/2\rangle$ & 721,8318\\
313 & \textbf{Tut}& $|6,1/2,5,11/2,9/2\rangle$ & 725,7550\\
314 & \textbf{Tuq}& $|6,1/2,5,11/2,11/2\rangle$ & 735,2848\\
\hline
315 & \textbf{Tup}& $|6,1/2,4,7/2,-7/2\rangle$ & 733,3181\\
316 & \textbf{Tuh}& $|6,1/2,4,7/2,-5/2\rangle$ & 734,2352\\
317 & \textbf{Tus}& $|6,1/2,4,7/2,-3/2\rangle$ & 735,6270\\
318 & \textbf{Tuo}& $|6,1/2,4,7/2,-1/2\rangle$ & 738,8942\\
319 & \textbf{Tue}& $|6,1/2,4,7/2,1/2\rangle$ & 743,1057\\
320 & \textbf{Tbn}& $|6,1/2,4,7/2,3/2\rangle$ & 746,3729\\
321 & \textbf{Tbu}& $|6,1/2,4,7/2,5/2\rangle$ & 747,7647\\
322 & \textbf{Tbb}& $|6,1/2,4,7/2,7/2\rangle$ & 748,6818\\
\hline
323 & \textbf{Tbt}& $|6,1/2,4,9/2,-9/2\rangle$ & 752,2449\\
324 & \textbf{Tbq}& $|6,1/2,4,9/2,-7/2\rangle$ & 756,1681\\
325 & \textbf{Tbp}& $|6,1/2,4,9/2,-5/2\rangle$ & 757,0852\\
326 & \textbf{Tbh}& $|6,1/2,4,9/2,-3/2\rangle$ & 758,4770\\
327 & \textbf{Tbs}& $|6,1/2,4,9/2,-1/2\rangle$ & 761,7442\\
328 & \textbf{Tbo}& $|6,1/2,4,9/2,1/2\rangle$ & 765,9557\\
329 & \textbf{Tbe}& $|6,1/2,4,9/2,3/2\rangle$ & 769,2229\\
330 & \textbf{Ttn}& $|6,1/2,4,9/2,5/2\rangle$ & 770,6147\\
331 & \textbf{Ttu}& $|6,1/2,4,9/2,7/2\rangle$ & 771,5318\\
332 & \textbf{Ttb}& $|6,1/2,4,9/2,9/2\rangle$ & 775,4550\\
\hline
\end{tabular}
}
\end{center}
\begin{center}
{\renewcommand{\arraystretch}{1.0}
\begin{tabular}{|c|c|l|c|}\hline
$Z$ & Element     & Vector $|\nu,s^\prime,\lambda,\iota_\lambda,\kappa\rangle$ & Mass  \\ \hline\hline
333 & \textbf{Ttt}& $|6,1/2,3,5/2,-5/2\rangle$ & 778,2352\\
334 & \textbf{Ttq}& $|6,1/2,3,5/2,-3/2\rangle$ & 779,6270\\
335 & \textbf{Ttp}& $|6,1/2,3,5/2,-1/2\rangle$ & 782,8942\\
336 & \textbf{Tth}& $|6,1/2,3,5/2,1/2\rangle$ & 787,1057\\
337 & \textbf{Tts}& $|6,1/2,3,5/2,3/2\rangle$ & 790,3729\\
338 & \textbf{Tto}& $|6,1/2,3,5/2,5/2\rangle$ & 791,7647\\
\hline
339 & \textbf{Tte}& $|6,1/2,3,7/2,-7/2\rangle$ & 794,8681\\
340 & \textbf{Tqn}& $|6,1/2,3,7/2,-5/2\rangle$ & 795,7852\\
341 & \textbf{Tqu}& $|6,1/2,3,7/2,-3/2\rangle$ & 797,1770\\
342 & \textbf{Tqb}& $|6,1/2,3,7/2,-1/2\rangle$ & 800,4442\\
343 & \textbf{Tqt}& $|6,1/2,3,7/2,1/2\rangle$ & 804,6557\\
344 & \textbf{Tqq}& $|6,1/2,3,7/2,3/2\rangle$ & 807,9229\\
345 & \textbf{Tqp}& $|6,1/2,3,7/2,5/2\rangle$ & 809,3147\\
346 & \textbf{Tqh}& $|6,1/2,3,7/2,7/2\rangle$ & 810,2318\\
\hline
347 & \textbf{Tqs}& $|6,1/2,2,3/2,-3/2\rangle$ & 812,6270\\
348 & \textbf{Tqo}& $|6,1/2,2,3/2,-1/2\rangle$ & 815,8942\\
349 & \textbf{Tqe}& $|6,1/2,2,3/2,1/2\rangle$ & 820,1057\\
350 & \textbf{Tpn}& $|6,1/2,2,3/2,3/2\rangle$ & 823,3729\\
\hline
351 & \textbf{Tpu}& $|6,1/2,2,5/2,-5/2\rangle$ & 823,4852\\
352 & \textbf{Tpb}& $|6,1/2,2,5/2,-3/2\rangle$ & 824,8770\\
353 & \textbf{Tpt}& $|6,1/2,2,5/2,-1/2\rangle$ & 828,1442\\
354 & \textbf{Tpq}& $|6,1/2,2,5/2,1/2\rangle$ & 832,3557\\
355 & \textbf{Tpp}& $|6,1/2,2,5/2,3/2\rangle$ & 835,6224\\
356 & \textbf{Tph}& $|6,1/2,2,5/2,5/2\rangle$ & 837,0147\\
\hline
357 & \textbf{Tps}& $|6,1/2,1,1/2,-1/2\rangle$ & 837,8942\\
358 & \textbf{Tpo}& $|6,1/2,1,1/2,1/2\rangle$ & 842,1057\\
\hline
359 & \textbf{Tpe}& $|6,1/2,1,3/2,-3/2\rangle$ & 841,5770\\
360 & \textbf{Thn}& $|6,1/2,1,3/2,-1/2\rangle$ & 844,8442\\
361 & \textbf{Thu}& $|6,1/2,1,3/2,1/2\rangle$ & 849,0557\\
362 & \textbf{Thb}& $|6,1/2,1,3/2,3/2\rangle$ & 852,3229\\
\hline
363 & \textbf{Tht}& $|6,1/2,0,1/2,-1/2\rangle$ & 848,8942\\
364 & \textbf{Thq}& $|6,1/2,0,1/2,1/2\rangle$ & 853,1057\\
\hline
\end{tabular}
}
\end{center}

\section{Homological Series}
All elements of the extended table, starting with hydrogen \textbf{H} ($Z=1$) and ending with \textbf{Thq} (Trihexiquadium, $Z=364$), form a single quantum system. The each element of the periodic system corresponds to the basis vector $|\nu,s^\prime,\lambda,\iota_\lambda,\kappa\rangle$, where $\nu$, $s^\prime$, $\lambda$, $\iota_\lambda$, $\kappa$ are quantum numbers of the symmetry group $G$ (Rumer-Fet group). Thus, we have the following set of state vectors:
\begin{eqnarray}
|\textbf{H}\rangle\phantom{bn}\hspace{0.25mm}&=&\left|1,-\frac{1}{2},0,\frac{1}{2},-\frac{1}{2}\right\rangle,\nonumber\\
|\textbf{He}\rangle\phantom{n}&=&\left|1,-\frac{1}{2},0,\frac{1}{2},\frac{1}{2}\right\rangle,\nonumber\\
|\textbf{Li}\rangle\phantom{bn}&=&\left|1,\frac{1}{2},0,\frac{1}{2},-\frac{1}{2}\right\rangle,\label{System}\\
&&\vdots\nonumber\\
|\textbf{Thq}\rangle&=&\left|6,\frac{1}{2},0,\frac{1}{2},\frac{1}{2}\right\rangle.\nonumber
\end{eqnarray}
In accordance with quantum mechanical laws, in the aggregate (\ref{System}), which forms a Hilbert space, we have linear superpositions of state vectors, as well as quantum transitions between different state vectors, that is, transitions between elements of the periodic system.

Let us now consider the operators that determine quantum transitions between the state vectors of the system (\ref{System}):
\begin{equation}\label{Operator1}
\boldsymbol{\Gamma}_+=\bsP_++\bsQ_+,\quad\boldsymbol{\Gamma}_-=\bsP_-+\bsQ_-.
\end{equation}
Operators (\ref{Operator1}) connect subspaces $\fF_n$ of the unitary representation $F^+$ of the conformal group $\SO(2,4)$ in the Fock space $\fF$. Indeed, the action of these operators on the basis vectors $|j,\sigma,\tau\rangle$ of $\fF$ has the form
\begin{multline}
\boldsymbol{\Gamma}_+|j,\sigma,\tau\rangle=i\sqrt{(j+\sigma+1)(j-\tau+1)}\left|j+\frac{1}{2},\sigma+\frac{1}{2},
\tau-\frac{1}{2}\right\rangle-\\
-i\sqrt{(j-\sigma+1)(j+\tau+1)}\left|j+\frac{1}{2},\sigma-\frac{1}{2},\tau+\frac{1}{2}\right\rangle,\nonumber
\end{multline}
\begin{multline}
\boldsymbol{\Gamma}_-|j,\sigma,\tau\rangle=-i\sqrt{(j+\sigma)(j-\tau)}\left|j-\frac{1}{2},\sigma-\frac{1}{2},
\tau+\frac{1}{2}\right\rangle+\\
+i\sqrt{(j-\sigma)(j+\tau)}\left|j-\frac{1}{2},\sigma+\frac{1}{2},\tau-\frac{1}{2}\right\rangle.\nonumber
\end{multline}
Hence it follows that $\boldsymbol{\Gamma}_+$ transforms vectors of the subspace $\fF_n$ into vectors of $\fF_{n+1}$, since for the Fock representation $\Phi_n=D_{\frac{n-1}{2},\frac{n-1}{2}}$ in the subspace $\fF_n$, where $j=\frac{n-1}{2}$, increasing the number $j$ by $1/2$ means increasing the number $n$ by 1 (see Appendix B). Analogously, the operator $\boldsymbol{\Gamma}_-$ transforms vectors of the subspace $\fF_n$ into vectors of $\fF_{n-1}$. Operators $\boldsymbol{\Gamma}_+$, $\boldsymbol{\Gamma}_-$ commute with the subgroup $G_2=\SO(3)\otimes\SU(2)$ belonging the group chain (\ref{Chain2}). Indeed, in virtue of commutation relations of the conformal group $\SO(2,4)$ (see section 3) it follows that
\[
\left[\bsP_\pm+\bsQ_\pm,\bsJ_++\bsK_+\right]=\left[\bsP_\pm+\bsQ_\pm,\bsJ_-+\bsK_-\right]=
\left[\bsP_\pm+\bsQ_\pm,\bsJ_3+\bsK_3\right]=0.
\]
Therefore, operators $\boldsymbol{\Gamma}_+$, $\boldsymbol{\Gamma}_-$ save quantum number $\mu$. Further, $\boldsymbol{\Gamma}_+$, $\boldsymbol{\Gamma}_-$ commute with a Casimir operator $(\bsJ_1+\bsK_1)^2+(\bsJ_2+\bsK_2)^2+(\bsJ_3+\bsK_3)^2$ of the subgroup $\SO(3)$ and thereby save quantum number $\lambda$. It is easy to see that $\boldsymbol{\Gamma}_+$, $\boldsymbol{\Gamma}_-$ commute with the operators $\boldsymbol{\tau}_k$ ($k=1,2,3$) of the subgroup $\SU(2)$ and, therefore, they save quantum number $s$. Since $\boldsymbol{\Gamma}_+$ and $\boldsymbol{\Gamma}_-$ commute with $\bsJ_k+\bsK_k$ and $\boldsymbol{\tau}_k$ separately, then they commute with the all subgroup $G_2=\SO(3)\otimes\SU(2)$. Further, the operators $\boldsymbol{\Gamma}_+$, $\boldsymbol{\Gamma}_-$ commute with the subgroup $\SU(2)^\prime$, which defines second ``doubling'', and, therefore, they save quantum number $s^\prime$. Since $\boldsymbol{\Gamma}_+$ transforms $\fF_n$ into $\fF_{n+1}$, and $\boldsymbol{\Gamma}_-$ transforms $\fF_n$ into $\fF_{n-1}$, then in the space $\fF^4=C(2)\otimes\fF^2=C(2)\otimes[C(2)\otimes\fF]$ of the representation $F^+_{ss^\prime}$ the operator $\boldsymbol{\Gamma}_+$, $\boldsymbol{\Gamma}_-$ raises, correspondingly lowers quantum number $\nu$ by 1. Thus, for the basis (\ref{RF4}) the operator $\boldsymbol{\Gamma}_+$ saves quantum numbers $s^\prime$, $\lambda$, $\mu$, $s$, raising $\nu$ by the unit, therefore, $\boldsymbol{\Gamma}_+\left|\nu,s^\prime,\lambda,\mu,s\right\rangle=
\eta|\nu+1,s^\prime,\lambda,\mu,s\rangle$, where $\eta\neq 0$. Analogously,  $\boldsymbol{\Gamma}_-\left|\nu,s^\prime,\lambda,\mu,s\right\rangle=
\eta^\prime|\nu-1,s^\prime,\lambda,\mu,s\rangle$, where $\eta^\prime\neq 0$. Since $\boldsymbol{\Gamma}_+$ (correspondingly $\boldsymbol{\Gamma}_-$) defines an isomorphic mapping of the space of $(\nu,s^\prime,\lambda)$ onto the space of $(\nu+1,s^\prime,\lambda)$ (corresp. $(\nu-1,s^\prime,\lambda)$), then $\eta$ (corresp. $\eta^\prime$) does not depend on quantum numbers $\mu$, $s$. Therefore, for the vectors $|\nu,s^\prime,\lambda,\iota_\lambda,\kappa\rangle$ of the basis (\ref{Basis2}) we have
\begin{eqnarray}
\boldsymbol{\Gamma}_+|\nu,s^\prime,\lambda,\iota_\lambda,\kappa\rangle&=&\eta
|\nu+1,s^\prime,\lambda,\iota_\lambda,\kappa\rangle,\label{Hom1}\\
\boldsymbol{\Gamma}_-|\nu,s^\prime,\lambda,\iota_\lambda,\kappa\rangle&=&\eta^\prime
|\nu-1,s^\prime,\lambda,\iota_\lambda,\kappa\rangle.\label{Hom2}
\end{eqnarray}
The equality (\ref{Hom2}) holds at $0\leq\lambda\leq\nu-2$. A visual sense of the operators $\boldsymbol{\Gamma}_+$, $\boldsymbol{\Gamma}_-$ is that they move basic vectors, represented by cells in Fig.\,4, to the right, correspondingly to the left through horizontal columns of the table. At this point, $\boldsymbol{\Gamma}_+$ always transfers basic vector of the column $(\nu,s^\prime)$ into basic vector of the same parity $(\nu+1,s^\prime)$ with multiplication by some non-null factor $\eta$. In turn, the operator $\boldsymbol{\Gamma}_-$ transfers basis vector of the column $(\nu,s^\prime)$ into basis vector of the same parity $(\nu-1,s^\prime)$ with multiplication by non-null factor $\eta^\prime$ when the column $(\nu-1,s^\prime)$ contains a vector on the same horizontal (otherwise we have zero).

Further, operators $\boldsymbol{\tau}^\prime_+=\boldsymbol{\tau}^\prime_1+i\boldsymbol{\tau}^\prime_2$, $\boldsymbol{\tau}^\prime_-=\boldsymbol{\tau}^\prime_1-i\boldsymbol{\tau}^\prime_2$ of the subgroup $\SU(2)^\prime$ also define quantum transitions between state vectors (\ref{System}). Since these operators commute with the subgroup $G_1=\SO(4)\otimes\SU(2)$, they save quantum numbers $\nu$, $\lambda$, $\iota_\lambda$, $\kappa$, related with $G_1$, and change only quantum number $s^\prime$:
\begin{eqnarray}
\boldsymbol{\tau}^\prime_+\left|\nu,-\tfrac{1}{2},\lambda,\iota_\lambda,\kappa\right\rangle\hspace{0.75mm}&=&
\left|\nu,\tfrac{1}{2},\lambda,\iota_\lambda,\kappa\right\rangle,\label{Hom3}\\
\boldsymbol{\tau}^\prime_-\left|\nu,\tfrac{1}{2},\lambda,\iota_\lambda,\kappa\right\rangle\phantom{-}&=&
\left|\nu,-\tfrac{1}{2},\lambda,\iota_\lambda,\kappa\right\rangle.\label{Hom4}
\end{eqnarray}
A visual sense of the operators $\boldsymbol{\tau}^\prime_+$, $\boldsymbol{\tau}^\prime_-$ is that $\boldsymbol{\tau}^\prime_+$ moves basis vectors of the each odd column (see Fig.\,4) on horizontal into basis vectors of neighboring right column; in turn, $\boldsymbol{\tau}^\prime_-$ moves basis vectors of the each even column on horizontal into basis vectors of neighboring left column. Thus, operators (\ref{Hom1})--(\ref{Hom4}) define quantum transitions between state vectors of the system (\ref{System}).

It is easy to see that on the horizontals of Fig.\,4 we have \textit{Mendeleev homological series}, that is, families of elements with similar properties. Therefore, operators (\ref{Hom1})--(\ref{Hom4}) define quantum transitions between elements of homological series. For example,
\begin{multline}
\boldsymbol{\Gamma}_+|\textbf{H}\rangle=\boldsymbol{\Gamma}_+\left|1,-\frac{1}{2},0,\frac{1}{2},-\frac{1}{2}\right\rangle=
\eta_1|\textbf{Na}\rangle=\eta_1\left|2,-\frac{1}{2},0,\frac{1}{2},-\tfrac{1}{2}\right\rangle\longmapsto\\
\eta_1\boldsymbol{\Gamma}_+|\textbf{Na}\rangle=\eta_1\eta_2|\textbf{Rb}\rangle=\eta_1\eta_2
\left|3,-\frac{1}{2},0,\frac{1}{2},-\frac{1}{2}\right\rangle\longmapsto\\
\eta_1\eta_2\boldsymbol{\Gamma}_+|\textbf{Rb}\rangle=\eta_1\eta_2\eta_3|\textbf{Fr}\rangle=\eta_1\eta_2\eta_3
\left|4,-\frac{1}{2},0,\frac{1}{2},-\frac{1}{2}\right\rangle\longmapsto\\
\eta_1\eta_2\eta_3\boldsymbol{\Gamma}_+|\textbf{Fr}\rangle=\eta_1\eta_2\eta_3\eta_4|\textbf{Uhe}\rangle=\eta_1\eta_2\eta_3\eta_4
\left|5,-\frac{1}{2},0,\frac{1}{2},-\frac{1}{2}\right\rangle.\nonumber
\end{multline}
Further, operators $\boldsymbol{\tau}^\prime_+$, $\boldsymbol{\tau}^\prime_-$ establish homology between lanthanides and actinides\footnote{This homology was first discovered by Seaborg. It is obvious that Seaborg homology is a particular case of the Mendeleev homology.}:
\[
\boldsymbol{\tau}^\prime_+|\textbf{La}\rangle=\boldsymbol{\tau}^\prime_+
\left|4,-\frac{1}{2},3,\frac{5}{2},-\frac{5}{2}\right\rangle=|\textbf{Ac}\rangle=
\left|4,\frac{1}{2},3,\frac{5}{2},-\frac{5}{2}\right\rangle,
\]
\[
\boldsymbol{\tau}^\prime_+|\textbf{Ce}\rangle=\boldsymbol{\tau}^\prime_+
\left|4,-\frac{1}{2},3,\frac{5}{2},-\frac{3}{2}\right\rangle=|\textbf{Th}\rangle=
\left|4,\frac{1}{2},3,\frac{5}{2},-\frac{3}{2}\right\rangle,
\]
\[
\vdots
\]
\[
\boldsymbol{\tau}^\prime_+|\textbf{Yb}\rangle=\boldsymbol{\tau}^\prime_+
\left|4,-\frac{1}{2},3,\frac{7}{2},\frac{7}{2}\right\rangle=|\textbf{No}\rangle=
\left|4,\frac{1}{2},3,\frac{7}{2},\frac{7}{2}\right\rangle,
\]
By means of operators $\boldsymbol{\Gamma}_+$, $\boldsymbol{\Gamma}_-$ we can continue the Seaborg homology to a superactinide group:
\[
\boldsymbol{\Gamma}_+|\textbf{La}\rangle=\boldsymbol{\Gamma}_+\left|4,-\frac{1}{2},3,\frac{5}{2},-\frac{5}{2}\right\rangle=
\eta|\textbf{Ute}\rangle=\eta\left|5,-\frac{1}{2},3,\frac{5}{2},-\frac{5}{2}\right\rangle,
\]
\[
\boldsymbol{\Gamma}_+|\textbf{Ce}\rangle=\boldsymbol{\Gamma}_+\left|4,-\frac{1}{2},3,\frac{5}{2},-\frac{3}{2}\right\rangle=
\eta|\textbf{Uqn}\rangle=\eta\left|5,-\frac{1}{2},3,\frac{5}{2},-\frac{3}{2}\right\rangle,
\]
\[
\vdots
\]
\[
\boldsymbol{\Gamma}_+|\textbf{Yb}\rangle=\boldsymbol{\Gamma}_+\left|4,-\frac{1}{2},3,\frac{7}{2},\frac{7}{2}\right\rangle=
\eta|\textbf{Upb}\rangle=\eta\left|5,-\frac{1}{2},3,\frac{7}{2},\frac{7}{2}\right\rangle.
\]
Correspondingly,
\[
\boldsymbol{\Gamma}_+|\textbf{Ac}\rangle=\boldsymbol{\Gamma}_+\left|4,\frac{1}{2},3,\frac{5}{2},-\frac{5}{2}\right\rangle=
\eta|\textbf{Uoe}\rangle=\eta\left|5,\frac{1}{2},3,\frac{5}{2},-\frac{5}{2}\right\rangle,
\]
\[
\boldsymbol{\Gamma}_+|\textbf{Th}\rangle=\boldsymbol{\Gamma}_+\left|4,\frac{1}{2},3,\frac{5}{2},-\frac{3}{2}\right\rangle=
\eta|\textbf{Uen}\rangle=\eta\left|5,\frac{1}{2},3,\frac{5}{2},-\frac{3}{2}\right\rangle,
\]
\[
\vdots
\]
\[
\boldsymbol{\Gamma}_+|\textbf{No}\rangle=\boldsymbol{\Gamma}_+\left|4,\frac{1}{2},3,\frac{7}{2},\frac{7}{2}\right\rangle=
\eta|\textbf{Bnb}\rangle=\eta\left|5,\frac{1}{2},3,\frac{7}{2},\frac{7}{2}\right\rangle.
\]

In conclusion of this paragraph we will say a few words about the \textit{principle of superposition} in relation to the system (\ref{System}). Apparently, the situation here is similar to the Wigner's \textit{superselection principle} \cite{Wigner} in particle physics, according to which not every superposition of physically possible states leads again to a physically possible state. The Wigner's principle limits (\textit{superselection rules}) the existence of superpositions of states. According to the superselection rules, superpositions of physically possible states exist only in the coherent subspaces of the physical Hilbert space. Thus, the problem of determining coherent subspaces for the system of states (\ref{System}) arises.

\section{Hypertwistors}
The Rumer-Fet group is constructed in many respects by analogy with the groups of internal (dynamic) symmetries, such as $\SU(3)$ and $\SU(6)$. Using quark model and $\SU(3)$-symmetry, we continue this analogy. As is known, quark is a vector of fundamental representation of the group $\SU(3)$. Let us define a vector of ``fundamental'' representation of the Rumer-Fet group.

The Rumer-Fet group
\[
\SO(2,4)\otimes\SU(2)\otimes\SU(2)^\prime
\]
is equivalent to
\[
\widetilde{\SO}(2,4)\otimes\SU(2)\simeq\SU(2,2)\otimes\SU(2),
\]
where $\SU(2,2)$ is a double covering of the conformal group (the group of psudounitary unimodular $4\times 4$ matrices). Further, in virtue of the isomorphism (\ref{ConfGroup2}) (see Appendix C) we will consider the double covering $\SU(2,2)$ as a \textit{spinor group}\footnote{The elements of group $\spin_+(2,4)$ are 15 bivectors $\e_i\e_j=\e_{ij}$, where $i,j=1,\ldots,6$. The explicit form of all fifteen generators leads through Cartan decomposition for group $\SU(2,2)$ to biquaternion angles, that is, to generalization of complex and quaternion angles for groups $\SL(2,\C)$ and $\Sp(1,1)$, where $\Sp(1,1)$ is a double covering of the de Sitter group \cite{Var04e,Var06,Var07}.}. Spintensor representations of the group $\spin_+(2,4)$ form a substratum of finite-dimensional representations $\boldsymbol{\tau}_{k/2,r/2}$, $\overline{\boldsymbol{\tau}}_{k/2,r/2}$ of the conformal group, which realized in the spaces $\Sym_{(k,r)}\subset\dS_{2^{k+r}}$ and $\overline{\Sym}_{(k,r)}\subset\overline{\dS}_{2^{k+r}}$, where $\dS_{2^{k+r}}$ is a spinspace. Twistor $\bsZ^\alpha=\left(\boldsymbol{s}^\alpha,\boldsymbol{s}_{\dot{\alpha}}\right)^\sT$ is a vector of fundamental representation of the group $\spin_+(2,4)$, where $\alpha,\dot{\alpha}=0,1$ and $\boldsymbol{s}^\alpha,\boldsymbol{s}_{\dot{\alpha}}$ are 2-component mutually conjugated spinors. Hence it immediately follows that a doubled twistor
\begin{equation}\label{Hyper}
\bsZ=\begin{bmatrix}
\bsZ_+\\
\bsZ_-
\end{bmatrix},
\end{equation}
or \textit{hypertwistor}, is a vector of fundamental representation of the group $\SU(2,2)\otimes\SU(2)$. Further, the twistor $\bsZ=[\boldsymbol{S},\overline{\boldsymbol{S}}]^\sT$ is a vector of general spintensor representation of the group $\spin_+(2,4)$, where $\boldsymbol{S}$ is a spintensor of the form (\ref{Spintensor}). Therefore, a general hypertwistor is defined by the expression of the form (\ref{Hyper}), where $\bsZ_+=[\boldsymbol{S},\overline{\boldsymbol{S}}]^\sT$, $\bsZ_-=[\overline{\boldsymbol{S}},\boldsymbol{S}]^\sT$.

Applying GNS-construction, we obtain vector states
\[
\omega_\Phi(H)=\frac{\langle\Phi\mid\pi(H)\Phi\rangle}{\langle\Phi\mid\Phi\rangle}=
\frac{\langle\Phi\mid F^+_{ss^\prime}(H)\Phi\rangle}{\langle\Phi\mid\Phi\rangle},
\]
where $H$ is an energy operator, $\left|\Phi\right\rangle$ is a cyclic vector of the Hilbert space $\sH_\infty$. A set of all pure states $\omega_\Phi(H)$ forms a \textit{physical Hilbert space} $\bsH_{\rm phys}=\bsH_8\otimes\bsH_\infty$\footnote{At the restriction of the group $G$ onto the Lorentz subgroup $\SO_0(1,3)$ and application of GNS-construction within double covering $\SL(2,\C)\simeq\spin_+(1,3)$, we obtain a  \textit{spinor} (vector of fundamental representation of the group $\spin_+(1,3)$), acting in a doubled Hilbert space $\bsH_2\otimes\bsH_\infty$ (Pauli space). Spinor is a particular case of hypertwistor.} and, correspondingly, a \textit{space of rays} $\hat{H}=\bsH_{\rm phys}/S^1$.

Further, with the aim of observance of electroneutrality and inclusion of discrete symmetries it is necessary to expand the double covering $\SU(2,2)\simeq\spin_+(2,4)$ up to an \textit{universal covering} $\pin(2,4)$. In general form (for arbitrary orthogonal groups) such extension has been given in the works \cite{Var01,Var04,Var05,Var15}. At this point, a pseudo-automorphism $\cA\rightarrow\overline{\cA}$ of the complex Clifford algebra $\C_n$ \cite{Ras55} plays a central role, where $\cA$ is an arbitrary element of the algebra $\C_n$. Since the real spinor structure appears as a result of reduction $\C_{2(k+r)}\rightarrow\cl_{p,q}$, then, as a consequence, the \textit{charge conjugation} $C$ (pseudo-automorphism $\cA\rightarrow\overline{\cA}$) for algebras $\cl_{p,q}$ over the real number field $\F=\R$ and the quaternion division ring $\K\simeq\BH$ (types $p-q\equiv 4,6\pmod{8}$) is reduced to a \textit{particle-antiparticle exchange} $C^\prime$. As is known, there are two classes of neutral particles: 1) particles that have antiparticles, such as neutrons, neutrinos, and so on; 2) particles that coincide with their antiparticles (for example, photons, $\pi^0$-mesons, and so on), that is, the so-called \textit{truly neutral particle}. The first class is described by neutral states $\omega^r_\Phi(H)$ with algebras $\cl_{p,q}$ over the field $\F=\R$ with the rings $\K\simeq\BH$ and $\K\simeq\BH\oplus\BH$ (types $p-q\equiv 4,6\pmod{8}$ and $p-q\equiv 5\pmod{8}$). To describe the second class of neutral particles we introduce \textit{truly neutral states} $\omega^{r_0}_\Phi(H)$ with algebras $\cl_{p,q}$ over the real number field $\F=\R$ and real division rings $\K\simeq\R$ and $\K\simeq\R\oplus\R$ (types $p-q\equiv 0,2\pmod{8}$ and $p-q\equiv 1\pmod{8}$). In the case of states $\omega^{r_0}_\Phi(H)$ the pseudo-automorphism $\cA\rightarrow\overline{\cA}$ is reduced to the identical transformation (the particle coincides with its antiparticle).

Following to \cite{Baez}, we define $\bsH_{\rm phys}=\bsH_8\otimes\bsH_\infty$ as a $\K$-Hilbert space, that is, as a space endowed with a $\ast$-ring structure, where $\ast$-ring is isomorphic to a division ring $\K=\R,\C,\BH$. Thus, the hypertwistor has a tensor structure (energy, mass) and a $\K$-linear structure (charge), and the connection of these two structures leads to a dynamic change in charge and mass.

\section*{Appendix A: Lorentz group and van der Waerden representation}
\setcounter{equation}{0}
\setcounter{section}{0}
\setcounter{subsection}{0}
\renewcommand{\thesubsection}{A.\arabic{subsection}}
\renewcommand{\theequation}{A.\arabic{equation}}

As is known, an universal covering of the proper orthochronous Lorentz group $\SO_0(1,3)$
(rotation group of the Minkowski space-time $\R^{1,3}$)
is the spinor group
\[\ar
\spin_+(1,3)\simeq\left\{\begin{pmatrix} \alpha & \beta \\ \gamma &
\delta
\end{pmatrix}\in\C_2:\;\;\det\begin{pmatrix}\alpha & \beta \\ \gamma & \delta
\end{pmatrix}=1\right\}=\SL(2,\C).
\]

Let $\fg\rightarrow T_{\fg}$ be an arbitrary linear
representation of the proper orthochronous Lorentz group
$\SO_0(1,3)$ and let $\sA_i(t)=T_{a_i(t)}$ be an infinitesimal
operator corresponding to the rotation $a_i(t)\in\SO_0(1,3)$.
Analogously, let $\sB_i(t)=T_{b_i(t)}$, where $b_i(t)\in\SO_0(1,3)$ is
the hyperbolic rotation. The elements $\sA_i$ and $\sB_i$ form a basis of the group algebra
$\mathfrak{sl}(2,\C)$ and satisfy the relations
\begin{equation}\label{Com1}
\left.\begin{array}{lll} \ld\sA_1,\sA_2\rd=\sA_3, &
\ld\sA_2,\sA_3\rd=\sA_1, &
\ld\sA_3,\sA_1\rd=\sA_2,\\[0.1cm]
\ld\sB_1,\sB_2\rd=-\sA_3, & \ld\sB_2,\sB_3\rd=-\sA_1, &
\ld\sB_3,\sB_1\rd=-\sA_2,\\[0.1cm]
\ld\sA_1,\sB_1\rd=0, & \ld\sA_2,\sB_2\rd=0, &
\ld\sA_3,\sB_3\rd=0,\\[0.1cm]
\ld\sA_1,\sB_2\rd=\sB_3, & \ld\sA_1,\sB_3\rd=-\sB_2, & \\[0.1cm]
\ld\sA_2,\sB_3\rd=\sB_1, & \ld\sA_2,\sB_1\rd=-\sB_3, & \\[0.1cm]
\ld\sA_3,\sB_1\rd=\sB_2, & \ld\sA_3,\sB_2\rd=-\sB_1. &
\end{array}\right\}
\end{equation}
Defining the operators
\begin{gather}
\sX_l=\frac{1}{2}i(\sA_l+i\sB_l),\quad\sY_l=\frac{1}{2}i(\sA_l-i\sB_l),
\label{SL25}\\
(l=1,2,3),\nonumber
\end{gather}
we come to a \textit{complex shell} of the group algebra $\mathfrak{sl}(2,\C)$.
Using the relations (\ref{Com1}), we find
\begin{equation}\label{Com2}
\ld\sX_k,\sX_l\rd=i\varepsilon_{klm}\sX_m,\quad
\ld\sY_l,\sY_m\rd=i\varepsilon_{lmn}\sY_n,\quad \ld\sX_l,\sY_m\rd=0.
\end{equation}
From the relations (\ref{Com2}) it follows that each of the sets of
infinitesimal operators $\sX$ and $\sY$ generates the group $\SU(2)$
and these two groups commute with each other. Thus, from the
relations (\ref{Com2}) it follows that the group algebra
$\mathfrak{sl}(2,\C)$ (within the complex shall) is algebraically isomorphic to the direct sum $\mathfrak{su}(2)\oplus\mathfrak{su}(2)$\footnote{In a sense, it allows one to represent the group
$\SL(2,\C)$ by a product $\SU(2)\otimes\SU(2)$ as it done by Ryder in his
textbook \cite{Ryd85}. Moreover, in the works \cite{AE93,Dvo96} the
Lorentz group is represented by a product $\SU_R(2)\otimes\SU_L(2)$,
where the spinors $\psi(p^\mu)=\begin{pmatrix}\phi_R(p^\mu)\\
\phi_L(p^\mu)\end{pmatrix}$ ($\phi_R(p^\mu)$ and $\phi_L(p^\mu)$ are
the right- and left-handed spinors) are transformed within
$(j,0)\oplus(0,j)$ representation space, in our case $j=l=\dot{l}$.
However, the isomorphism $\SL(2,C)\simeq\SU(2)\otimes\SU(2)$ is not correct from group-theoretical viewpoint. Indeed, the groups $\SO_0(1,3)$ and $\SO(4)$ are real forms of the complex 6-dimensional Lie group $\SO(4,\C)$ with complex Lie algebra $D_2=A_1+A_1$. Real Lie algebras are compact iff the Killing form is negative definite \cite{Knapp}. That is the case for Lie algebra of $\SO(4)$, not for $\SO_0(1,3)$.}.

Further, introducing operators of the form (`rising' and `lowering' operators of the group $\SL(2,\C)$)
\begin{equation}\label{SL26}
\left.\begin{array}{cc}
\sX_+=\sX_1+i\sX_2, & \sX_-=\sX_1-i\sX_2,\\[0.1cm]
\sY_+=\sY_1+i\sY_2, & \sY_-=\sY_1-i\sY_2,
\end{array}\right\}
\end{equation}
we see that
\[
\ld\sX_3,\sX_+\rd=\sX_+,\quad\ld\sX_3,\sX_-\rd=-\sX_-,\quad\ld\sX_+,\sX_-\rd=2\sX_3,
\]
\[
\ld\sY_3,\sY_+\rd=\sY_+,\quad\ld\sY_3,\sY_-\rd=-\sY_-,\quad\ld\sY_+,\sY_-\rd=2\sY_3.
\]
In virtue of commutativity of the relations (\ref{Com2}) a space of an irreducible finite-dimensional representation of the group $\SL(2,\C)$ can be spanned on the totality of
$(2l+1)(2\dot{l}+1)$ basis ket-vectors $|l,m;\dot{l},\dot{m}\rangle$ and basis bra-vectors
$\langle l,m;\dot{l},\dot{m}|$, where $l,m,\dot{l},\dot{m}$ are integer
or half-integer numbers, $-l\leq m\leq l$, $-\dot{l}\leq
\dot{m}\leq \dot{l}$. Therefore,
\begin{eqnarray}
&&\sX_-|l,m;\dot{l},\dot{m}\rangle= \sqrt{(l+m)(l-m+1)}|l,m-1;\dot{l},\dot{m}\rangle
\;\;(m>-l),\nonumber\\
&&\sX_+|l,m;\dot{l},\dot{m}\rangle= \sqrt{(l-m)(l+m+1)}|l,m+1;\dot{l},\dot{m}\rangle
\;\;(m<l),\nonumber\\
&&\sX_3|l,m;\dot{l},\dot{m}\rangle=
m|l,m;\dot{l},\dot{m}\rangle,\nonumber\\
&&\langle l,m;\dot{l},\dot{m}|\sY_-=
\langle l,m;\dot{l},\dot{m}-1|\sqrt{(\dot{l}+\dot{m})(\dot{l}-\dot{m}+1)}\;\;(\dot{m}>-\dot{l}),\nonumber\\
&&\langle l,m;\dot{l},\dot{m}|\sY_+=
\langle l,m;\dot{l},\dot{m}+1|\sqrt{(\dot{l}-\dot{m})(\dot{l}+\dot{m}+1)}\;\;(\dot{m}<\dot{l}),\nonumber\\
&&\langle l,m;\dot{l},\dot{m}|\sY_3= \langle l,m;\dot{l},\dot{m}|\dot{m}.\label{Waerden}
\end{eqnarray}
In contrast to the
Gelfand-Naimark representation
for the Lorentz group \cite{GMS,Nai58},
which does not find a wide application in physics,
a representation (\ref{Waerden}) is a most useful in theoretical physics
(see, for example, \cite{AB,Sch61,RF,Ryd85}). This representation for the
Lorentz group was first given by van der Waerden in his brilliant book
\cite{Wa32}.
It should be noted here that the representation basis, defined by the
formulae (\ref{SL25})--(\ref{Waerden}), has an evident physical meaning.
For example, in the case of $(1,0)\oplus(0,1)$-representation space
there is an analogy with the photon spin states. Namely, the operators
$\sX$ and $\sY$ correspond to the right and left polarization states of the
photon. For that reason we will call the canonical basis consisting of the
vectors $\mid lm;\dot{l}\dot{m}\rangle$ as
{\it a helicity basis}.

Thus, the complex shell of the group algebra $\mathfrak{sl}(2,\C)$, generating complex momentum, leads to a \textit{duality} which is mirrored in the appearance of the two spaces: a space of ket-vectors $|l,m;\dot{l},\dot{m}\rangle$ and a dual space of bra-vectors $\langle l,m;\dot{l},\dot{m}|$.

\section*{Appendix B: The group $\SO(4)$ and Fock representation}
\setcounter{equation}{0}
\setcounter{section}{0}
\setcounter{subsection}{0}
\renewcommand{\thesubsection}{B.\arabic{subsection}}
\renewcommand{\theequation}{B.\arabic{equation}}

As is known, the group $\SO(4)$ is a maximal compact subgroup of the conformal group $\SO_0(2,4)$. $\SO(4)$ corresponds to basis elements $\bsJ=(\sJ_1,\sJ_2,\sJ_3)$ and $\bsK=(\sK_1,\sK_2,\sK_3)$ of the algebra $\mathfrak{so}(2,4)$:
\[
\left[\sJ_k,\sJ_l\right]=i\varepsilon_{klm}\sJ_m,\quad\left[\sJ_k,\sK_l\right]=i\varepsilon_{klm}\sK_m,\quad
\left[\sK_k,\sK_l\right]=i\varepsilon_{klm}\sJ_m.
\]
Introducing linear combinations $\sV=(\sJ+\sK)/2$ and $\sV^\prime=(\sJ-\sK)/2$, we obtain
\[
\left[\sV_k,\sV_l\right]=i\varepsilon_{klm}\sV_m,\quad\left[\sV^\prime_k,\sV^\prime_l\right]=i\varepsilon_{klm}
\sV^\prime_m.
\]
Generators $\sV$ and $\sV^\prime$ form bases of the two independent algebras $\mathfrak{so}(3)$. It means that the group $\SO(4)$ is isomorphic to the product $\SO(3)\otimes\SO(3)$\footnote{Such state of affairs is explained by the following definition: the group $\SO(4)$ is \textit{locally} decomposed into a direct product of subgroups $\SO(3)$. \textit{In whole} (that is, without the supposition that all the matrices similar the unit matrix) this decomposition is ambiguous. $\SO(4)$ is an unique group (among the all orthogonal groups $\SO(n)$) which admits such local decomposition.}.

An universal covering of the rotation group $\SO(4)$ of the four-dimensional Euclidean space $\R^4$ is a spinor group
\[\ar
\spin(4)\simeq\left\{\begin{pmatrix} \alpha & \beta \\ \gamma &
\delta
\end{pmatrix}\in\BH\oplus\BH:\;\;\det\begin{pmatrix}\alpha & \beta \\ \gamma & \delta
\end{pmatrix}=1\right\}=\SU(2)\otimes\SU(2).
\]
Let $\SO(3)_\bsJ$ and $\SO(3)_\bsK$ be the subgroups of $\SO(4)$ with the generators $\sJ_k$ and $\sK_k$ ($k=1,2,3$), respectively. Then the each irreducible representation $T$ of the group $\SO(4)$ has the following structure: a space $\fR$ of the representation $T$ is a tensor product of spaces $\fR_1$ and $\fR_2$ in which we have irreducible representations $D_{j_1}$ and $D_{j_2}$ of the subgroups $\SO(3)_\bsJ$ and $\SO(3)_\bsK$ with dimensionality $2j_1+1$ and $2j_2+1$. Thus, a dimensionality of $T$ is equal to $(2j_1+1)(2j_2+1)$, where $j_1$ and $j_2$ are integer or half-integer numbers. An action of $\sJ_k$, $\sK_k$ on the basis vectors is defined by the formulas
\begin{eqnarray}
&&\sJ_-|\sigma,\tau\rangle= \sqrt{(j_1+\sigma)(j_1-\sigma+1)}|\sigma-1,\tau\rangle,\nonumber\\
&&\sJ_+|\sigma,\tau\rangle= \sqrt{(j_1-\sigma)(j_1+\sigma+1)}|\sigma+1,\tau\rangle,\nonumber\\
&&\sJ_3|\sigma,\tau\rangle=\sigma|\sigma,\tau\rangle,\nonumber\\
&&\sK_-|\sigma,\tau\rangle= \sqrt{(j_2+\tau)(j_2-\tau+1)}|\sigma,\tau-1\rangle,\nonumber\\
&&\sK_+|\sigma,\tau\rangle= \sqrt{(j_2-\tau)(j_2+\tau+1)}|\sigma,\tau+1\rangle,\nonumber\\
&&\sK_3|\sigma,\tau\rangle=\tau|\sigma,\tau\rangle.\label{Fock}
\end{eqnarray}
This representation of the group $\SO(4)$, denoted via $D_{j_1j_2}$, is irreducible and unitary.

A structure of the Fock representation $\Phi$ of $\SO(4)$ is defined by the decomposition onto irreducible components $\Phi_n$ in the spaces $\fF_n$. As is known \cite{Fet}, at the reduction on the subgroup $\SO(3)$ an irreducible representation $\Phi_n$ is decomposed into a sum of irreducible representations of $\SO(3)$ with dimensionality $1,\,3,\,\ldots,\,2n-1$. Taking into account the structure of irreducible representations of the group $\SO(4)$, we see that a smallest dimensionality 1 should be equal to $2|j_1-j_2|+1$, from which it follows that $j_1=j_2$. Therefore, the representation $\Phi_n$ has the form $D_{j_1j_2}$, and since for the biggest dimensionality should be $2n-1=2(j_1+j_2)+1=4j+1$, then $j=(n-1)/2$. Thus, for the Fock representation we have the following structure:
\[
\Phi=\Phi_1\oplus\Phi_2\oplus\ldots\oplus\Phi_n\oplus\ldots,\quad\text{где}\;\Phi_n=D_{\frac{n-1}{2},\frac{n-1}{2}}.
\]
It means that in the each space $\fF_n$ there is an orthonormal basis $|j,\sigma,\tau\rangle$, where $j=(n-1)/2$. All these bases form together an orthonormal basis of the Fock space $\fF$:
\begin{multline}
|j,\sigma,\tau\rangle\quad(j=0,1/2,1,3/2,\ldots;\;\\
\sigma=-j,-j+1,\ldots,j-1,j;\;\tau=-j,-j+1,\ldots,j-1,j),\label{Fock2}
\end{multline}
in which Lie algebra of the group $\SO(4)$ acts via the formulas (\ref{Fock}) with $j_1=j_2=j$.

\section*{Appendix C: Twistor structure and group $\SU(2,2)$}
\setcounter{equation}{0}
\setcounter{section}{0}
\setcounter{subsection}{0}
\renewcommand{\thesubsection}{C.\arabic{subsection}}
\renewcommand{\theequation}{C.\arabic{equation}}

The main idea of the Penrose twistor program \cite{Pen77,PM72} consists in the representation of classical space-time as some secondary construction obtained from more primary notions. As more primary notions we have here two-component (complex) spinors, moreover, the pairs of two-component spinors. In the Penrose program they called \textit{twistors}.

So, a twistor $\bsZ^\alpha$ is defined by the pair of two-component spinors: spinor $\boldsymbol{\omega}^s$ and covariant spinor $\boldsymbol{\pi}_{\dot{s}}$ from a conjugated space, that is,  $\bsZ^\alpha=\left(\boldsymbol{\omega}^s,\boldsymbol{\pi}_{\dot{s}}\right)$. In twistor theory momentum ($\vec{\omega}$) and impulse ($\vec{\pi}$) of the particle are constructed from the quantities $\boldsymbol{\omega}^s$ and $\boldsymbol{\pi}_{\dot{s}}$. One of the most important aspects of this theory is a \textit{transition from twistors to coordinate space-time}. Penrose described this transition by means of so-called \textit{basic relation of twistor theory}
\begin{equation}\label{Twistor}
\boldsymbol{\omega}^s=ix^{s\dot{r}}\boldsymbol{\pi}_{\dot{s}},
\end{equation}
where $x^{s\dot{r}}$ is a mixed spintensor of second rank. In more detailed record this relation has the form
\[
\begin{bmatrix}
\omega_1\\
\omega_2
\end{bmatrix}=\frac{i}{\sqrt{2}}\begin{bmatrix}
x^0+x^3 & x^1+ix^2\\
x^1+ix^2 & x^0-x^3
\end{bmatrix}\begin{bmatrix}
\pi_{\dot{1}}\\
\pi_{\dot{2}}
\end{bmatrix}.
\]
From (\ref{Twistor}) it immediately follows that points of space-time $\R^{1,3}$ are reconstructed over the twistor space $\C^4$ (these points correspond to linear subspaces of the twistor space $\C^4$). Therefore, points of $\R^{1,3}$ present secondary (derivative) construction with respect to twistors.

In fact, twistors can be defined as ``reduced spinors'' of pseudounitary group $\SO_0(2,4)$ acting in a six-dimensional space with the signature $(+,+,-,-,-,-)$. These reduced spinors are derived as follows. General spinors are elements of minimal left ideal of a \textit{conformal algebra} $\cl_{2,4}$:
\[
I_{2,4}=\cl_{2,4}f_{24}=\cl_{2,4}\frac{1}{2}(1+\e_{15})\frac{1}{2}(1+\e_{26}).
\]
Reduced spinors (twistors) are formulated within an even subalgebra $\cl^+_{2,4}\simeq\cl_{4,1}$ (\textit{de Sitter algebra}). The minimal left ideal of the algebra $\cl_{4,1}\simeq\C_4$ is defined by the following expression \cite{Var00}:
\[
I_{4,1}=\cl_{4,1}f_{4,1}=\cl_{4,1}\frac{1}{2}(1+\e_0)\frac{1}{2}
(1+i\e_{12}).
\]
Therefore, after reduction $I_{2,4}\rightarrow I_{4,1}$, generated by the isomorphism $\cl^+_{2,4}\simeq\cl_{4,1}$, we see that twistors $\bsZ^\alpha$ are elements of the ideal $I_{4,1}$ which leads to the group $\SU(2,2)\simeq\spin_+(2,4)\in\cl^+_{2,4}$ (see further (\ref{ConfGroup0}) and (\ref{ConfGroup})).
Indeed, let us consider the algebra $\cl_{2,4}$ associated with a six-dimensional pseudoeuclidean space $\R^{2,4}$. A double covering $\spin_+(2,4)$ of the rotation group $\SO_0(2,4)$ of the space $\R^{2,4}$ is described within the even subalgebra $\cl^+_{2,4}$. The algebra $\cl_{2,4}$ has the type $p-q\equiv 6\pmod{8}$, therefore, according to $\cl^+_{p,q}\simeq\cl_{q,p-1}$ we have $\cl^+_{2,4}\simeq\cl_{4,1}$, where $\cl_{4,1}$ is the de Sitter algebra associated with the space $\R^{4,1}$. In its turn, the algebra $\cl_{4,1}$ has the type $p-q\equiv 3\pmod{8}$ and, therefore, there is an isomorphism $\cl_{4,1}\simeq\C_4$, where $\C_4$ is a \textit{Dirac algebra}. The algebra $\C_4$ is a comlexification of the space-time algebra: $\C_4\simeq\C\otimes\cl_{1,3}$. Further, $\cl_{1,3}$ admits the following factorization: $\cl_{1,3}\simeq\cl_{1,1}\otimes\cl_{0,2}$. Hence it immediately follows that $\C_4\simeq\C\otimes\cl_{1,1}\otimes\cl_{0,2}$. Thus,
\begin{equation}\label{ConfGroup0}
\spin_+(2,4)=\left\{s\in\C\otimes\cl_{1,1}\otimes\cl_{0,2}\;|\;N(s)=1\right\}.
\end{equation}
On the other hand, in virtue of $\cl_{1,3}\simeq\cl_{1,1}\otimes\cl_{0,2}$ a general element of the algebra $\cl_{1,3}$ can be written in the form
\[
\cA_{\cl_{1,3}}=\cl^0_{1,1}\e_0+\cl^1_{1,1}\phi+\cl^2_{1,1}\psi+\cl^3_{1,1}\phi\psi,
\]
where $\phi=\e_{123}$, $\psi=\e_{124}$ are quaternion units. Therefore,
\begin{equation}\label{ConfGroup}
\spin_+(2,4)=
{\renewcommand{\arraystretch}{1.2}
\left\{s\in\left.\begin{bmatrix} \C\otimes\cl^0_{1,1}-i\C\otimes\cl^3_{1,1} &
-\C\otimes\cl^1_{1,1}+i\C\otimes\cl^2_{1,1}\\
\C\otimes\cl^1_{1,1}+i\C\otimes\cl^2_{1,1} & \C\otimes\cl^0_{1,1}+i\C\otimes\cl^3_{1,1}\end{bmatrix}\right|\;N(s)=1
\right\}.}
\end{equation}
Mappings of the space $\R^{1,3}$, generated by the group $\SO_0(2,4)$, induce linear transformations of the twistor space $\C^4$ with preservation of the form $\bsZ^\alpha\overline{\bsZ}_\alpha$ of the signature $(+,+,-,-)$. Hence it follows that a corresponding group in the twistor space is $\SU(2,2)$ (the group of pseudo-unitary unimodular $4\times 4$ matrices, see (\ref{ConfGroup})):
\begin{equation}\label{ConfGroup2}
\SU(2,2)=\left\{\ar\begin{bmatrix} A & B\\ C & D\end{bmatrix}
\in\C_4:\;\det\begin{bmatrix} A & B \\ C & D\end{bmatrix}=1
\right\}\simeq\spin_+(2,4).
\end{equation}

\end{document}